\documentclass[a4paper]{article}
\usepackage{amssymb}

\usepackage{graphicx}
\usepackage{amsmath}
\usepackage[widemargins]{a4}

\newtheorem{theorem}{Theorem}

\input{tcilatex}
\begin{document}

\author{J. H. Misguich$^{1}\thanks{%
E-mail : j.misguich@cea.fr}$, J.-D.\ Reuss$^{1}$, D.\ Constantinescu$^{2}$,
G.\ Steinbrecher$^{2}$, \and M.\ Vlad $^{3}$, F.\ Spineanu $^{3}$, B.\
Weyssow$^{4}$, R.\ Balescu$^{4}$ \\
$^{1}${\small Association Euratom-C.E.A. sur la Fusion, CEA/DSM/DRFC,
C.E.A.-Cadarache, }\\
{\small \ \ F-13108 Saint-Paul-lez-Durance, France.}\\
$^{2}${\small Association Euratom-N.A.S.T.I., Dept of Physics, University of
Craiova, }\\
{\small \ Str. A.I.\ Cuza No 13, Craiova-1100, Romania.}\\
$^{3}${\small Association Euratom-N.A.S.T.I., National Institute of Laser,
Plasma }\\
{\small and Radiation Physics, P.O.\ Box MG-36, Magurele, Bucharest, Romania.%
}\\
$^{4}${\small Association Euratom-Etat Belge sur la Fusion, Universit\'{e}
Libre de Bruxelles, }\\
{\small CP 231, Campus Plaine, Boulevard du Triomphe, B-1050 Bruxelles,
Belgium.}}
\title{Noble internal transport barriers and radial subdiffusion of toroidal
magnetic lines.}
\date{\textbf{Submitted (March 2002).}\\
}
\maketitle

\begin{abstract}

\medskip

Internal transport barriers (ITB's) observed in tokamaks are described by a
purely magnetic approach. Magnetic line motion in toroidal geometry with
broken magnetic surfaces is studied from a previously derived Hamiltonian
map in situation of incomplete chaos. This appears to reproduce in a
realistic way the main features of a tokamak, for a given safety factor
profile and in terms of a single parameter $L$ representing the amplitude of
the magnetic perturbation. New results are given concerning the Shafranov
shift as function of $L$. The phase space ($\psi ,\theta $) of the
''tokamap'' describes the poloidal section of the line trajectories, where $%
\psi $ is the toroidal flux labelling the surfaces. For small values of $L$,
closed magnetic surfaces exist (KAM tori) and island chains begin to appear
on rational surfaces for higher values of $L$, with chaotic zones around
hyperbolic points, as expected. Island remnants persist in the chaotic
domain for all relevant values of $L$ at the main rational q-values.

Single trajectories of magnetic line motion indicate the persistence of a
central protected plasma core, surrounded by a chaotic shell enclosed in a
double-sided transport barrier : the latter is identified as being composed
of two Cantori located on two successive ''most-noble'' numbers values of
the perturbed safety factor, and forming an internal transport barrier
(ITB). Magnetic lines which succeed to escape across this barrier begin to
wander in a wide chaotic sea extending up to a very robust barrier (as long
as $L\precsim 1)$ which is identified mathematically as a robust KAM surface
at the plasma edge.\ In this case the motion is shown to be intermittent,
with long stages of pseudo-trapping in the chaotic shell, or of sticking
around island remnants, as expected for a continuous time random walk.

For values of $L\succeq 1$, above the escape threshold, most magnetic lines
succeed to escape out of the external barrier which has become a permeable
Cantorus. Statistical analysis of a large number of trajectories,
representing the evolution of a bunch of magnetic lines, indicate that the
flux variable $\psi $ asymptotically grows in a diffusive manner as $%
(L^{2}t) $ with a $L^{2}$ scaling as expected, but that the average radial
position $r_{m}(t)$ asymptotically grows as $(L^{2}t)^{1/4}$ while the mean
square displacement around this average radius asymptotically grows in a
subdiffusive manner as $(L^{2}t)^{1/2}$. This result shows the slower
dispersion in the present incomplete chaotic regime, which is different from
the usual quasilinear diffusion in completely chaotic situations. For
physical times $t_{\varphi }$ of the order of the escape time defined by $%
x_{m}(t_{\varphi })\sim 1$, the motion appears to be superdiffusive,
however, but less dangerous than the generally admitted quasi-linear
diffusion. The orders of magnitude of the relevant times in Tore Supra are
finally discussed.

\end{abstract}

\tableofcontents
\listoffigures

\section{\label{Ch1}Introduction}

\bigskip

The ideal picture of perfect axisymmetric magnetic surfaces in a toroidal
magnetic confinement device like a tokamak, appears to be strongly modified
either in presence of field inhomogeneities (e.g. divertors) or of tearing
instabilities which result in the appearance of magnetic \emph{island chains}%
, with a helical symmetry around the magnetic surfaces. The careful
experimental analysis performed by N.J.\ Lopes Cardozo \emph{et al. }\cite%
{Lop-Cardozo}, with a very high spatial resolution on electron temperature $%
T_{e}$ and density $n_{e}$ radial profiles \cite{Lop-Cardozo2}, up to a few
times the width of an electron banana orbit, has shown that ''\emph{small
structures appear across the entire profile}, with large magnetic islands
occurring when the density disruption limit is approached'' as a result of
plasma filamentation \cite{Lop-Cardozo3}. He concludes that ''the structures
are interpreted as \emph{evidence that the magnetic topology in the tokamak
discharges is not the paradigmatic nest of perfect flux surfaces}, but more
complex than that''. Similar results have been obtained on the TJ-II
stellarator \cite{TJIIstellerator} where the very detailed structure of the $%
T_{e}$ profile measured along a chord has been observed with a $k^{-4}$
spectrum .

\bigskip

In presence of several island chains, it is well-known that overlapping may
occur \cite{Chir}, resulting in the appearance of \emph{chaotic zones} near
hyperbolic points, and even \emph{chaotic seas} with only \emph{island
remnants}. Equations describing magnetic lines in a torus are expressed in
terms of the toroidal coordinate $\zeta $ along the line \cite{RaduLiv88}.\
This variable $\zeta $ is usually interpreted in analogy with a ''time''
variable, so that the Hamiltonian equations for magnetic lines are
interpreted as ''equations of motion''. In a situation where island
overlapping occurs, the classical picture describes a given magnetic line as
''percolating'' through the plasma, wandering in the chaotic sea, remaining
almost trapped around island remnants \emph{(stickiness) }and possibly
reaching the plasma edge for increasing values of $\zeta $. In either case,
this radial motion of the perturbed magnetic lines is responsible for an
increased transport of particles and energy, since in the lowest
approximation (with vanishing Larmor radii and vanishing magnetic drifts)
charged particles just follow magnetic lines in the real time variable which
is $t\sim \zeta $ in this simple case. In the following description of
magnetic lines, we mainly describe their so-called ''motion'', for
simplicity, in spite of the fact that a magnetic line of course remains
static and is just followed in $\zeta $ along the toroidal direction.

\bigskip

It is known that a perturbed magnetic field configuration may sometimes be
associated with the appearance of \emph{\ transport barriers} (TB)
characterized by a jump in the slope of density or temperature profiles, and
a strong anomalous transport in the outside zone. TB\ could also appear in
simulations due a strong velocity shear flow (''zonal flows''), even without
magnetic turbulence \cite{Beyer2000}. In Ref. \cite{Lop-Cardozo2} it is
recalled that a strongly reduced ion thermal conductivity has been obtained
in various tokamaks, which remains at the neoclassical level over part or
even the entire plasma. This is due to the presence of an \emph{ion TB},
associated with a strong velocity shear and a reduction of the density
fluctuation level, which is not the case for \emph{electron TB} observed due
to the thin structures measured on $T_{e}$ \cite{Lop-Cardozo2}. Such
structures could be associated with alternative layers of good and bad
confinement, localized between local barriers. It is important to stress the
fact that transport barriers have been shown to be coupled to the safety
factor profile ($q$-profile) and to exist also in ohmic plasmas.

\medskip

Experiments in Tore-Supra, JT-60U, JET, TFTR and other tokamaks (see review
in Ref. \cite{TS1}) clearly exhibit the influence of the safety factor
profile ($q$-profile) on the appearance of \emph{internal transport barriers}
(ITB's) in tokamaks. For this reason we study here TB in a purely magnetic
description. Electron ITB has been maintained during $2$ $s.$ in Tore-Supra %
\cite{TS Hoang}. Generally ITB's are obtained in presence of a \emph{%
reversed magnetic shear}, \emph{i.e.} with a $q$ -profile which presents a
local maximum near the magnetic axis, and a minimum $q_{\min }$ typically at
a normalized radius of $0.3$ - $0.4$, and then a regular increase towards
the edge of the plasma. It is generally believed that ITB's appear around 
\emph{rational} $q$-values, and may even follow the time-evolution of a
given magnetic surface $q=2$ for instance in JET \cite{JET1}. Such ITB's may
have a finite width and one finds an ''ITB layer'' \cite{JT60U1}, defined as
a thin layer with large $T_{e}$ gradients \cite{Lop-Cardozo2} inside the
''ITB foot''.

\bigskip

On the other hand, even with a \emph{monotonic} $q$ -profile , a reduced
heat diffusivity $\chi $ has been observed in the core region of the plasma,
leading to the idea that \emph{ITB may appear even without reversed magnetic
shear} \cite{JET2}. This is the simple case we will mainly consider here.

\medskip

The experimental observation that ITB appear \emph{''near''} rational
surfaces may seem surprising from a theoretical point of view. Rational
surfaces are not densely covered by magnetic lines and thus are the most
sensitive ones to plasma instabilities. It is a generic property that
rational surfaces, on which field lines close back on themselves after a
finite number of toroidal turns, are topologically unstable \cite%
{RosSagTayZas1966}. Irrational surfaces, on the other hand, are covered by a
single magnetic line, in an ergodic way, and appear to be more resistant. In
the same way, the appearance of large scale chaotic motion is described in
dynamical systems theory by successive destructions of KAM\ tori, and their
transformation into permeable dense sets, called \emph{Cantori}. The most
resistant KAM torus in a chaotic system is of course of crucial importance
since it is the last inner barrier preventing large scale motion when the
stochasticity parameter is increased. In simple cases it corresponds to a
rotational transform (winding number) given by the ''most irrational''
number, the Golden number. From theoretical grounds, one can thus expect
that \emph{irrational surfaces} are more resistant to chaos \cite%
{RMacKayBook}\ and thus more likely to form a transport barrier, if any. In
either case, stability or chaos, the breakup of magnetic surfaces\emph{\
''is a problem of number theory''} \cite{Greene}, as will be verified here
again.

\medskip

In the present work we are mainly concerned with the location of ITB's in a
monotonic $q$-profile which appear naturally in a realistic model for
toroidal magnetic lines, called ''tokamap'' \cite{Tmap1}. The unusual point
we find here is that the magnetic perturbation, responsible for the
appearance of the magnetic island chains, not only creates a non-vanishing
Shafranov shift of the magnetic axis as expected, but also build a \emph{%
locally non-monotonic} $q$-profile in the equatorial plane, with a
spontaneous \emph{local maximum} $q_{\max }$ on the magnetic axis, and two
local minima. The positions of the ITB found here appear to correspond
rather exactly, in the perturbed $q$ -profile , to \emph{''most noble''}
values of $q$ which are the ''next most'' irrational numbers beyond the
Golden number \cite{Percival} \cite{Meiss 92} \footnote{%
It is well known that the continuous fraction expansion of the golden number
(the ''most irrational'' number) $G=(\sqrt{5}+1)/2$ $\ =1+1/\left[
1+1/(1+...)\right] =1.618033989...$ is $\left[ 1,1,1,1,1....\right] $. By
changing the first $1$ to the left into an integer $i>1$, one simply add
unities to the golden number : $G+(i-1)=\left[ i,1,1,1,1....\right] $. By
changing the second $1$ to the left into an integer $n>1$, one obtains the 
\emph{''most noble'' numbers }$N(i,n)=\left[ i,n,1,1,1....\right] $ which
are the next most irrational numbers beyond the Golden one.\ }. These will
appear to correspond to the $q-$position of the internal barriers of the
tokamap.

\bigskip

It has already been shown that the magnetic surface corresponding to a $q$%
-value given by the golden KAM torus is \emph{not} the most robust barrier
in the tokamap \cite{Tmap1}. In other systems too, the golden mean is \emph{%
not} found to be associated with the last KAM curve and the transition to
global stochasticity \cite{LichtenLieberman1983}, \cite{EscandeDoveil1981}.
Most noble values of $q$ are however the location where we may expect, from
KAM theory \cite{KAM} that the most robust tori are finally destroyed and
changed into hardly permeable Cantori, which are thus good candidates to be
identified with ITB's. That is what we will check to occur in the tokamap.
This result does not fully agree with the generally admitted idea that ITB's
in tokamaks would always be associated with rational $q$-values, but we have
to note that the two Cantori forming the barrier found below are
nevertheless observed on both side of a low order rational.

\bigskip

Very interesting theoretical models based on transport across chaotic layers
and internal barriers have been proposed to explain precise measurements
performed on the RTP\ tokamak. A model of radial transport in a series of
chaotic layers has been developed \cite{De Rover PhD} where the standard
magnetic equilibrium, with monotonously increasing $q-$profile, is perturbed
by small closed \emph{current filaments}: a number of filaments are
localized on low order rational $q-$values, with suitably chosen values for
their finite width and current. These current filaments break the topology
of nested flux surfaces and are of course responsible for the appearance of 
\emph{magnetic islands and chaotic regions}. Test particle transport is
computed and is found to be subdiffusive in such perturbed magnetic field,
with a mean square displacement growing like the toroidal angle to the power 
$\frac{1}{2}$\ \cite{deRover99}.

\bigskip

A model for inhomogeneous heat transport in a stratified plasma has also
been developed. \emph{Electron heat transport} might of course be locally
enhanced across each chain of magnetic islands (corresponding to rational $q$%
-value) and this could cause the appearance of the plateaux observed in the
temperature profile \cite{Lop-Cardozo} at rational $q$-values, explaining
some jumps in the slope of the temperature profile. Such stochastic zones
around low order rational chains, could also be limited by permeable
Cantori. A simple analysis of heat transport in an inhomogeneous stratified
medium shows that the measured values may deviate dramatically from simple
linear averages \cite{Lop-Cardozo}: the global transport description should
take into account ''insulating'' regions but can ignore ''turbulent''
regions of high diffusivity. In such models for electron heat transport \cite%
{Hogeweij 1998} \cite{Schilham PhD 2001}, a number of transport barriers,
with suitably chosen width and local heat conductivity, are assumed to be
localized on surfaces with low order rational $q-$values.\ As a result the
large electron heat conductivity, assumed to be constant in the conductivity
zone, presents a series of depletions (\emph{''}$q$\emph{-comb model''}).\
This model succeeds to reproduce the changes observed in the temperature
profile when scanning the deposition radius $\rho _{dep}$\ of electron
cyclotron heating (ECH) from \ central to far off-axis deposition.\ The main
feature of these measurements \cite{Mantica 2000} is the discontinuous
response of the $T_{e}$ profile to a continuous variation of the deposition
radius $\rho _{dep}$: five plateaux, in which $T_{e}$ is rather insensitive
to changes in $\rho _{dep}$, are separated by sudden transitions occurring
for small changes of $\rho _{dep}$. \ 

\bigskip

In all these models, the perturbations are assumed to be localized around
low-order rational $q-$values, and even if several parameters have to be
adjusted, the assumptions of these\ successful models are fully compatible %
\cite{Schilham PhD 2001}\ with the result obtained here, i.e. the magnetic
structure of an ITB as being composed of two noble Cantori.

\bigskip

In Section (\ref{Map}) we present this simple Hamiltonian twist map
(''tokaMAP'' \cite{Tmap1}) which has been proved to describe toroidal
magnetic lines in a realistic way for tokamaks. Its derivation is
summarized. We restrict ourselves mainly to the case of a monotonic $q$
-profile in the unperturbed configuration. We derive some results concerning
the localization of the \emph{fixed points}, and determine the Shafranov
shift. The \emph{bifurcations} determining the number of fixed points are
recalled, with an interesting example related to Kadomtsev's mechanism of
sawtooth instability.

\bigskip

Individual magnetic lines are calculated in Section (\ref{IndiviDUAL}) for
very large numbers of iterations, and a threshold region of the
stochasticity parameter $L$ is found, above which most lines from the
central region actually reach the edge of the plasma (\emph{global internal
chaos}). The motion of a single magnetic line is found to be intermittent,
with very long periods of trapping in different regions. The motion can
indeed be localized between some layers separated by \emph{Cantori }which
correspond rather precisely to \emph{noble numbers} in the profile of the
perturbed $q$-values around the magnetic axis. The spatial localization of
these barriers is discussed in Section (\ref{LocalisationITB}).\textbf{\ }We
also present in Section (\ref{Dana})\ the calculation of the flux through
the Cantori barriers. A Cantorus is a fractal set of points, of fractal
dimension zero in the poloidal plane \cite{DimCantor} (or of dimension $1$
in the torus: a single magnetic line).\ Cantori are known to generally
represent local permeable barriers in dynamical systems.

\bigskip

In Section (\ref{AsymptRadial}) we introduce an set of magnetic lines
starting from a very small region (or a constant initial radius) and perform
averages over this ensemble of lines. Iterations of the tokamap are
interpreted in terms of ''time evolution'' describing the toroidal motion of
a magnetic line. The average radial motion of the lines is described by
calculating the ''time''- dependent \emph{average radius} and average
poloidal flux 
\begin{equation}
r_{m}(t)\equiv <r(t)>\text{ \ \ \ \ , \ \ \ }\psi _{m}(t)=<\psi (t)>\text{\
\ \ }  \label{rm(t)}
\end{equation}%
reached by the lines at each time, along with the mean square deviation of
the flux coordinate $\psi \sim r^{2}$, 
\begin{equation}
MSD_{\psi }(t)\equiv <\left[ \psi (t)-\psi _{m}(t)\right] ^{2}>
\label{MSDpsi(t)}
\end{equation}%
and the dispersion of the radial coordinate with respect to this average
radius 
\begin{equation}
MSD_{rm}(t)\equiv <\left[ r(t)-r_{m}(t)\right] ^{2}>  \label{MSDRm(t)}
\end{equation}%
Here the brackets $\left\langle ...\right\rangle $\ indicate an average over
the initial conditions.

\bigskip \bigskip

The time dependence of these three quantities are analyzed and the existence
of an \emph{asymptotic regime} is exhibited: in this regime we observe a
diffusion of the flux coordinate (the variance grows linearly in time) , but
a \emph{subdiffusion of the spatial motion} (the variance $MSD_{rm}(t)$
grows like $t^{1/2}$), and a still slower behavior of the average radius,
which grows asymptotically like $r_{m}(t)\sim $ $t^{1/4}$. Simple scaling
laws are found to describe the dependence of the corresponding ''diffusion''
coefficients as function of the stochasticity parameter $L$. A quasilinear
scaling in $L^{2}$ is obtained for the flux diffusion $MSD_{\psi }$ ,
similarly to what is found in the standard map \cite{Chiri79}, \cite%
{Dif-Sta-Map}, for which additional oscillations deeply modify this simple
quadratic growth. As expected, the subdiffusive radial motion $MSD_{rm}(t)$
has a coefficient characterized by a scaling in $L^{1}$, while the slower
radial motion $r_{m}(t)$ is characterized by a scaling in $L^{1/2}$.

\bigskip \bigskip

We finally discuss the order of magnitude of the relaxation time, the time
necessary to reach this asymptotic regime and discuss which regime could
describe the magnetic line motion before reaching the plasma edge.{\LARGE \ }%
The conclusions are summarized in Section (\ref{Conclus}). Partial results
have already been presented in a EPS-ICPP conference and published in\ \cite%
{Quebec2000}.

\bigskip

\section{\label{Map}Hamiltonian map for toroidal magnetic lines}

\smallskip

The unperturbed magnetic field realized in tokamaks is ideally represented
by a set of nested toroidal \emph{magnetic surfaces} wound around a circular
magnetic axis. The condition $\mathbf{\nabla .B}=0$ allows us to express the
magnetic field in the Clebsch form \cite{RaduLiv88} in terms of the
dimensionless toroidal flux $\psi $ and poloidal flux $F$. 
\begin{equation}
\mathbf{B=\nabla }\psi \mathbf{\times \nabla }\theta \mathbf{-\nabla }F%
\mathbf{\times \nabla }\zeta  \label{Clebsch}
\end{equation}

\subsection{Equations of motion for magnetic lines}

We use traditional toroidal coordinates $(\psi ,\theta ,\zeta )$ where $%
\theta $ and $\zeta $ are the poloidal and toroidal angles, respectively,
and $\psi $ is the flux coordinate. From (\ref{Clebsch}) the ''equations of
motion'' for the magnetic lines are easily derived: 
\begin{equation}
\frac{d\psi }{d\zeta }=-\frac{\partial F}{\partial \theta }~\text{ , ~}\frac{%
d\theta }{d\zeta }=\frac{\partial F}{\partial \psi }  \label{Hamilton}
\end{equation}
These equations obviously have a Hamiltonian structure: the toroidal angle $%
\zeta $ plays the role of ''time'', and the poloidal flux $F$ the role of
the Hamiltonian.

\bigskip

In the \emph{unperturbed case}, $F$ is simply a ''surface function'' $%
F=F(\psi )$ which represents an unperturbed Hamiltonian with one degree of
freedom and corresponds to an integrable system: 
\begin{equation}
\frac{d\psi }{d\zeta }=0~\text{ , }~\text{~ }\frac{d\theta }{d\zeta }=\frac{%
\partial F(\psi )}{\partial \psi }\equiv W(\psi )  \label{lineEq}
\end{equation}
where 
\begin{equation}
W(\psi )\equiv 1/q(\psi )=\iota (\psi )/2\pi  \label{WindingNumber}
\end{equation}
is the \emph{winding number}, the inverse of the \emph{safety factor} $%
q(\psi )$. Here the action variable $\psi \sim r^{2}$ labels the magnetic
surfaces, it is canonically conjugated to the angle variable $\theta $ .

\bigskip

When a magnetic perturbation is applied, due to internal factors
(instabilities, fluctuations) or to external causes (imperfection, divertor
coils...), the poloidal flux $F$ becomes in general a function of the three
coordinates: 
\begin{equation}
F(\psi ,\theta ,\zeta )=F_{0}(\psi )+L\text{ }\delta F(\psi ,\theta ,\zeta )
\label{FdF}
\end{equation}
where $L$ is the \emph{stochasticity parameter} (). The field line equations
become 
\begin{equation}
\frac{d\psi }{d\zeta }=-L\frac{\partial \delta F(\psi ,\theta ,\zeta )}{%
\partial \theta }~\text{ , }~\text{~ }\frac{d\theta }{d\zeta }=W(\psi )+L%
\frac{\partial \delta F(\psi ,\theta ,\zeta )}{\partial \psi }
\label{pert line Eq}
\end{equation}
corresponding to a Hamiltonian dynamical system with $1\frac{1}{2}$ degrees
of freedom, generically non integrable. The perturbation is responsible for
the appearance of chaos.

\subsection{\label{SecHmaps}\protect\bigskip Hamiltonian maps, twist maps
and standard map}

In order to avoid long symplectic integration in computing the magnetic line
motion from (\ref{pert line Eq}), discrete iterative maps have been
introduced, specially to study the plasma edge. Many examples in the
literature have been quoted in \cite{Tmap1}. An explicit iterative
two-dimensional map consists in discrete transformations of the form 
\begin{equation}
\begin{array}{ccc}
\psi _{\nu +1}=P(\psi _{\nu },\theta _{\nu }) & \text{,} & \theta _{\nu
+1}=\Theta (\psi _{\nu },\theta _{\nu })%
\end{array}
\label{Hmap}
\end{equation}
where $\nu $ is a non-negative integer which represents physically the
number of large turns around the torus, and where the functions $P(\psi
_{\nu },\theta _{\nu })$ and $\Theta (\psi _{\nu },\theta _{\nu })$ are
explicit in the ''previous'' values $\psi _{\nu }$ and $\theta _{\nu }$.

\medskip

Such transformations must of course conserve the Hamiltonian structure of
the equations (\ref{pert line Eq}) : the model should be a \emph{Hamiltonian
map} (i.e. area-preserving or symplectic) and therefore the transformation (%
\ref{Hmap}) has to be a \emph{canonical transformation} of the canonical
variables $(\psi $, $\theta )$. In order to derive the map, one thus
introduces a general \emph{generating function} $F(\psi _{\nu +1},\theta
_{\nu })$ which allows to write down the map as 
\begin{equation}
\begin{array}{ccc}
\psi _{\nu }=\frac{\partial \text{ }F(\psi _{\nu +1},\theta _{\nu })}{%
\partial \theta _{\nu }} & \text{,} & \theta _{\nu +1}=\frac{\partial \text{ 
}F(\psi _{\nu +1},\theta _{\nu })}{\partial \psi _{\nu +1}}%
\end{array}
\label{map from gen fct}
\end{equation}
which is in an implicit form. (Note that we choose here a generating
function of the new momentum $\psi $ and the old angle $\theta $, but the
inverse choice is also possible and would lead to another family of maps).

\bigskip

Other choices are possible for the same twist map, for instance with another
kind of generating function $F_{a}(\theta _{\nu },\theta _{\nu +1})$ called
the \emph{action generating function }used in Section (\ref{Dana}), from
which the map can be written as 
\begin{equation}
\begin{array}{ccc}
\psi _{\nu }=-\frac{\partial \text{ }F_{a}(\theta _{\nu },\theta _{\nu +1})}{%
\partial \theta _{\nu }} & \text{,} & \psi _{\nu +1}=\frac{\partial \text{ }%
F_{a}(\theta _{\nu },\theta _{\nu +1})}{\partial \theta _{\nu +1}}%
\end{array}
\label{F a Eqs}
\end{equation}
The relation between $F$ and $F_{a}$ is thus (from (\ref{map from gen fct})
and (\ref{F a Eqs})): 
\begin{equation}
F_{a}(\theta _{\nu },\theta _{\nu +1})=\psi _{\nu +1}(\theta _{\nu },\theta
_{\nu +1}).\theta _{\nu +1}-F\left[ \psi _{\nu +1}(\theta _{\nu },\theta
_{\nu +1}),\theta _{\nu }\right]  \label{eqC}
\end{equation}
(see Ref. \cite{g}).

\medskip

\subsubsection{General form of a twist map}

In order to proceed, the following general form of the generating function
has been chosen: 
\begin{equation}
F(\psi _{\nu +1},\theta _{\nu })=\psi _{\nu +1}.\theta _{\nu }+F_{0}(\psi
_{\nu +1})+L\text{ }\delta F(\psi _{\nu +1},\theta _{\nu })
\label{general gen fct}
\end{equation}
where $L$ is the perturbation parameter and $F_{0}(\psi _{\nu +1})$ the
unperturbed term taking into account the winding number (see(\ref%
{WindingNumber})) 
\begin{equation}
W(\psi )=\frac{\partial F_{0}(\psi )}{\partial \psi }\equiv \frac{1}{q(\psi )%
}  \label{W de psi}
\end{equation}
As a consequence of (\ref{map from gen fct}) the map takes the following
form 
\begin{equation*}
\psi _{\nu +1}=\psi _{\nu }+L\text{ }h(\psi _{\nu +1},\theta _{\nu })
\end{equation*}
\begin{equation}
\theta _{\nu +1}=\theta _{\nu }+W(\psi _{\nu +1})+L\text{ \ }j(\psi _{\nu
+1},\theta _{\nu })  \label{R14}
\end{equation}
which is an Hamiltonian form because the two functions $h$ and $j$ defined
by 
\begin{equation}
\begin{array}{ccc}
h(\psi _{\nu +1},\theta _{\nu })=-\frac{\partial \text{ }\delta F(\psi _{\nu
+1},\theta _{\nu })}{\partial \theta _{\nu }} & \text{,} & j(\psi _{\nu
+1},\theta _{\nu })=\frac{\partial \text{ }\delta F(\psi _{\nu +1},\theta
_{\nu })}{\partial \psi _{\nu +1}}%
\end{array}
\label{R15}
\end{equation}
automatically insure that 
\begin{equation}
\frac{\partial \text{ }h(\psi _{\nu +1},\theta _{\nu })}{\partial \text{ }%
\psi _{\nu +1}}+\frac{\partial \text{ }j(\psi _{\nu +1},\theta _{\nu })}{%
\partial \text{ }\theta _{\nu }}=0  \label{R16}
\end{equation}
Equations (\ref{R14}) has the form of a general Hamiltonian map, from which
simple cases can be recovered.

\bigskip\ 

When variables are clearly separated in (\ref{R14}) \emph{i.e.} for $%
h=h(\theta _{\nu })$ and $j=j(\psi _{\nu +1})$ one finds a general \emph{%
twist map} in the case where the profile $q(\psi )$ is monotonous (see
section \ref{twist maps} below).

\subsubsection{The standard map}

The Chirikov-Taylor \emph{standard map}\cite{Chiri79} \cite{Dif-Sta-Map}
belongs to that family and corresponds to 
\begin{equation}
\begin{array}{ccccc}
h(\theta )=-\sin 2\pi \theta & \text{,} & j(\psi )=0 & \text{,} & W(\psi
)=\psi%
\end{array}
\label{hjW sta map}
\end{equation}
and has the following form 
\begin{equation}
\begin{array}{cccc}
\psi _{\nu +1}=\psi _{\nu }-L\sin 2\pi \theta _{\nu } &  & \text{,} & \text{ 
}\theta _{\nu +1}=\theta _{\nu }+\psi _{\nu +1}%
\end{array}
\label{SM}
\end{equation}
corresponding to 
\begin{equation}
\begin{array}{ccc}
F_{0}(\psi )=\frac{1}{2}\psi ^{2} & \text{,} & \delta F(\psi ,\theta )=-%
\frac{1}{2\pi }\cos (2\pi \theta )%
\end{array}
\label{dF SM}
\end{equation}
The generating functions for the standard map are thus 
\begin{equation}
F(\psi _{\nu +1},\theta _{\nu })=\psi _{\nu +1}.\theta _{\nu }+\frac{1}{2}%
\psi _{\nu +1}^{2}-\frac{L}{2\pi }\cos (2\pi \theta _{\nu })  \label{19a}
\end{equation}
and 
\begin{equation}
F_{a}(\theta _{\nu },\theta _{\nu +1})=\frac{1}{2}\left[ \theta _{\nu
}-\theta _{\nu +1}\right] ^{2}+\frac{L}{2\pi }\cos (2\pi \theta _{\nu })%
\text{ }  \label{19b}
\end{equation}

\subsubsection{The Wobig map}

The Wobig map \cite{Wobig} on the other hand corresponds to : 
\begin{equation}
\begin{array}{ccccc}
h(\psi ,\theta )=-\psi \sin 2\pi \theta & \text{,} & j(\psi )=-\frac{1}{%
(2\pi )}\cos 2\pi \theta & \text{,} & W(\psi )=\psi%
\end{array}
\label{Wob 1}
\end{equation}
and has the form 
\begin{equation}
\begin{array}{ccc}
\psi _{\nu +1}=\psi _{\nu }-L\psi _{\nu +1}\sin 2\pi \theta _{\nu } & \text{,%
} & \theta _{\nu +1}=\theta _{\nu }+\psi _{\nu +1}-\frac{L}{(2\pi )}\cos
2\pi \theta _{\nu }%
\end{array}
\label{WM}
\end{equation}
corresponding to 
\begin{equation}
\begin{array}{ccc}
F_{0}(\psi )=\frac{1}{2}\psi ^{2} & \text{,} & \delta F(\psi ,\theta )=-%
\frac{1}{2\pi }\psi \cos (2\pi \theta )%
\end{array}
\label{dF WM}
\end{equation}
The generating functions for the Wobig map are thus 
\begin{equation}
F(\psi _{\nu +1},\theta _{\nu })=\psi _{\nu +1}.\theta _{\nu }+\frac{1}{2}%
\psi _{\nu +1}^{2}-\frac{L}{2\pi }\psi _{\nu +1}\cos (2\pi \theta _{\nu })
\label{22a}
\end{equation}
and 
\begin{equation}
F_{a}(\theta _{\nu },\theta _{\nu +1})=\frac{1}{2}\left[ \theta _{\nu
+1}-\theta _{\nu }+\frac{L}{2\pi }\cos (2\pi \theta _{\nu })\right] ^{2}%
\text{ }  \label{22b}
\end{equation}

\bigskip

These last two maps are actually not suitable to represent magnetic lines in
a tokamak first because they do not insure that a nonnegative value of $\psi
\sim r^{2}$ remains nonnegative after iteration (as it should to represent a
real value of the radial position) and, second, because they do not involve
any realistic profile of the winding number $W(\psi )$ (they correspond to a 
$q$ -profile everywhere decreasing: $q\sim 1/r^{2}).$ In order to satisfy
these two necessary properties, another model is described in the next
section.

\subsection{The tokamap}

\bigskip

A specific model, the ''tokamap'' has been derived in \cite{Tmap1} from
general properties of Hamiltonian \emph{twist maps}. The advantage consists
in succeeding to describe the whole body of a tokamak plasma, including
chains of magnetic islands in a realistic way. This map describes the basic
motion of the magnetic lines in the two dimensional poloidal plane $(\psi
,\theta )$ by the winding number $W(\psi )$ which is here modified by an
additional contribution from the magnetic perturbation.

\bigskip

The general expression of the tokamap results from the following choice in
the generating function (\ref{general gen fct}) : 
\begin{equation}
\delta F(\psi _{\nu +1},\theta _{\nu })=-\frac{1}{2\pi }\frac{\psi _{\nu +1}%
}{1+\psi _{\nu +1}}\cos 2\pi \theta _{\nu }\text{ \ (tokamap)}
\label{gen fct Tokamap}
\end{equation}%
which involves an additional $\psi $ dependence as compared to Eq.(\ref{dF
WM}).\ We thus consider the following generating function for the Tokamap
(see (\ref{general gen fct}) and (\ref{gen fct Tokamap})): 
\begin{equation}
F(\psi _{\nu +1},\theta _{\nu })=\psi _{\nu +1}.\theta _{\nu }+F_{0}(\psi
_{\nu +1})-L\frac{\psi _{\nu +1}}{1+\psi _{\nu +1}}\cos 2\pi \theta _{\nu }%
\text{ \ (tokamap)}  \label{generating fct Tokamap}
\end{equation}%
This immediately leads to the following implicit form of the tokamap (use (%
\ref{map from gen fct})): 
\begin{equation}
\psi _{\nu +1}=\psi _{\nu }-L\frac{\psi _{\nu +1}}{1+\psi _{\nu +1}}\sin
(2\pi \theta _{\nu })  \label{Tok psi}
\end{equation}%
\begin{equation}
\theta _{\nu +1}=\theta _{\nu }+W(\psi _{\nu +1})-\frac{L}{2\pi }\frac{1}{%
(1+\psi _{\nu +1})^{2}}\cos (2\pi \theta _{\nu })  \label{Tok angle}
\end{equation}%
where $\theta $ denotes the poloidal angle divided by $2\pi $ . In this
nonlinear map a unique root is chosen for Eq. (\ref{Tok psi}): 
\begin{equation}
\psi _{\nu +1}=\frac{1}{2}\left( P(\psi _{\nu },\theta _{\nu })+\sqrt{\left[
P(\psi _{\nu },\theta _{\nu })\right] ^{2}+4\psi _{\nu }}\right)
\label{Tok psi explicit}
\end{equation}%
where 
\begin{equation}
P(\psi ,\theta )=\psi -1-L\sin (2\pi \theta )  \label{Tok psi explicit 2}
\end{equation}%
This explicit map (\ref{Tok psi}-\ref{Tok psi explicit 2}) has been shown to
be compatible with minimal toroidal geometry requirements. The polar axis $%
(\psi =0)$ represents the magnetic axis in the unperturbed configuration.
This map depends on one parameter, the stochasticity parameter $L$ (%
\footnote{%
The stochasticity parameter used here for convenience is related to the
parameter $K$ of the original reference by Balescu, Vlad \& Spineanu [16]
by: $L=K/2\pi $.}), and on one arbitrary function $W(\psi )$ which is chosen
according to the $q$ -profile we want to represent. For increasing values of 
$L$, chaotic regions appear mostly near the edge of the plasma.

\medskip

It is a simple matter to check that the symmetries of the tokamap imply
that, for negative values of $L$, the phase portrait is the same as for
positive values, but with the simple poloidal translation : $\theta
\Rightarrow \theta +1/2$ which means that the original phase portrait for $%
\theta =0$ to $1$ is recovered identically but for $\theta =-1/2$ to $+1/2$ .

\medskip

It has been shown that, contrary to what occurs in the Chirikov-Taylor
standard map \cite{Chiri79} or even in the Wobig map \cite{Wobig}, a
nonnegative value of $\psi $ always yields a nonnegative iterated value,
insuring that the radius remains a real number (see Eq.(\ref{A8 Psider})).
In the domain $L<1$, however, the Wobig map also conserves nonnegativity of $%
\psi $ .

\medskip

The tokamap has recently been deduced in a very different way \cite{Bo-map 1}%
, as a particular case of a \emph{particle map} describing guiding centre
toroidal trajectories in a perturbed magnetic field given in general by (\ref%
{Clebsch}), and to lowest order by the standard magnetic field model \cite%
{Knorr}, \cite{MisWeyBal92}: 
\begin{equation}
\mathbf{B}=\frac{B_{0}}{1+\varepsilon _{T}\text{ }x\cos \theta }\left[ 
\mathbf{e}_{\zeta }+\varepsilon _{T}\frac{x}{q(x)}\mathbf{e}_{\theta }\right]
\label{BKnorr}
\end{equation}%
where $\varepsilon _{T}=a/R_{0}$ is the inverse aspect ratio of the torus, $%
R_{0}$ the large radius of the torus, $B_{0}$ the magnetic field strength on
the axis and $x=\rho /a$ the normalized radial coordinate of the plasma of
small radius $a$ . This magnetic field\ (\ref{BKnorr}) is checked to be
divergenceless.\ By comparing the poloidal flux expressions in the general
Clebsch formula (\ref{Clebsch}) and in the above standard model (\ref{BKnorr}%
), it is easy to deduce that the toroidal flux $\psi $ in this case is given
by%
\begin{equation}
\psi =\frac{x^{2}}{2}  \label{newEq35bis}
\end{equation}%
in the dimensionless units used in the tokamap.

\bigskip

In \cite{Bo-map 1} canonical coordinates for guiding centre have been
derived, allowing for the symplectic integration of the equations of motion
and the derivation of a Hamiltonian map for guiding centre in a perturbed
toroidal geometry. As a particular case, the tokamap is deduced from this
particle map when one applies a simple $m=0$ nonresonant magnetic
perturbation.\ In order to describe magnetic line motion only, the magnetic
moment is considered to be zero, and one keeps terms to the lowest order in
the inverse aspect ratio $\varepsilon _{T}$. Moreover, in order to go from
the time dependence of the particle trajectory to the toroidal $\zeta $%
-dependence of the position $(\psi ,\theta )$ of the magnetic line, all
equations are simply divided by the equation for $\partial \zeta /\partial t$
. As a result, Eqs.(\ref{Tok psi}-\ref{Tok psi explicit 2}) are exactly
recovered. This derivation also allows to write down explicitly the form of
the magnetic perturbation involved in the tokamap : The generating function (%
\ref{general gen fct}) takes into account a magnetic divergence-free
perturbation $\delta \mathbf{B}$\ with the following components:

\begin{equation}
\frac{\delta B_{\theta }}{B_{0}}=L.\frac{x.\epsilon _{T}}{h(x,\theta )}.%
\frac{\partial \text{ }\delta F(\psi ,\theta )}{\partial \psi }
\label{gen magn perturb}
\end{equation}%
and 
\begin{equation}
\frac{\delta B_{\rho }}{B_{0}}=-L.\frac{\epsilon _{T}}{x.h(x,\theta )}.\frac{%
\partial \text{ }\delta F(\psi ,\theta )}{\partial \theta }
\label{gen magn perturb2}
\end{equation}%
where $h(x,\theta )=\left( 1+\epsilon _{T}.x.\cos \theta \right) $
represents the toroidal effect.

\bigskip

\bigskip A different but related method of derivation of symplectic maps has
been used by Abdullaev \cite{Abdu}.

\subsubsection{Safety factor profile}

In order to take magnetic shear into account, a specific $q$ -profile has to
be introduced to describe the unperturbed equilibrium. We consider here the
same profile as in Ref. \cite{Tmap1} : 
\begin{equation}
q(\psi )=\frac{4}{w.(2-\psi ).(2-2\psi +\psi ^{2})}  \label{profilq(psi)}
\end{equation}
which corresponds to a classical cylindrical equilibrium \cite{MisBoprof} in
which the value on the axis ($\psi =0$) is given by the parameter $w$ : 
\begin{equation}
q(\psi =0)=\frac{1}{w}  \label{q0}
\end{equation}
and on the edge ($\psi =1$) by : 
\begin{equation}
q(\psi =1)=4q(0)  \label{q(1)}
\end{equation}
which is four times the central value. The derivation of this $q$ -profile (%
\ref{profilq(psi)}) is presented in \textbf{Appendix A}.

\medskip

\subsubsection{One example with reversed magnetic shear : the ''Rev-tokamap''%
}

If a non monotonic $q$ -profile is used instead, a reversed magnetic shear
is introduced in the unperturbed magnetic field, and it has been shown that
the above mapping becomes a \emph{nontwist map} : this ''Rev-tokamap'' \cite%
{RexTokamap} has very different properties namely because the
Kolmogorov-Arnold-Moser (KAM)\ theorem does not apply anymore. It has been
found \cite{RexTokamap} that a critical surface appears in the plasma near
the minimum of the $q$ -profile , separating an external, globally
stochastic region from a central, robust nonstochastic core region. Such a
phenomena of ''semiglobal chaos'' has been shown to be analogous to the
appearance of ITB in reversed shear experiments. Later, similar theoretical
results have been obtained in the Lausanne group \cite{EPFL} in a reversed
shear TEXTOR equilibrium; by studying a symplectic perturbed map for the
magnetic topology, they confirmed that the transport barrier is indeed
localized near the minimum of the $q$ -profile .

\bigskip

Coming back on the reversed shear model of Ref. \cite{RexTokamap} we will
show below in Section (\ref{SectITBREV})\textbf{\ }that the ITB found in
that work is also localized with a good precision on a $q$-value which will
appear to be a \emph{noble} number.

\subsubsection{Fixed points and bifurcations}

The general structure of the phase portrait of the tokamap has been
described in detail in \cite{Tmap1}. \emph{Fixed points} of the map should
not be confused with secondary magnetic axes inside the islands: the latter
are not fixed but wander upon iteration from one island to another island of
the chain and appear as \emph{periodic points}.

\bigskip

Fixed points, defined by $\psi _{\nu +1}=\psi _{\nu }$ and $\theta _{\nu
+1}=\theta _{\nu }+n$ where $n$ is an integer, can easily be calculated by
using first Eq. (\ref{Tok psi}) which yields either $\psi =0$ (the polar
axis) or $\sin (2\pi \theta _{\nu })=0$. This proves that \emph{all fixed
points are either }

- \emph{on the polar axis (}$\psi =0$\emph{), }for instance \emph{\ }$\theta
=0.25$ and $0.75$ which are two hyperbolic points on the polar axis, and
exist as long as $L<1$,

- \emph{or in the equatorial plane} ($\theta =0$ or $0.5$). By using next
Eq. (\ref{Tok angle}), one finds that the values of $\psi $ at the fixed
points with $\theta =0$ are solutions of 
\begin{equation}
F_{1}(\psi ,K,w,n)\equiv W(\psi )-\frac{L}{2\pi }\frac{1}{(1+\psi )^{2}}-n=0
\label{F1}
\end{equation}
while values of $\psi $ at the fixed points with $\theta =0.5$ are solutions
of 
\begin{equation}
F_{2}(\psi ,K,w,n)\equiv W(\psi )+\frac{L}{2\pi }\frac{1}{(1+\psi )^{2}}-n=0
\label{F2}
\end{equation}

It has been shown \cite{Tmap1} that the origin, the \emph{polar axis} $(\psi
=0)$ is a fixed point as long as $L<1$. Fixed points have been discussed in
great detail in \cite{Tmap1}, including consideration of the ''ghost space'' 
$\psi <0$ which appears necessary in order to check that the Poincar\'{e}%
-Birkhoff theorem and the conservation of stability index (the difference
between the number of elliptic and hyperbolic points which remains constant
as $w$ varies) are indeed satisfied. Let us simply recall here that
bifurcations occur according to the value of $w$ representing the central
winding number (see \ref{q0}) as compared to the following $L$-dependent
parameters 
\begin{equation}
w_{m}(L)\equiv 1-L  \label{wm(K)}
\end{equation}%
and 
\begin{equation}
w_{M}(L)=2-w_{m}=1+L>w_{m}(L)  \label{wM(K)}
\end{equation}%
As long as $w<w_{m}=1-L$, the only invariant point is the polar axis $\psi
=0 $, which obviously has to be interpreted in this case as the magnetic
axis of the tokamak (see Fig. 15 in \cite{Tmap1}).

\medskip

For higher values of $w$,\ when $w_{m}<w<w_{M}\equiv 2-w_{m}$ , \emph{i.e.} $%
1-L<w<1+L$, a first bifurcation has occurred, with the appearance of an
elliptic point off the polar axis. This latter elliptic point has to be
interpreted as the \emph{magnetic axis}, displaced by the Shafranov shift %
\cite{Shafranov 62}, \cite{Mercier Luc 74}. In the case $w=1$ (to which we
will restrict ourselves in the next Sections), the position $\psi _{MA}$ of
this magnetic axis is obtained by using Eq. (\ref{F2}) as function of $L$
as: 
\begin{equation}
F_{2}(\psi _{MA},L,w=1,n=1)=0  \label{Shafran shift}
\end{equation}%
The curve giving $\psi =\psi _{MA}(K=2\pi L)$ is plotted in Fig.(\ref%
{Shafran}).

\bigskip 

\begin{figure}
\centering
\includegraphics[width=12.0221cm]{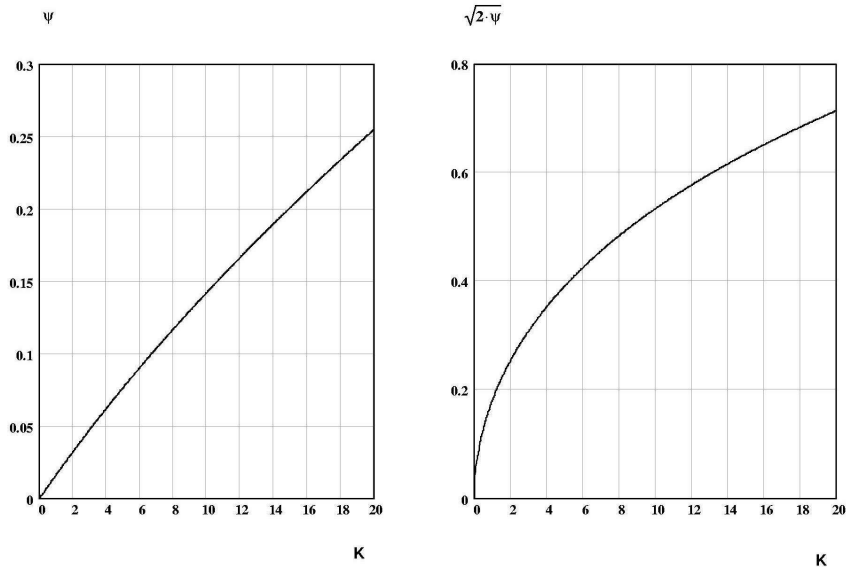}
\caption{Analytic solution for the Shafranov shift: position $\protect\psi $\ of the magnetic axis as function of $K=2\protect\pi L$\ in the case $w=1$. For $L=4.875/2\protect\pi $\ this formula yields $\protect\psi (L=4.875/2\protect\pi )=0.07492$\ . Here $r=\protect\sqrt{\protect\psi }$\ represents the radial coordinate, with $r=\protect\sqrt{2}$ on the edge.}
\label{Shafran}
\end{figure}

\bigskip

In the absence of any perturbation ($L=0$), the magnetic surfaces are
centered around the polar axis.

\medskip

A first typical phase portrait is given in Fig. 5 of Ref. \cite{Tmap1} for
the simple case where $q(0)=1$, and $L=4.5/2\pi \sim 0.716$ , hence $w=1$
and $w_{m}<w$. One can see the formation of a \emph{chaotic belt} which
surrounds several large islands chains. This belt is clearly confined
between two surfaces which were interpreted as KAM surfaces in this case.
Several other examples are given below (see Figs.\ref{ExFig2}-\ref{ExFig4b}%
). We will discuss why, for slightly higher values of $L$, such boundary
surfaces can be identified as \emph{fractal barriers} or \emph{Cantori},
which can be crossed by a magnetic line after a very long time.

\medskip

\subsubsection{Second bifurcation for $q(0)<1$}

\begin{figure}
\centering
\includegraphics[width=5.9221cm]{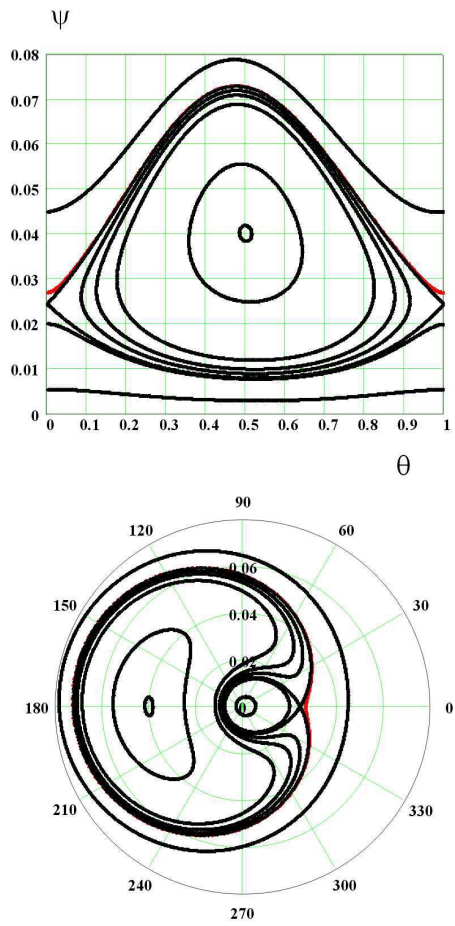}
\caption{Tokamap
phase portrait for $L=0.5/2\protect\pi $ with $q(0)<1$ $(w=1.05$ and $N=8190$%
). The polar axis $(0,0)$ is the main magnetic axis of the plasma core. The
position of the elliptic point in the $m=1$ island is $\protect\psi %
=0.040246 $, $\protect\theta =0.5$ and the hyperbolic point is located in $%
\protect\psi =0.024481$, $\protect\theta =0$.}
\label{ExFig2}
\end{figure}

A second bifurcation occurs for still higher values of $w$ when $%
w_{m}<w_{M}<w$, \emph{i.e.} when $1+L<w$\ \ which implies $q(0)<1$ : the
magnetic axis remains displaced from the origin and corresponds to the
center of a $m=1$ island in the equatorial plane at $\theta =0.5$, resulting
in the appearance of an hyperbolic point (''$X$ point'') for $\theta =0$
with a separatrix enclosing the origin.\ The latter appears to be the
''main'' magnetic axis, ''expelled'' by the presence of the $m=1$ island.\
This can be seen in Fig.(\ref{ExFig2}) which presents the tokamap phase
portrait for $L=1/4\pi $, $w=1.05$ (thus $q(0)<1$ and $w_{m}<w$) over $%
N=8193 $ iterations.

\medskip \bigskip

This example
shows the final situation after a big island $(m=1)$ has grown and expelled
a small region around the original magnetic axis near the polar axis; this
occurs when the unperturbed $q$-value on the axis is smaller than unity, $%
q(0)=1/w<1$, allowing for the existence of a surface $q=1$ inside the
plasma. It has been stressed that this process occurs as a result of a \emph{%
reconnection} of magnetic lines, and is typical of Kadomtsev's theory of 
\emph{sawtooth instabilities.} The position of the $m=1$ magnetic axis (''$O$
point''), as well as the position of the hyperbolic $X$ point can be
calculated analytically from Eqs. (\ref{F1}, \ref{F2}), in agreement with
Fig. 17 of Ref. \cite{Tmap1}. (We note that two misprints appeared in the
text of that paper for the exact numerical values which should be written $%
\psi 1=$ $0.040 $ of the $O$ point, and $\psi 2=0.024$ of the $X$ point, as
indicated by the Fig.\ 17, instead of $0.179$ and $0.167$ written in the
main text).

\bigskip

These points can be calculated as follows. For $w_{M}<w$ the position $\psi
_{H}$ of the hyperbolic $X$ point is given by the solution of $F_{1}(\psi
_{H},L,w,1)=0$ (with $\theta =0$) as defined in Eq. (\ref{F1}).\ For a given 
$w$, this solution exists only for the smaller values of $L$. For instance
for $w=1.05$ it exists only for $K=2\pi L\leq 1.97392$ , which means that
for $q(0)<1$ this $X$ point only appears for small values of the
perturbation parameter.

\bigskip

The position $\psi _{E}$ of the $m=1$ magnetic axis (elliptic or $O$ point)
is given by the solution of $F_{2}(\psi _{E},L,w,1)=0$ (with $\theta =0.5$).
We remark that this last solution $\psi _{E}(L)$ \emph{does not vanish for} $%
L\Rightarrow 0$, which would seem to indicate the existence of a fixed point 
$\left[ \psi _{E}(L=0),\theta =0.5\right] $ \emph{out of} the polar axis $%
\psi =0$, even in the unperturbed case where all magnetic surfaces are
circles centered around the origin... In this case we have $w_{M}(L=0)=1<w$
which means $q(0)<1$ and this fixed point $\left[ \psi _{E}(L=0),\theta =0.5%
\right] $ is simply a standard point of the circular surface $q=1$ : this is
just a special feature of this unperturbed surface on which every point
appears as stationary since it exactly comes back after one toroidal
rotation (one iteration). This specific point with $\theta =0.5$ in the
unperturbed case is however important since it acts as a seed for the $m=1$
island which appears when $L$ becomes different from zero. We note that this
position $\psi _{E}(L)$ is such that $q(\psi _{E}(L=0))=1$.

\bigskip

In order to simplify the discussion about dispersive motion of the magnetic
lines in the tokamap, we will restrict ourselves in the following Sections
to the case 
\begin{equation}
w=1  \label{w 1}
\end{equation}
which means $q(0)=1$ . The unperturbed $q$ -profile is a continuous and
monotonous function, with a value growing from $1$ in the center to $4$ on
the edge of the plasma, a rather standard profile in most ohmic discharges.
In this case for any $L>0$, we have from Eq. (\ref{wm(K)}) $w_{m}(L)=1-L\leq 
$ $w$, \emph{i.e.} we are\emph{\ }beyond the first bifurcation and the
position of the magnetic axis is given by Fig. (\ref{Shafran})\textbf{.}

\section{\label{IndiviDUAL}Individual trajectories: tokamap phase space
portrait}

In this section we first present the phase portrait of the tokamap in the
case $w=1$, $q(0)=1$ for increasing values of the stochasticity parameter
and then study very long trajectories.

\subsection{ Increasing stochasticity}

For increasing values of the stochasticity parameter the tokamap exhibits
all the main features of chaotic systems, in a way which is very realistic
for tokamaks.

\subsubsection{Weak stochasticity regime : confinement by KAM tori}

\begin{figure}
\centering
\includegraphics[width=5.2631cm]{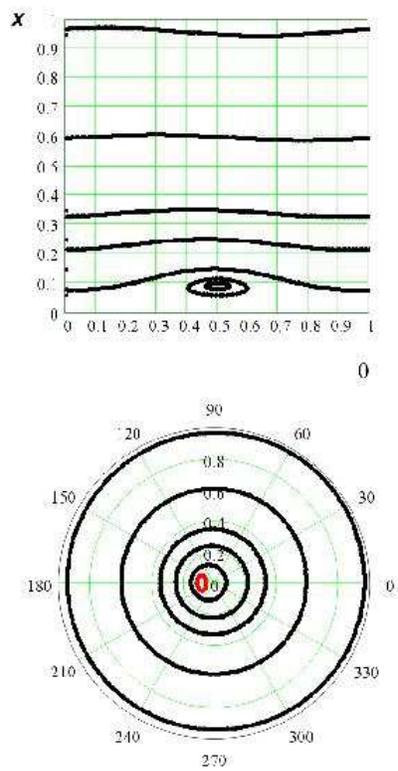}
\caption{Regular KAM magnetic surfaces with a Shafranov shift for a small value of the stochasticity parameter $L=0.08$, with $w=1$. The angle $\protect\theta $ is represented modulo $2\protect\pi $.}
\label{EsFig3}
\end{figure}

For small values of the stochasticity parameter $L=K/2\pi <<1$, most of the
KAM surfaces are preserved and the phase portrait of the tokamap appears to
be described by embedded tori, around a magnetic axis displaced from the
origin $(\psi =0,\theta =0)$ by the Shafranov shift (\ref{Shafran shift}),
as seen in Fig.(\ref{EsFig3}).

\bigskip 

In this case all magnetic lines remain confined, no magnetic island is seen.
From the position of the shifted magnetic axis, we note that, for positive
values of $L$, the weak field side of the torus is in the direction $\theta
=0.5$ .

\subsubsection{Appearance of island chains on rational $q$-values}

\begin{figure}
\centering
\includegraphics[width=6.3548cm]{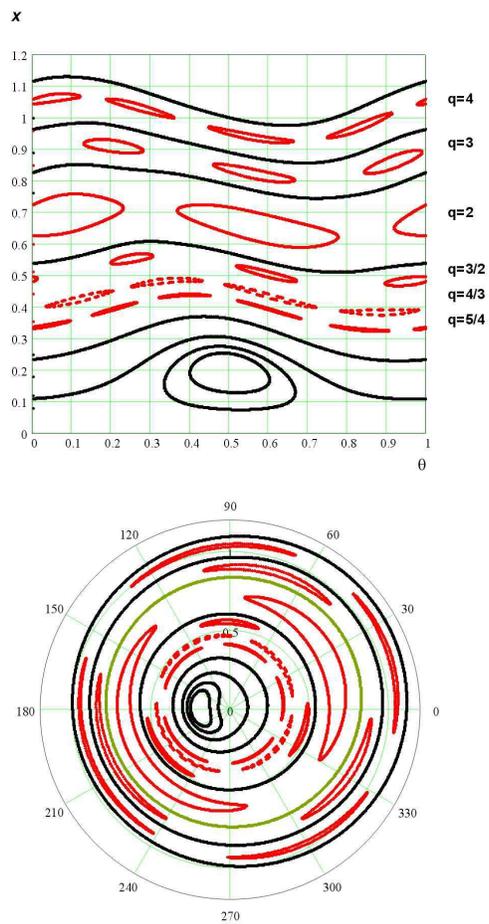}
\caption{Regular KAM tori for  $L=0.4$, $w=1$, and main rational island chains corresponding to $q=5/4$, $ 4/3$, $3/2$, $2$, $3$ and $4$}
\label{ExFig4}
\end{figure}

For an increased value of $L$, \emph{regular chains of magnetic islands}
appear; their number $m$ and the order in which these islands are visited by
a magnetic line allow us to deduce their $q$-value in a simple way. For
instance a chain of $m$ islands visited one by one at each iteration, in the
direction of increasing values of $\theta $, has a rotational transform or
winding number $\iota (\psi )/2\pi $ equal to $1/m$, and a $q$-value $q=m$.
These are the main island chains, as seen in Fig.(\ref{ExFig4}).

\bigskip

\bigskip

A direct measurement of the exact value of $q$ can be performed, specially
for almost undestroyed magnetic surfaces, by computing the average \emph{%
winding number} along a trajectory, see (\ref{WindingNumber}). \ For small
circles around the magnetic axis (which do not encircle the polar axis), a
change of coordinate is necessary to evaluate correctly the rotation around
the magnetic axis. On the other hand, for all surfaces encircling the polar
axis, the knowledge of the iterated values $\theta _{\tau }$ allows us to
calculate the following surface quantity (average on a given magnetic
surface) 
\begin{equation}
\left\langle \iota /2\pi \right\rangle =\lim_{N\rightarrow \infty }\frac{1}{N%
}\sum_{\tau =1}^{\tau =N}\left( \theta _{\tau }-\theta _{\tau -1}\right)
\label{Iota mesure}
\end{equation}%
which is the average increase of the poloidal angle over a large number $N$
of iterations. The $q-$value of the corresponding trajectory is immediately
obtained from (\ref{WindingNumber}).

\subsubsection{Overlapping chains, secondary islands and chaotic regions}

\begin{figure}
\centering
\includegraphics[width=7.027cm]{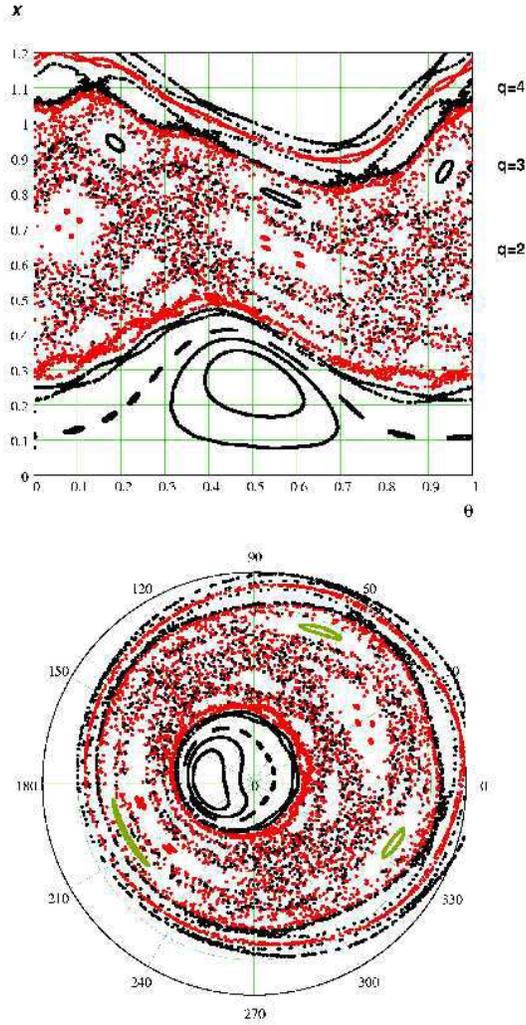}
\caption{With the same initial
conditions as in Fig. 4 $(w=1)$ with a higher value of the stochasticity
parameter $L=4.875/2\protect\pi \sim 0.776$, regular KAM tori are observed
in the central part, while the main rational island chains corresponding to $%
q=$ $3/2$, $2$, $3$ and $4$ are stochastized.}
\label{ExFig4b}
\end{figure}

For an increased stochasticity parameter $L=4.875/2\pi \sim 0.776$ , the
same initial conditions as in Fig.(\ref{ExFig4}) actually describe a wide
chaotic zone surrounding a central part with regular KAM\ tori : see Fig.(%
\ref{ExFig4b}).

\bigskip 

The main rational chains $q=3/2$, $2$, $3$ and $4$ are partly stochastized.

\bigskip

\begin{figure}
\centering
\includegraphics[width=9.4345cm]{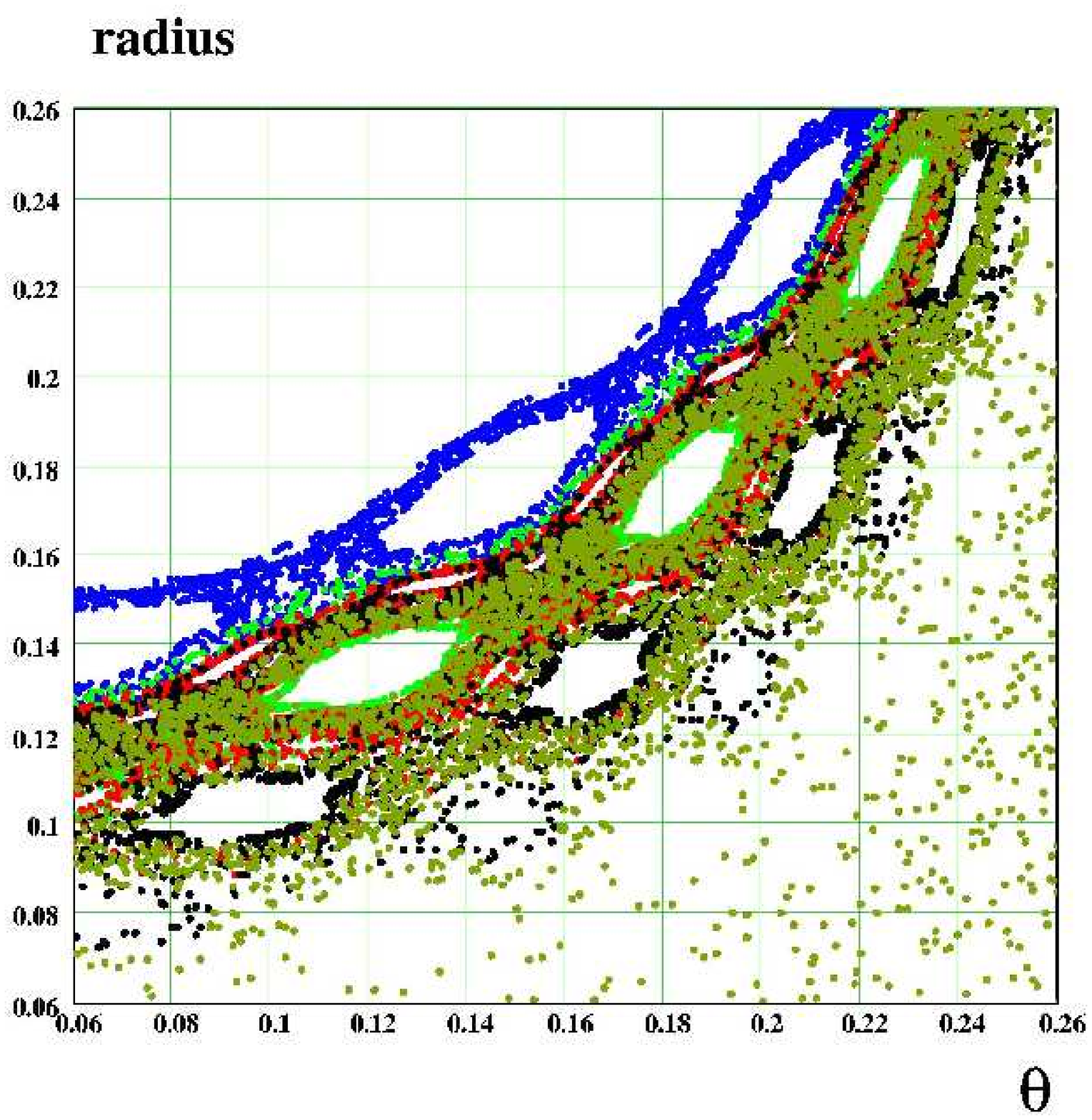}
\caption{Partial vue of the
phase portrait with six trajectories for $L=4.875/2\protect\pi \sim 0.776$,
which shows the structures of rational chains of island remnants and the
stochastic web expanding from the hyperbolic points.}
\label{ExFig10}
\end{figure}

We show on Fig.(\ref{ExFig10}) a small part of the phase portrait obtained
by following $6$ trajectories along $16$ $380$ iterations: the main chain of
primary islands is observed, with intermediate higher rational chains,
surrounded by \emph{stochastic regions} in the vicinity of the hyperbolic
points (as usual in nonlinear dynamical systems).

\bigskip

\subsubsection{Island remnants in the chaotic sea}

\begin{figure}
\centering
\includegraphics[width=10.1067cm]{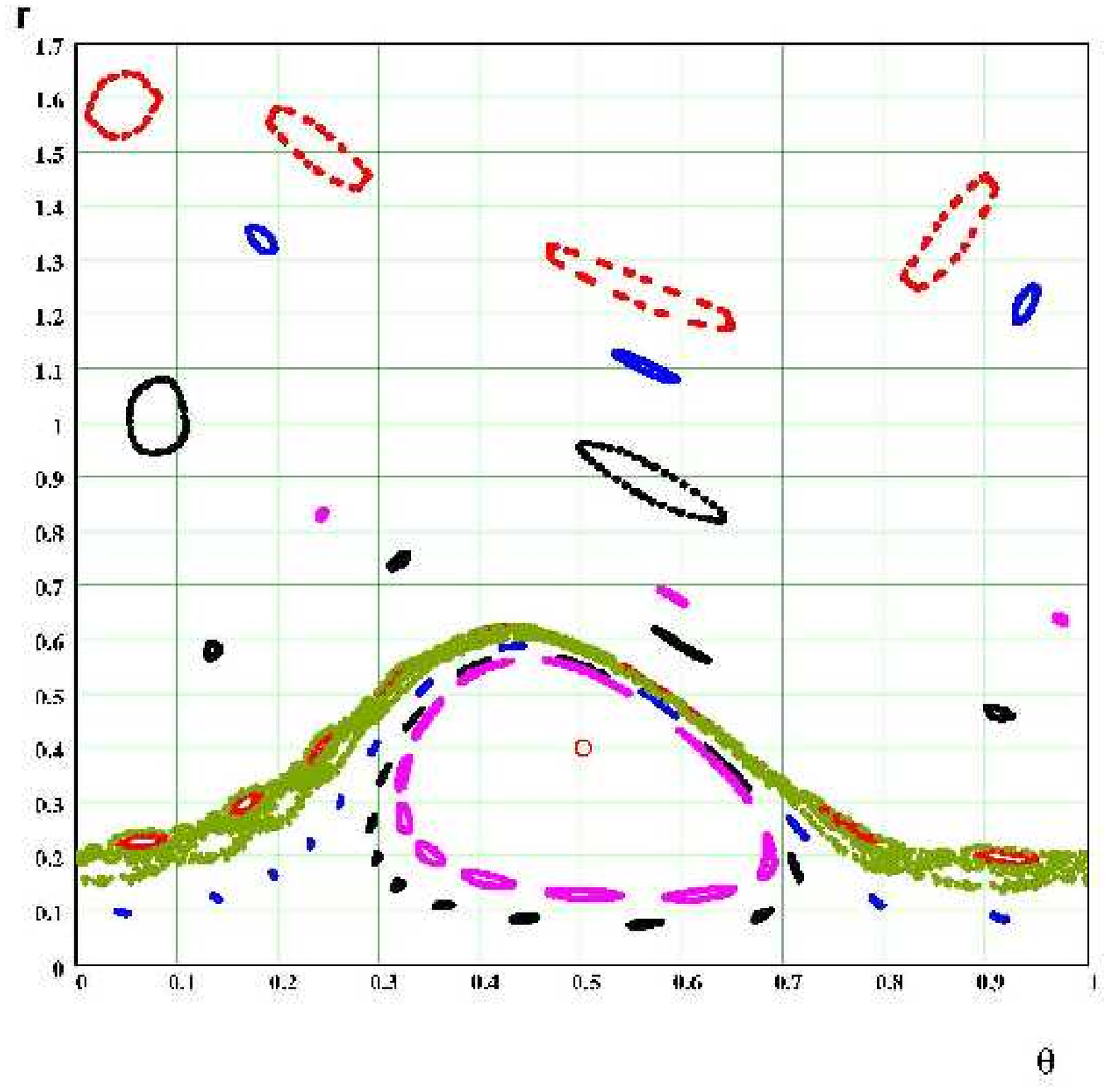}
\caption{Part of the skeleton of
island remnants in the phase portait ($r=\protect\sqrt{2\protect\psi }=x%
\protect\sqrt{2}$ as function of \ $\protect\theta $)\ \ for $L=4.875/2%
\protect\pi \sim 0.776$\ : rational chains can be seen corresponding to $q=4$%
, $3$, $2$, $3/2$, $4/3$, $8/7$, $9/8$, $12/11$ .}
\label{ExFig4c}
\end{figure}

\smallskip The general aspect of the phase portrait is determined mainly by
the $q$ -profile . The structure of the stochastic sea is analyzed in terms
of the various island chains. The islands are partly ''destroyed'' by
chaotic regions on their edge, so that we mainly observe ''\emph{island
remnants}'' which form a \emph{''skeleton''} of the phase space\emph{. }%
Several island chains among the largest, which belong to the ''dominant''
classes $n=m$ and $n=m-1$ with 
\begin{equation}
q=\frac{m}{n}  \label{q Islands}
\end{equation}%
are represented in Fig.(\ref{ExFig4c}).\ This skeleton structure is then
invaded by the chaotic sea (not represented here).

This last series $n=m-1$\ has some importance here : it corresponds to
chains of $m$ islands visited one by one at each iteration, but with \emph{%
decreasing} values of $\theta $, thus with a rotational transform $\iota
/2\pi $ equal to $-1/m$ (modulo 1), or equivalently to $(m-1)/m$, hence $%
q=m/(m-1)$.

\bigskip

This \emph{global island structure} (or \emph{geographic chart of the
stochastic sea }or \emph{''skeleton''}) is only slightly perturbed for
larger values of $L$. In Fig.(\ref{ExFig4c}) we have represented \textbf{\ }%
for $L=5.25/2\pi \sim 0.836$ a series of \ small island remnants appearing
as resistant KAM surfaces around the secondary magnetic axis inside the
islands (''vibrational KAM'' \cite{RMacKayBook}). From top to bottom we
observe

- chains of island corresponding to $\left[ m=4,q=4\right] $, $\left[ m=3,q=3%
\right] $, $\left[ m=2,q=2\right] $,

- along with smaller islands: $\left[ m=3,q=3/2\right] $, $\left[ m=4,q=4/3%
\right] $, $\left[ m=8,q=8/7\right] $,

- a stochastized region around the $\left[ m=9,q=9/8\right] $ island
remnants,

- and also $\left[ m=12,q=12/11=1.09\right] $.

All these chains are seen to encircle both the magnetic axis and also the
polar axis (since they cover the whole interval of $\theta $). We also
represented two chains encircling the magnetic axis only : $\left[
m=13,q=13/12=1.083\right] $ and $\left[ m=11,q=11/10=1.1\right] $ which is
inside the former ones.

\medskip

From the respective positions of the two latter chains, we note the
unexpected fact that, locally, the \emph{perturbed }$q$\emph{-value }is
actually \emph{growing towards the magnetic axis}, a very important feature
for transport properties. This proves that a non-monotonous $q$ -profile
around the magnetic axis, and a \emph{negative magnetic shear} has been
spontaneously created by the magnetic perturbation. The precise perturbed $q$
-profile is presented below in Section \ref{Perturbedq} (see Figs.\ref%
{ExFig17}, \ref{ExFig18} ).\medskip

\subsubsection{Good confinement of the plasma core}

\begin{figure}
\centering
\includegraphics[width=13.0479cm]{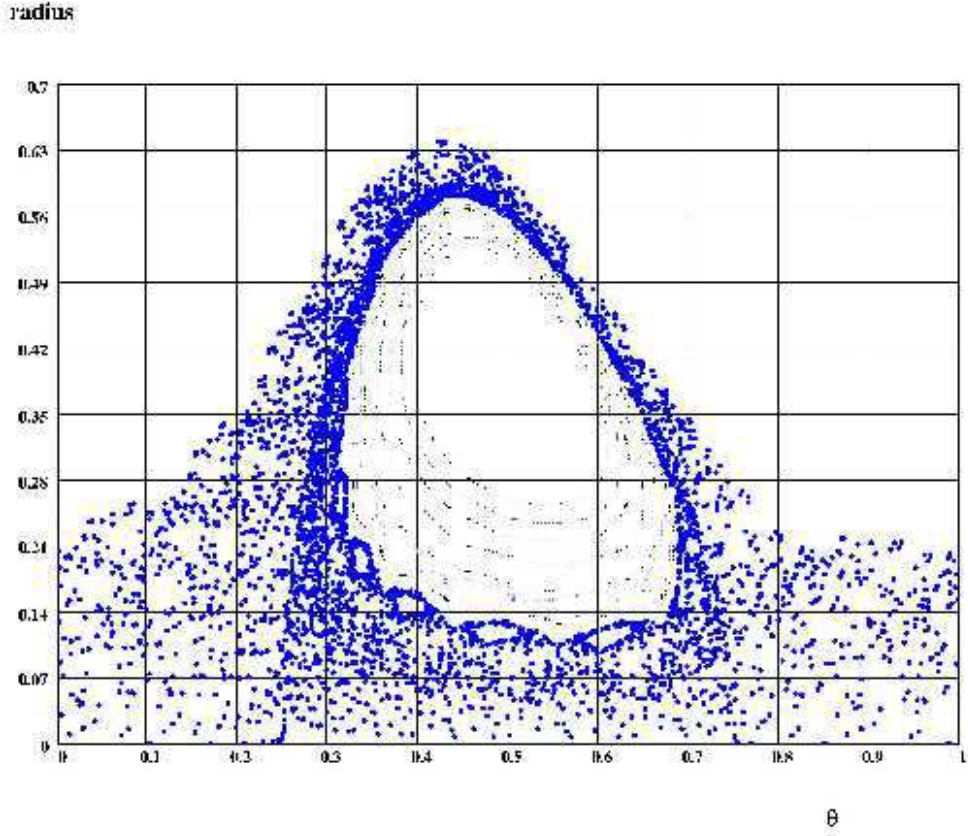}
\caption{Several
trajectories for $L=5.5/2\protect\pi \sim 0.875..$ , yielding a central
protected plasma core, surrounded by chains of regular islands and KAM
surfaces, and, outside, a stochastized zone allowing the line to escape from
this central chaotic shell.}
\label{ExFig6}
\end{figure}

Between these island remnants are large \emph{chaotic regions}. Of
particular interest is a circular chaotic shell (or belt) around the
magnetic axis.\ For instance with $L=5.5/2\pi \sim 0.875$ we have
represented in Fig.(\ref{ExFig6}) several trajectories filling a circular
shell surrounding the magnetic axis, and wandering in a chaotic layer around
island remnants. We stress the fact that this \emph{chaotic central shell}
surrounds a regular central part (around the magnetic axis), which
represents \emph{a quiet central plasma core} protected from chaos, thus a
good confinement zone.

\bigskip 

\bigskip We note that the chaotic zone represented here has an obvious
internal separation, with a visible trajectory passing very near $r\sim 0$, $%
\theta =0.25$, where there is actually a fixed (unstable) hyperbolic point %
\cite{Tmap1}. The separation is actually nothing else than the chaotized
separatrix in the $(\psi ,\theta )$ plane, joining the two hyperbolic points
located in $\psi =0$, $\ \theta =0.25$ and $0.75$. In a polar representation
this separatrix is the circular magnetic ''surface'' encircling the magnetic
axis and tangent to the polar axis $\psi =0$.

\medskip

\subsubsection{\label{acrossEDGE KAM}Stochasticity threshold for escaping
lines out of the plasma bulk}

The aim of the present work consists, first, to determine, for a given $q$
-profile , the threshold domain of the values of the stochasticity parameter 
$L$ for which the plasma boundary becomes permeable, allowing the magnetic
line to escape across this broken barrier to the edge and corresponding to a
disruption of the plasma towards the wall. This happens between $%
L=4.875/2\pi \sim 0.776$ and $L=5/2\pi \sim 0.796$. The next question is
(see Sections \ref{RoughLOCAL} and \ref{convergents}) : for lower values of $%
L$, in the confinement domain, which is the most robust barrier able to
inhibit the motion of the magnetic lines up to the edge ? In other words : 
\emph{which is the most robust KAM surface inside the plasma, the last one
to be broken ?} Which is its $q$ value ?

\bigskip

\begin{figure}
\centering
\includegraphics[width=10.2582cm]{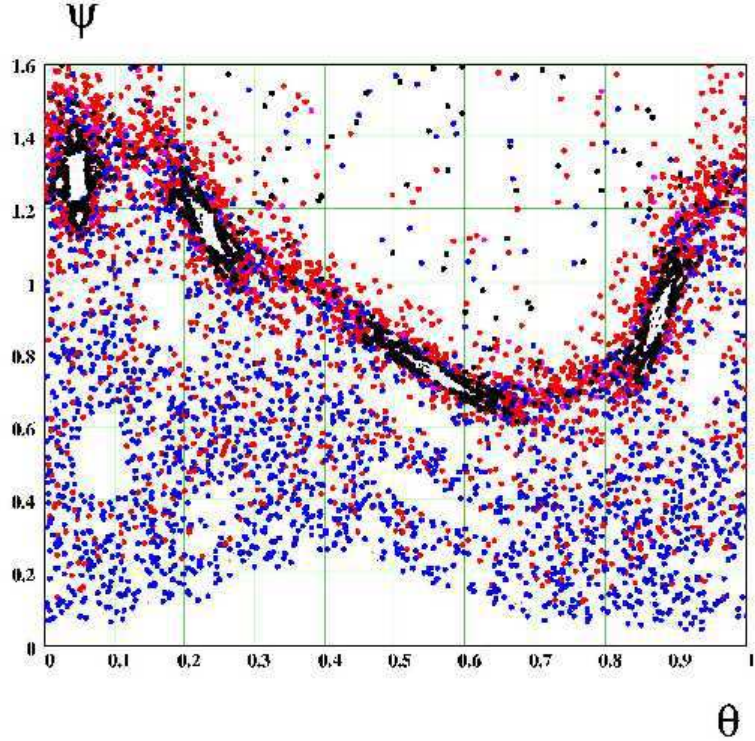}
\caption{Phase portrait of
the plasma edge, after complete destruction of the KAM surfaces.\ Several
trajectories are chosen to represent the details of the four island remnants 
$q=m=4$ at a value $L=6/2\protect\pi \sim 0.954...$ beyond the escape
threshold.\ The plasma edge has become permeable and is strongly deformed as
compared to the unperturbed circle $\protect\psi =1$. One remarks the
presence of four satellites or daughter islands around each main island.}
\label{ExFig9}
\end{figure}

For $L=6/2\pi $ we remark that the edge of the plasma has become permeable
and is strongly deformed as compared to the unperturbed circle $\psi =1$.
The four islands $q=m=4$ can be seen on Fig.(\ref{ExFig9}), along with their
four satellites or ''daughter islands''. This value of the stochasticity
parameter is obviously larger than that of an escape threshold.

\bigskip 

\bigskip By performing a very long iteration up to $2.10^{9\text{\ }}$time
steps on a Alpha workstation, with a value $L=4.875/2\pi \sim 0.776$, we did
not reach the time where the particle could possibly escape from the plasma
: up to this time the trajectory remains confined by what can be considered
as a KAM surface.

\medskip

\begin{figure}
\centering
\includegraphics[width=9.5048cm]{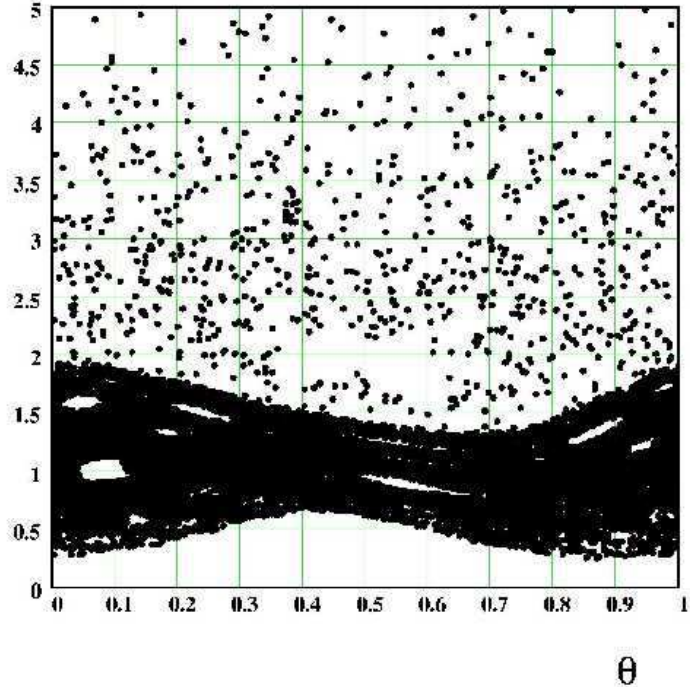}
\caption{Single trajectory at 
$L=5.5/2\protect\pi \sim 0.875$\ followed during $N=16380$\ iterations.
After a long trapping period in the chaotic sea, this trajectory finally
escapes out of the plasma edge. The initial conditions are $r=\protect\sqrt{2%
\protect\psi }=0.4$\ and $\protect\theta =0.025$.\protect\bigskip .}
\label{ExFig19}
\end{figure}

On the other hand, for slightly larger values of $L$ of the order of $5/2\pi
\sim 0.796$, most of the magnetic lines are found to rapidly escape from the
plasma (see for instance Fig.(\ref{ExFig19}) for $L=5.5/2\pi \sim 0.875$). A
threshold region of the stochasticity parameter has thus been found slightly
below $L=5/2\pi \sim 0.796$. For larger $L$ values, magnetic lines escape
across the plasma edge even when starting from the central chaotic shell.

\bigskip 

\subsection{\label{Interm confinmt}Intermittent motion inside the confined
plasma: crossing internal barriers}

\medskip

For values of $L$ smaller than this escape threshold, magnetic lines are
wandering inside the plasma. We mainly consider here the case $L=4.875/2\pi
\sim 0.776$ at which the edge barrier is not yet broken. We follow one
single magnetic line along a large number of iterations, starting near the
central region. With this value of the stochasticity parameter, the
corresponding phase portrait is particularly rich, even with only one point
represented every $10009$ iterations \footnote{%
In selecting a finite number of points on the graph, in order to avoid a
fully black drawing, we have represented in this case only one point every $%
10009$, a large prime number to avoid lower order graphical stroboscopic
effects (by using any multiple of $5$ for instance we would have drawn only
one of the five islands $q=5/2$ since the same magnetic line visits the
neighborhood of these five islands one after the other - two by two - and
comes back in the first island neighborhood after $5$ iterations. In order
to represent the whole set of $5$ daughter islands around the two islands $%
q=2$ we need to avoid multiples of $10$, a.s.o... so that we choose a large
prime number).}.

\bigskip

\begin{figure}
\centering
\includegraphics[width=14.1221cm]{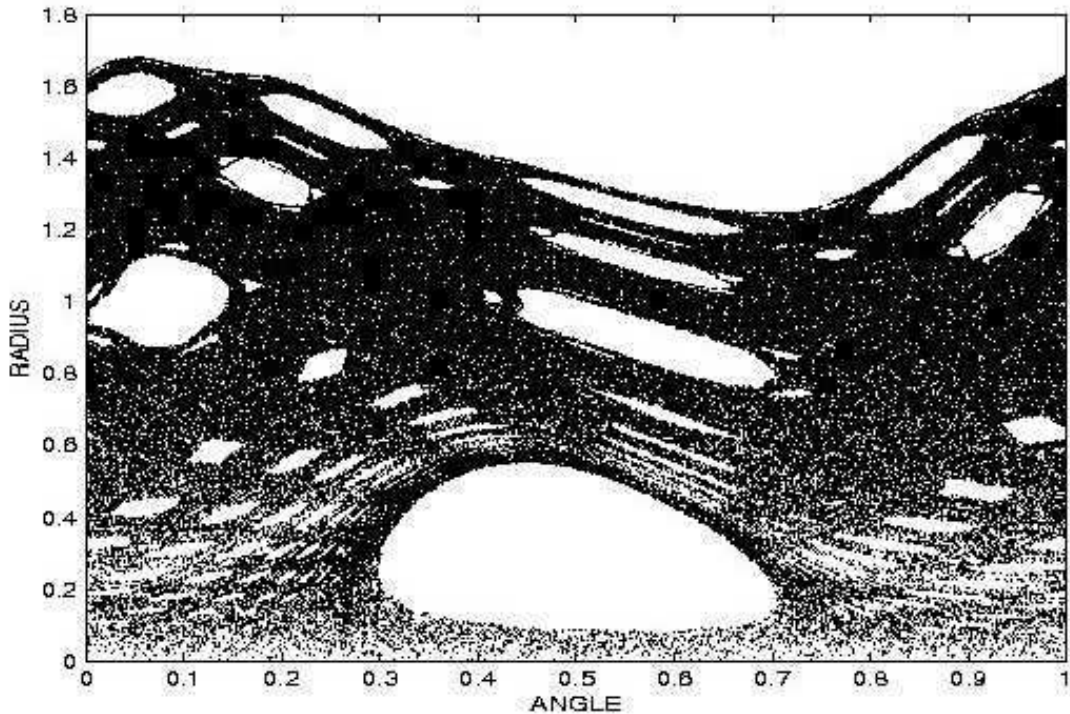}
\caption{Phase portrait for $%
L=4.875/2\protect\pi \sim 0.776$ as drawn in coordinates $\protect\theta $
and $r=\protect\sqrt{2\protect\psi }$ by a single trajectory followed during 
$2.10^{9}$\ iterations, with one point represented every $10009$ iterations.
The\emph{\ inner shell }appears slightly darker, surrounding the \emph{%
plasma core} (white) and the \emph{magnetic axis}. Long \emph{sticking}
occurs here around the five small $q=5/2$\ islands (dark lines in the upper
part of the \emph{chaotic sea}).}
\label{ExFig12}
\end{figure}

The chaotic sea is represented on the Fig.(\ref{ExFig12})\textbf{\ }where we
recognize the protected zones corresponding to the island remnants with $q=4$%
, $3$, $2$ and $1$ (the central core) beside all other main rational chains.

\bigskip 

\begin{figure}
\centering
\includegraphics[width=9.0325cm]{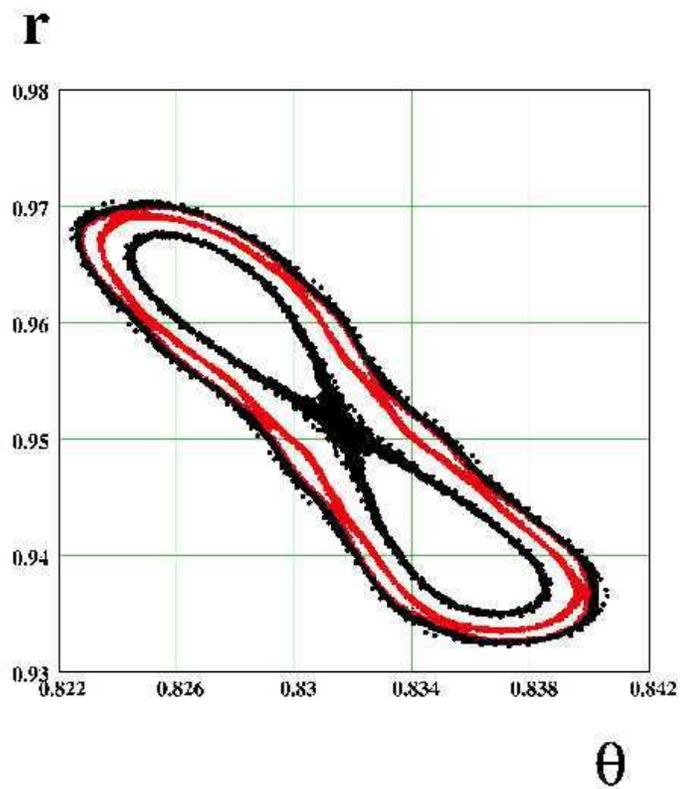}
\caption{Detail of a small
structure of the phase portrait around one of the five $q=5/2$\ islands,
where a long sticking stage occurs during $10^{8}$\ iterations along the
trajectory represented in Figs. 11 and 12.}
\label{ExFig13}
\end{figure}

\bigskip For this trajectory the \emph{sticking} stage is particularly long
around the boundary of the $q=5/2$ island remnant, and lasts for several $%
10^{8}$ iterations. Starting from points in the neighborhood, and performing
a small number of iterations ($N=16380$) we find that this dense stochastic
zone takes a figure-eight form (which indicates that a period doubling has
occurred) with 9 daughter islands around (see Fig.(\ref{ExFig13}))\textbf{. }%
Each of the central elliptic points in these $q=5/2$ islands has already
bifurcated, giving rise to an \emph{inverse hyperbolic point} and to two new
elliptic points.

\bigskip

It is most interesting to determine the \emph{time behavior} of the magnetic
line position in its complicated motion across the chaotic sea of \ the
poloidal plane. The full information has been represented on an animated
movie \cite{GIFmovie}. We have followed a very long trajectory on $2.10^{9%
\text{\ }}$time steps starting from $\psi =0.0136125$, $\theta =0.033$. As
discussed in \textbf{Appendix A }(Eq. (\ref{A8 Psider})), this initial
condition corresponds to a radius $r=0.165$ , slightly out of the central
protected core of the plasma (we recall that in these notations, the edge of
the plasma $\psi =1$ is at $r=\sqrt{2}$ ). From such initial conditions, the
road is completely open up to the edge of the plasma, which is a resistant
KAM barrier for this $L$ value, but the path is far from trivial: the
magnetic line describes a path ''\emph{percolating}'' among the rich variety
of island remnants which are known to form a \emph{hierarchical fractal
structure}. In that movie one remarks that the line remains for a very long
''time'' in a \emph{chaotic layer}, between the central protected core and
some visible transport barrier \emph{''around''} $q=9/8$, and then crosses
this barrier to explore the upper \emph{chaotic sea} (extending up to the
plasma edge). Such transitions across the barrier occur repeatedly, inward
and outward, in an intermittent way and after random \emph{residence times}.
Shorter periods of \emph{sticking} inside this barrier are also observed.

\bigskip

\begin{figure}
\centering
\includegraphics[width=14.1221cm]{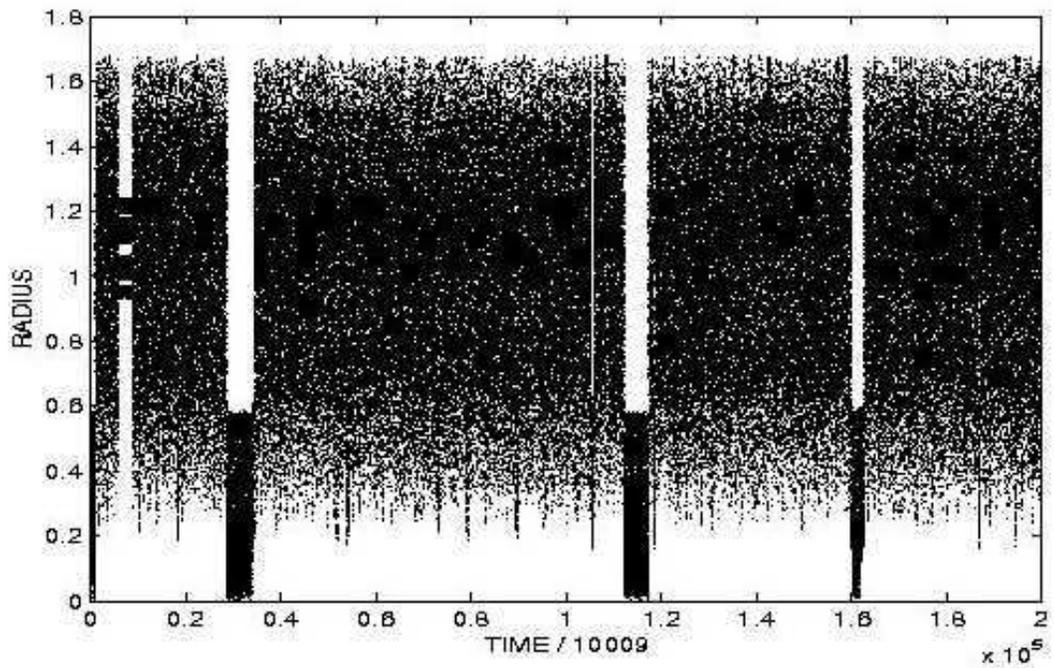}
\caption{Radial position of a single
trajectory followed during $2.10^{9}$\ iterations, as function of time for $%
L=4.875/2\protect\pi \sim 0.776,$with initial condition $\protect\psi %
=0.0136125$, $\protect\theta =0.033$\ : different stages of trapping in
different zones indicate an intermittent behaviour. Note that trapping
stages are very long at may last for at least several $10^{7}$\ iterations
for this value of the stochasticity parameter, near the escape threshold.\
(Such long times near the critical threshold for large scale chaotic motion
could be reminiscent of critical phenomena.}
\label{ExFig11}
\end{figure}

This behavior can also be represented\emph{\ }in a $(r,t)$ graph shown in
Fig.(\ref{ExFig11}) where we present the time variation of the radial
position : this behavior is not only stochastic, but presents different
stages :

- a first period of \emph{pseudo-trapping} or \emph{temporary confinement}
in an inner shell below $r\precsim 0.6$,

- an escape in an external shell, and a wandering stage in a \emph{wide
chaotic sea}, avoiding island remnants, up to $r\precsim 1.6$,

- a period of \emph{deep sticking} around the five island remnants $q=5/2$
(only four bands can be seen since two of the five islands overlap in
radius),

- a new wandering in the chaotic sea of the external shell, then a second
stage of pseudo-trapping in the inner shell,

- and so one and so forth...

In other words the oscillating radial position of a single line is seen to
wander randomly from the (inner) stochastic shell, to the stochastic sea
(sometimes around $q=5/2$ island remnants) with small downward peaks
indicating excursion towards and \emph{inside} the transport barrier, like
flood tide and ebb tide.

\bigskip

Such an \emph{intermittent behavior} is typical of the phenomena of
alternative trapping into different ''basins'' of a \emph{Continuous Time
Random Walk }(CTRW) \cite{Raduliv97}, as previously observed \cite%
{SymbolDynam98} in a stochastic layer of the Chirikov-Taylor standard map.
One may wonder why these average residence times (trapping times) are so
long in the present case. This could be related with the fact that the value 
$L=4.875/2\pi $ considered in Fig.(\ref{ExFig11}) is not far from the escape
threshold $L\sim 5/2\pi $, and that the latter could be the analogous of a 
\emph{critical point of percolation}, in the vicinity of which cluster
lengths and diffusion characteristic times are generally diverging as some
inverse power of the deviation from the critical parameter value, with a
critical exponent.

\bigskip

This conjecture however remains to be proved. In the present problem, such
long characteristic times seem to exclude the possibility to compute the
histogram (the probability distribution of these residence times), which
would need awfully long iterations times in order to obtain a good
statistics.

\smallskip

\section{\protect\medskip \label{LocalisationITB}Localization of transport
barriers}

\medskip

The intermittent motion described in Fig.(\ref{ExFig11})\textbf{\ }clearly
exhibits intermittent periods of confined motion between structures playing
the role of internal transport barriers (ITB), the most resistant curves
inside the confined plasma. It is interesting to identify such barriers and
to note their positions along the $q$ -profile .

\subsection{\label{RoughLOCAL}Rough localization}

\subsubsection{Central core barrier}

An initial period of pseudo trapping is observed, during which the
trajectory remains in the inner shell of Fig.(\ref{ExFig12}) for several $%
10^{6}$ iterations. This shell is separated from the protected central core
of the plasma by a very strong barrier on the plasma edge which has not been
crossed during $2.10^{9}$ iterations. This barrier can thus be considered as
a KAM torus (or a very robust Cantorus if it could have been crossed by
performing still more iterations). By analyzing the innermost points reached
by the long trajectory, and performing short iteration series on several of
them, we can determine which part of the trajectory is a good candidate to
localize the barrier. A rational estimate of the value of the safety factor
can be determined on this trajectory by carefully analyzing the (almost)
periodic repetition of the variation of the poloidal angle $\theta $ in
time, along with an estimation of the average poloidal rotation at each
iteration, or by a direct calculation of the winding number (\ref{Iota
mesure}) after a possible change of coordinates for innermost trajectories
not encircling the origin. We find that a very rough estimate of the inner
barrier protecting the plasma core from invasion from the inner shell is
characterized by: 
\begin{equation}
q_{C1}=\frac{27}{25}=1.080  \label{qC1}
\end{equation}

\medskip

\subsubsection{Internal barrier preventing outward motion in the chaotic
shell (lower Cantorus)}

In this inner shell, the outward motion is limited by a series of points
which has been analyzed in a similar way. Measurement of the $q$-value of
the outermost trajectory in a short sample yields the following rough
rational estimate: 
\begin{equation}
q_{C2}=\frac{10}{9}=1.111  \label{qC2}
\end{equation}

\medskip

\subsubsection{External barrier on the plasma edge}

\begin{figure}
\centering
\includegraphics[width=8.5339cm]{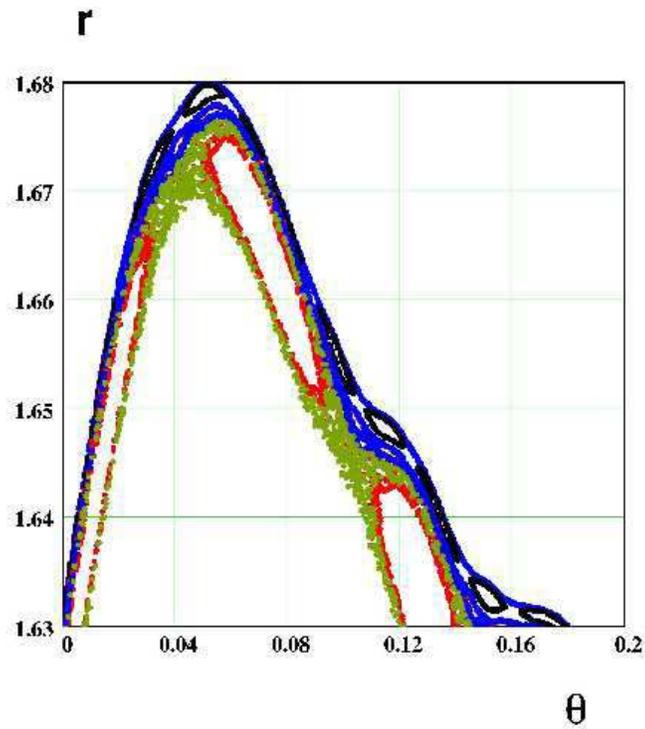}
\caption{Structure of rational
surfaces (islands) and chaotic zones below the edge KAM torus $N(4,2)$
(boundary circle) and the three convergents from below to $N(4,2$): $q=13/3$%
, $q=35/8$, $q=92/21$. This detailed drawing for $L=4.875/2\protect\pi \sim
0.776$\ represents short iteration times ($N=8192$), performed on the most
external points selected in the long trajectory of Figs.(\textbf{11} and 
\textbf{13}). This graph reveals different structures near the plasma edge,
allowing to choose the most external part of the trajectory, and a rational
estimate of the $q$-value of the external barrier : a KAM surface (or a
robust Cantorus).}
\label{ExFig14}
\end{figure}

The upper edge of the chaotic sea in Fig.(\ref{ExFig12}) has also been
analyzed in a similar way. Iteration of some of the outermost points allows
us to draw Fig.(\ref{ExFig14}) which allows us to select a good candidate
for the most external local trajectory. The determination of a rational
approximate for its $q$-value yields: 
\begin{equation}
q_{C4}=\frac{92}{21}=4.381  \label{qC4}
\end{equation}
For this external barrier we have successively identified on Fig.(\ref%
{ExFig14}) surfaces with $q=13/3$, $35/8$ and the outermost one : $q=92/21$.
Of course a still better precision is possible, but this is enough for the
present purpose.

\bigskip

\medskip

\subsubsection{\label{int barrier preventing}\label{doublesided}\label%
{double sided}Internal barrier preventing inward motion in the chaotic sea:
a two-sided internal barrier (upper Cantorus)}

\begin{figure}
\centering
\includegraphics[width=13.3774cm]{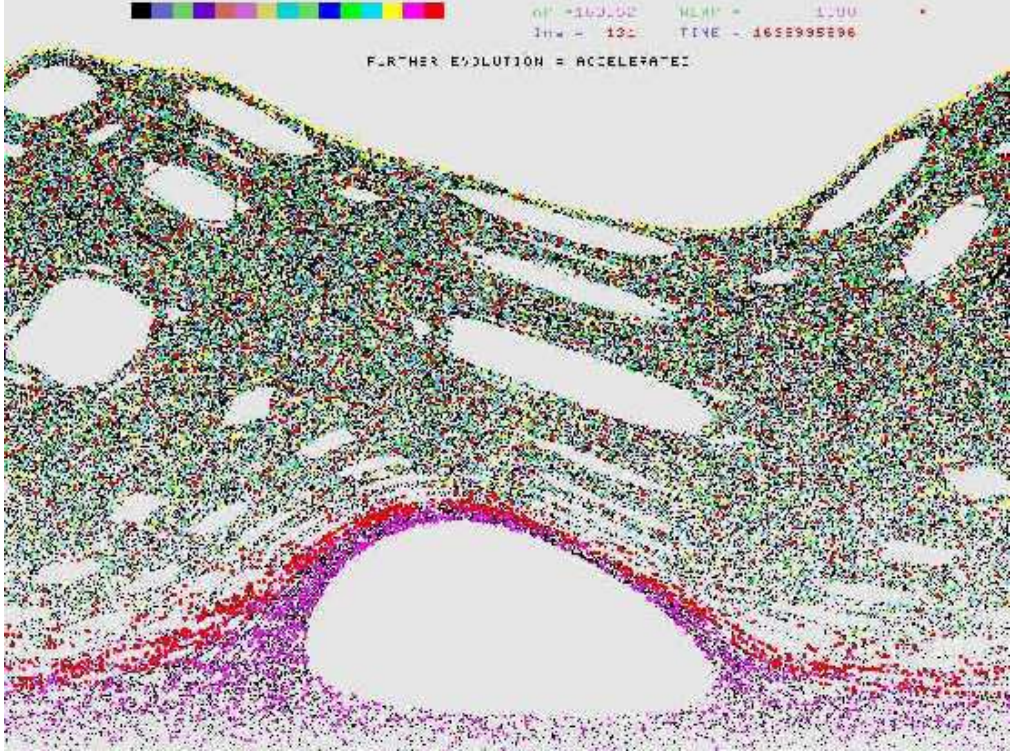}
\caption{One of the last snapshot in
the java animation. Bold points in this phase portrait $(\protect\psi ,%
\protect\theta )$ represent the most recent iterated points along a single
magnetic line trajectory. Here, after filling the whole chaotic sea, the
line is partly confined in the chaotic shell and then is sticking around the
internal barrier.}
\label{ExFig15}
\end{figure}

During long intermediate stages in the intermittent history of Fig.(\ref%
{ExFig11}), the trajectory remains above what appears as an \emph{internal
barrier preventing the inner motion}. A similar analysis allows us to
determine a rough rational approximate for the $q$-value of the innermost
part of this trajectory: 
\begin{equation}
q_{C3}=\frac{17}{15}=1.133  \label{qC3}
\end{equation}%
which appear to be different from the neighboring surface $q_{C2}$
preventing upward motion from the inner shell. We are thus in presence of a 
\emph{two-sided transport barrier}. It is quite remarkable that both sides
could have been distinguished. Actually this difference clearly appears to
the eyes on the movie \cite{GIFmovie} which has been realized on this
simulation, in which we present a succession of snapshots to which new
groups of points are added in a discontinuous way on each new picture : it
can clearly be observed on some pictures how new, ''fresh'' or ''recent''
points are really aligned along a barrier which is different from the lower
one. One of these snapshots is presented in Fig.(\ref{ExFig15}).

\subsection{\label{Section 4.2}Expectations from theory of nonlinear
dynamical systems}

From classical theory of chaos in simple nonlinear dynamical systems \cite%
{RMacKayBook}, it is expected that the most resistant KAM\ torus in a $q-$%
interval of values is either the Golden number or at least a \emph{''noble''
number} (after Percival \cite{Percival}) . This is known to be the case in
the standard map. When represented in a continuous fraction expansion \ 
\begin{equation}
\left[ a_{1},a_{2},a_{3},....,a_{j}\right] =1/\left( a_{2}+1/\left(
a_{3}+1/...+1/a_{j}\right) \right)  \label{ContinFRACTExpans}
\end{equation}%
the Golden number has a simple $\left[ \left( 1,\right) ^{\infty }\right] $
coding: 
\begin{equation}
G\equiv 1+\frac{1}{1+\frac{1}{G}}=\frac{\sqrt{5}+1}{2}=\left[ 1,1,1,1,......1%
\right] =\left[ \left( 1,\right) ^{\infty }\right] =1.61803399....
\label{Queb 2}
\end{equation}%
and other $\emph{noble}$ $\emph{numbers}$ have a $\left( 1,\right) ^{\infty
} $ tail, with other integers before. Most noble numbers are the ''most
irrational'', defined as those with the smallest number of integers before
the $\left( 1,\right) ^{\infty }$ tail.

\bigskip

In presence of magnetic shear the situation can be different: for instance,
to the best of our knowledge, nobody knows why the Golden number actually
plays no special role in the tokamap with monotonous $q$ -profile \cite%
{Tmap1}\textbf{. }In other words, the special role played by the Golden KAM
in the standard map does not seem to be conserved in other maps in presence
of shear.\ We denote the most noble numbers of interest here by 
\begin{equation}
N(i,j)\equiv \left[ i,j,\left( 1,\right) ^{\infty }\right] =i+\frac{1}{j+%
\frac{1}{G}}  \label{Queb 3}
\end{equation}%
($i,j\in Z,j>1)$ which represent the \emph{''next most irrational'' numbers}
after the Golden one $G\equiv N(1,1).$ For $i=1$ we note that these noble
numbers are ''good milestones'' in the $q$ -profile since they are rather
well distant, and even \emph{of measure zero} \cite{Meiss 92}. It is simple
to prove that the values of $N(1,j)$ are actually inserted in the $q$
-profile between successive dominant island chains $Q(j)\equiv j/(j-1)$
since it is easy to demonstrate that:

\begin{equation}
Q(1+j)>N(1,j)>Q(2+j)  \label{nobles inserted}
\end{equation}%
Better and better rational approximants to a noble number $\omega $ (in the
sense of the Diophantine approximation)\ \ are obtained by truncating this
infinite series at a higher and higher level : these rational approximants
are known to converge towards the noble number $\omega $ and are called the 
\emph{''convergents''}, with the property ''to be the closest rational to
the irrational $\omega $, compared to rationals with the same or smaller
denominator'' \cite{RMacKayBook}. The successive convergents to an
irrational number 
\begin{equation}
\omega =\left[ a_{1},a_{2},a_{3},....,a_{j,}...\right]  \label{irrat}
\end{equation}%
are defined by\ (see (\ref{ContinFRACTExpans})) the series : 
\begin{equation*}
\begin{array}{cccccc}
\lbrack a_{1}]\equiv a_{1} & \text{,} & [a_{1},a_{2}]=a_{1}+\frac{1}{a_{2}}
& \text{,} & \left[ a_{1},a_{2},a_{3}\right] =a_{1}+\frac{1}{a_{2}+\frac{1}{%
a_{3}}} & \text{,}%
\end{array}%
\end{equation*}

\begin{equation}
\begin{array}{ccccc}
\left[ a_{1},a_{2},a_{3},a_{4}\right] & \text{,} & \left[
a_{1},a_{2},a_{3},a_{4},a_{5}\right] & \text{,} & \text{etc...}%
\end{array}
\label{CONvergents}
\end{equation}

\bigskip

When looking at the \emph{most noble} $q-$values in any given interval
between $q_{0}=m_{0}/n_{0}$ and $q_{1}=m_{1}/n_{1}$ (with $\left|
m_{0}n_{1}-m_{1}n_{0}\right| =1$ [\footnote{%
As a direct consequence of this fundamental recurrence relation for
continued fractions, this relation simply expresses the fact that the two
limits $m_{0}/n_{0}$ and $m_{1}/n_{1}$of the considered interval are
actually two successive convergents (or approximants) of some continued
fraction.}]), it is known \cite{GrMKayStark86} that the most irrational
number is 
\begin{equation}
\alpha _{1}=\frac{m_{0}+G.m_{1}}{n_{0}+G.n_{1}}  \label{Queb 4}
\end{equation}%
and is expected to correspond to the most robust barrier in that interval.
If we now look at the successive intervals $\left\{ Q(m+1),Q(m+2)\right\} $
between the main rational chains of the dominant series $q=Q(m+1)=(m+1)/m$
(as observed in Fig.(\ref{ExFig12}), we see that these intervals are \emph{%
Farey intervals} \cite{McKaySt92} covering the real axis and the most noble $%
q-$values in each of these intervals are precisely given by Eq. (\ref{Queb 3}%
) : 
\begin{equation}
\alpha _{m}=\frac{m+1+G(m+2)}{m+G(m+1)}=1+\frac{1}{m+\frac{G}{G+1}}=1+\frac{1%
}{m+\frac{1}{G}}=\left[ 1,m,\left( 1,\right) ^{\infty }\right] =N(1,m)
\label{Queb 5}
\end{equation}%
(where we used $G+1=G^{2}$). These most noble numbers are not frequent and
simply alternate with the main rational chains. In the barrier found
''around'' $q=9/8$ the candidates for robust circles are thus $q=N(1,7)$ and 
$N(1,8)$ since we have $8/7>N(1,7)>9/8>N(1,8)>10/9$.

\bigskip

\begin{figure}
\centering
\includegraphics[width=14.0518cm]{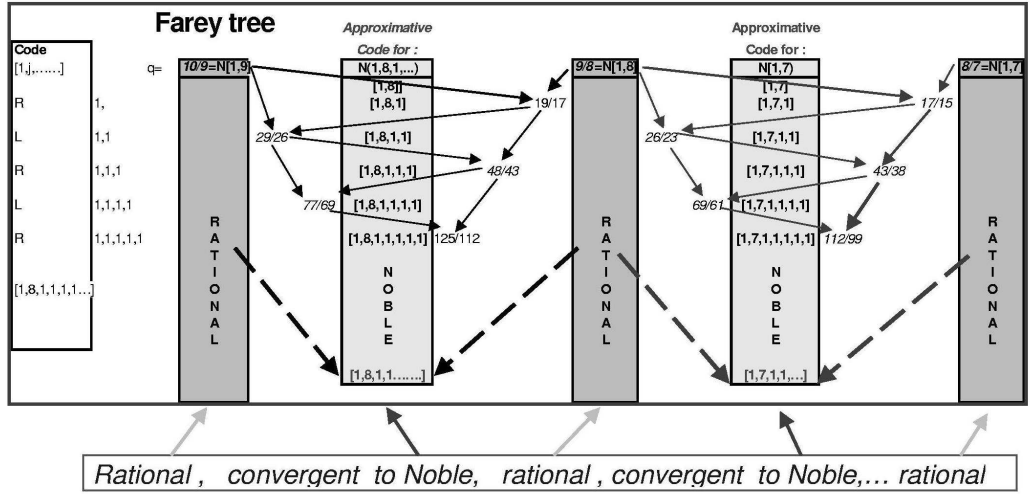}
\caption{Farey tree indicating how the
most noble number in a given interval between rationals ($10/9$ and $9/8$,
or $9/8$ and $8/7$) can be built with results $N(1,8)$ or $N(1,7)$.}
\label{NewFigA}
\end{figure}

It is worth mentioning that the convergents towards these two noble numbers
are, for the lower Cantorus\ (see the lowest approximation $10/9$ found in (%
\ref{qC2})) : 
\begin{equation}
\frac{10}{9}\rightarrow \frac{10+9}{9+8}=\frac{19}{17}\rightarrow \frac{29}{%
26}\rightarrow \frac{48}{43}\rightarrow \frac{77}{69}\rightarrow \frac{125}{%
112},...\Rightarrow \left[ 1,8,\left( 1,\right) ^{\infty }\right]
=N(1,8)=1.116<\frac{9}{8}  \label{Queb 6}
\end{equation}
and for the upper Cantorus (see the lowest approximation $17/15$ found in (%
\ref{qC3})) :

\begin{equation}
\frac{9}{8}\rightarrow \frac{9+8}{8+7}=\frac{17}{15}\rightarrow \frac{26}{23}%
\rightarrow \frac{43}{38}\rightarrow \frac{69}{61}\rightarrow \frac{112}{99}%
,...\Rightarrow \left[ 1,7,\left( 1,\right) ^{\infty }\right] =N(1,7)=1.131<%
\frac{8}{7}  \label{Queb 7}
\end{equation}%
where successive approximants are alternatively from below and from above,
due to the $\left( 1,\right) ^{\infty }$ series in the \emph{Farey coding} %
\cite{McKaySt92}, see Fig.(\ref{NewFigA}).

\bigskip It is well known that the numerators and denominators $p_{\mu }$ of
the successive approximants actually grow as a \emph{Fibonacci series} $%
p_{\mu +2}=p_{\mu +1}+p_{\mu }$, with an asymptotic growth rate $%
\lim_{\lambda \rightarrow \infty }$ $p_{\mu +1}/p_{\mu }=G$ given by the
golden number $G$.

\bigskip

\begin{figure}
\centering
\includegraphics[width=8.8436cm]{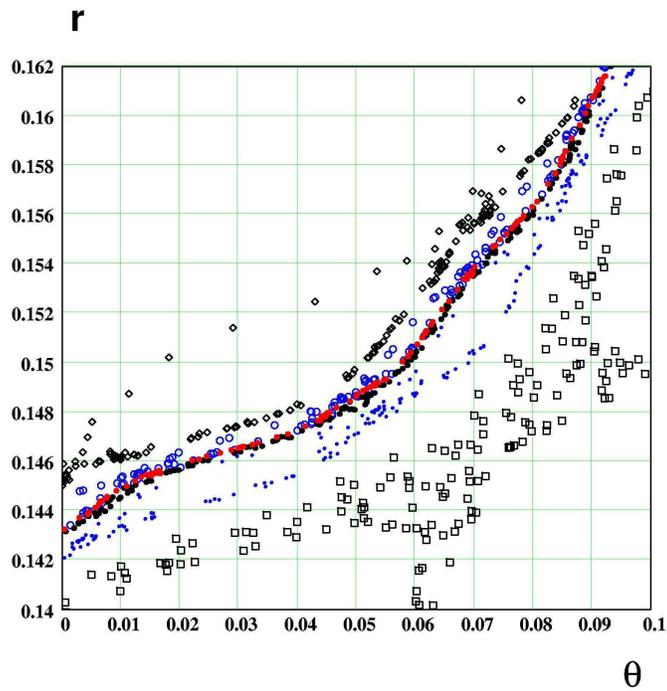}
\caption{The upper Cantorus
for $L=4.875/2\protect\pi $ can be localized in the central region, between
bold dots.\ Here we have plotted series of $2046$ iterations along each of
the island chains with rational $q$ values corresponding to the following
approximants to the upper Cantorus $N(1,7)$ : $q=17/15$, $43/38$ and $112/99$
from above, and $q=9/8$, $26/23$ and $69/61$ from below.}
\label{FigUpperCantorus}
\end{figure}

\bigskip

In Fig.(\ref{FigUpperCantorus}), we have plotted a series of $2046$
iterations along the island chains with rational $q$-values corresponding to
the following approximants to the upper Cantorus (\ref{Queb 7}) : $q=17/15$, 
$43/38$ and $112/99$ from above, and $q=9/8$, $26/23$ and $69/61$ from
below. This allows us to localize the upper Cantorus on this graph as being
located in the very thin interval located between bold dots.

\subsection{\label{convergents}Identification of ''convergents'' towards
these most noble Cantori}

It is simple but numerically delicate to check that the barrier is composed
of two Cantori localized precisely on these two most noble numbers : a \emph{%
lower Cantorus }on $N(1,8)$ and an \emph{upper Cantorus }on $N(1,7)$. Why
are the most resistant KAM tori located around $q=9/8$ for this value of $L$
could probably not be deduced on theoretical grounds. Of course, beside the
most robust barrier described here, other, less robust barriers also exist.
A detailed analysis of the various thin downward peaks in Fig.(\ref{ExFig11}%
) reveals that the lower boundary of the magnetic line motion in the
stochastic sea is actually oscillating in time, and does not always
correspond to the same upper Cantorus above the island chain $q=9/8$ as
defined in Eq. (\ref{Queb 7}). The lower boundary of the stochastic sea can
be found on different Cantori located most often above the island chain $%
q=6/5$, but also on the Cantori above $q=7/6$ and $8/5.$ This seems to
indicate that a \emph{global barrier} could rather be composed of \emph{%
different sticking regions} around island chains with main rational $q-$%
values, and that these sticking regions are limited successively by the
Cantori between them, in some kind of cascading process.\ Global motion in
these boundary regions could then be described \cite{GIFmovie} as a
succession of approaches, like \emph{flood tide and ebb tide}, allowing the
magnetic line to pass through the successive barriers.

\bigskip

During the upward motion starting from the chaotic layer, the most important
barrier (but not the last one) nevertheless remains the upper Cantorus $%
q=N(1,7)$ above the islands $q=9/8$ : once this one has been passed, the
magnetic line rapidly invades the whole chaotic sea. As seen in Fig.(\ref%
{ExFig11}), the inverse, downward motion from the chaotic sea however
encounters several barrier crossing, back and forth, before crossing the
upper Cantorus and entering in the zone between the two Cantori composing
the ITB described in this paper.

\bigskip

\subsubsection{Convergent island chains towards the two Cantori}

Localization of island chains with given $q-$value can be obtained
numerically with great precision by searching hyperbolic and/or elliptic
periodic points, by a numerical algorithm derived from a generalization of
the Fletcher-Reeves method, involving the Jacobi matrix of the tokamap,
explained in \textbf{Appendix B}. Localizing the position of a noble
Cantorus can only be achieved as a limiting procedure, by localizing the
series of its convergents, as given by Eqs. (\ref{Queb 6}, \ref{Queb 7}).

In order to check the above predictions we have sampled (in a very long
trajectory of $2.10^{9}$ iterations) various sections of trajectory \emph{%
(a) }reaching the highest $\psi -$values in the\emph{\ chaotic shell} (below
the lower Cantorus), and \emph{(b)} reaching the lowest $\psi -$values in
the \emph{chaotic sea }(above the upper Cantorus): such trajectories remain
indeed well separated, by the width of the barrier. Then we have checked
that any of these sections of trajectory \emph{(a)} remains indeed \emph{%
below} the limit of convergence of $N(1,8)$ of the lower Cantorus, more
precisely below the limit of the convergents from below : 
\begin{equation}
\begin{array}{ccc}
\frac{10}{9},\frac{29}{26},\frac{77}{69},\frac{202}{181},\frac{529}{474},%
\frac{1385}{1241},...\Rightarrow N(1,8) &  & \text{(from below)}%
\end{array}
\label{LowCfromBELOW}
\end{equation}
and that any of these sections of trajectory \emph{(b)} remains indeed \emph{%
above} the limit of convergence of $N(1,7)$ of the upper Cantorus, more
precisely above the limit of the convergents from above : 
\begin{equation}
\begin{array}{ccc}
\frac{17}{15},\frac{43}{38},\frac{112}{99},\frac{293}{259},\frac{767}{678},%
\frac{2008}{1775}...\Rightarrow N(1,7) &  & \text{(from above)}%
\end{array}
\label{UpperCANTfromABOVE}
\end{equation}
In other words we have checked that observed trajectories in the chaotic
shell and in the chaotic sea remain actually always on their own side of the
pair of Cantori, which localizes the barrier.

\subsubsection{Convergent island chains towards noble KAM's on the edge and
around the plasma core}

By the same method we have also identified the $q-$value of the robust
boundary circle forming the plasma edge as being equal to $N(4,2)$ which
appears to be the most irrational between $q=4$ and $q=4.5$, according to
Eq. (\ref{Queb 4})\textbf{\ }: 
\begin{equation*}
\frac{4}{1}\rightarrow \frac{4+9}{1+2}=\frac{13}{3}\rightarrow \frac{22}{5}%
\rightarrow \frac{35}{8}\rightarrow \frac{57}{13}\rightarrow \frac{92}{21}%
\rightarrow \frac{149}{34},...\Rightarrow
\end{equation*}
\begin{equation}
\left[ 4,2,\left( 1,\right) ^{\infty }\right] =N(4,2)=4.381<\frac{9}{2}
\label{Queb 8}
\end{equation}
in agreement with (\ref{qC4}). There should probably exist a stronger
barrier out of the ''plasma edge'' on $N(4,1)\sim 4.618,$ but this one
remains out of reach for trajectories starting from the plasma bulk at the
considered value $L=4.875/2\pi $ since the KAM on the edge at $%
q=N(4.2)<N(4,1)$ has not yet been broken.

\bigskip

We have also identified the $q-$value of the robust boundary circle
protecting the plasma core as being equal to $N(1,11)$, the most irrational
between $q=1$ and $q=12/11$, according to Eq. (\ref{Queb 4}): 
\begin{equation*}
\frac{1}{1}\rightarrow \frac{1+12}{1+11}=\frac{13}{12}\rightarrow \frac{25}{%
23}\rightarrow \frac{38}{35}\rightarrow \frac{63}{58}\rightarrow \frac{101}{%
93},...\Rightarrow
\end{equation*}
\begin{equation}
\Rightarrow \left[ 1,11,\left( 1,\right) ^{\infty }\right] =N(1,11)=1.086<%
\frac{12}{11}  \label{Queb 9}
\end{equation}
which is in poor agreement with the above rapid estimate (\ref{qC1}), but
correct to the first three digits.

\subsubsection{\label{SecITB9s8}The ITB: a double sided barrier around $%
q=9/8 $}

\bigskip The final scheme which results from the above measurements on a
very long tokamap trajectory at $L=4.875/2\pi $ is the following.\ A
magnetic line with an initial condition inside the \emph{inner shell }(IS)
actually has an \emph{inward} motion limited by a robust \emph{KAM torus}
protecting the plasma core (or a Cantorus ?) at 
\begin{equation}
q_{IS}>N(1,11)=1.086>\frac{13}{12}  \label{qIS min}
\end{equation}
and an \emph{outward} motion limited by a \emph{semi-permeable Cantorus }at%
\emph{:} 
\begin{equation}
q_{IS}<N(1,8)=1.116  \label{qIS max}
\end{equation}
This numerical analysis shows that noble Cantori are good candidates to be
identified with internal transport barriers, as could have been anticipated
from the relation between KAM theory and number theory \cite{Meiss 92}.

\medskip

\medskip Once arrived in the main \emph{chaotic sea} (CS) extending up to
the plasma edge, the magnetic line wanders around island remnants, remains
stuck around some of them (here in Figs.(\ref{ExFig11}, \ref{ExFig12})
around the $q=5/2$ chain), but its inward motion is limited by a
semi-permeable Cantorus 
\begin{equation}
q_{CS}>N(1,7)=1.131>\frac{9}{8}  \label{qCS min}
\end{equation}
The measurements indicate that this inward-motion limit in the chaotic sea
is different from (and located above) the outward motion limit of the IS : 
\begin{equation}
N(1,8)=1.116<\frac{9}{8}=1.125<N(1,7)=1.131  \label{twosidebarrier}
\end{equation}
We have thus observed the existence of \emph{a two-sided transport barrier}
around $q=9/8$ and limited on each side by a Cantorus, respectively at $%
q=N(1,8)$ and $q=N(1,7)$.

On the other end, the outward motion in the chaotic sea is limited by a
curve which will be proved in Section (\ref{Dana}) to be a KAM surface at 
\begin{equation}
q_{CS}<N(4,2)=4.382  \label{qCS max}
\end{equation}
which is the most irrational number between $4$ and $4.5$. This external KAM
surface thus appears indeed a good candidate for the external barrier
observed here.

\bigskip

We are thus in presence of \emph{a double-sided transport barrier}, composed
of \emph{two Cantori}, the lower Cantorus with noble $q-$value $N(1,8)=1.116$%
, the upper Cantorus with noble value $N(1,7)=1.131.$ These Cantori appear
on Fig.(\ref{Figb3Quebec}). From (\ref{nobles inserted}) we have $%
N(1,7)>Q(9)>N(1,8)$ , which shows that the rational surface $q=Q(9)=9/8$ is
actually \emph{between} these two Cantori and thus inside the transport
barrier. In this sense, one can say that \emph{this internal transport
barrier is actually located around a dominant rational surface}, in spite of
the fact that it is actually composed of two irrational, noble Cantori.

\subsubsection{Experimental localization of barriers in the Tokamak $q$
-profile}

From the experimental point of view, transport barriers in tokamaks are
indeed generally observed \emph{''around''} the main rational $q$-values %
\cite{Lop-Cardozo2}.

\medskip

In the RTP tokamak, with a wide range of $q$-values ($0.8\Longrightarrow 5$)
in a reversed sheared profile, it has been observed, by varying the heat
deposition radius $\rho _{dep}$ of off-axis ECH heating, that the central
electron temperature $T_{e}(0)$ decreases by a series of \emph{plateaux} %
\cite{Lop-Cardozo2}. It is observed that the values of \ $q_{m}$ of the
different plateaux fall in half-integer bands, \emph{i.e.} $q_{m}$ crosses a
half-integer value each time the discharge transits from one plateau to the
next. Since these transitions correspond to the loss of a TB, these authors
deduced that ''the barriers are associated with half-integer values of $q$
.''

\bigskip

It has been reported that some ''ears'' appear in the $T_{e}$ profile
(appearance of one bump on each side of the central value) when the heat
deposition radius $\rho _{dep}$ is localized inside a transition between two
plateaux \cite{Lop-Cardozo2}. We note that the existence of such ''ears''
could be an indication in favor of the existence of a double-sided
semi-permeable TB around $\rho _{dep},$ as found in the previous Section (%
\ref{SecITB9s8}). Such ''ears'' appear to be unstable and to crash in a
repetitive fashion, showing that the barrier can indeed be crossed and
appears to be permeable.

\medskip

The central sawteeth allows to place the first barrier ''near'' $q=1$ ;
off-axis sawteeth indicate other barriers near $q=3/2$, $2$ and $3$.
Remaining barriers are attributed to $q=4/3$ and $5/2$.

\medskip

In all cases the barriers observed in experiments are associated with those
''dominant rational'' $q$-values, but a specific experimental work could
hardly have been done in order to determine more precisely the possible role
of noble or other irrational values around these ''dominant rationals''.

\medskip

Up to now it has been generally admitted \cite{Lop-Cardozo2}, \cite{TS1}, %
\cite{JET1} that internal transport barriers could correspond to \emph{%
rational} $q$-values, where primary island appear. Chaotic motion can be
observed between these primary rational islands but mainly around the
hyperbolic points. The picture we obtain here is rather different.\ It is
well known that, in presence of several island chains, secondary islands
appear and accumulate around \emph{irrational }surfaces where KAM surfaces
finally broke themselves into discrete pieces forming a Cantorus. This is
precisely the location where we have found internal barriers: they appear as 
\emph{Cantori on irrational surfaces rather than rational surfaces}.

\bigskip

This finding can in turn be helpful to build transport models like the $q$%
-comb model \cite{Hogeweij 1998}, \cite{Schilham PhD 2001}.\ From
experimental considerations, such models define indeed barriers are
localized around main rational $q$-values, but the width if the barriers has
still to be defined.\ We propose to localize the edges of each barrier on
the most irrational value in the prescribed domain.

\medskip

\subsection{Last barrier in the standard map: the golden KAM}

\begin{figure}
\centering
\includegraphics[width=8.725cm]{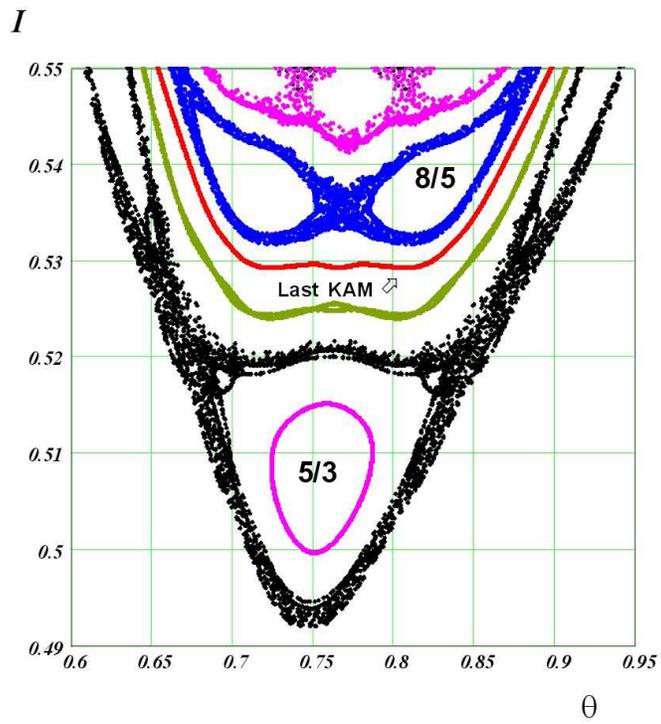}
\caption{Part of the phase
portrait of the standard map for $K=K_{C}$. Two series of islands $q=8/5$
and $q=5/3$ can \ be seen, corresponding to two convergents towards the
Golden number $G$. \ The last KAM with $q=G$ is exhibited inbetween, and it
appears as the last existing non-chaotized KAM surface; its initial
conditions are $I=0.52969$, $\protect\theta =0.75$.}
\label{GoldenKAM}
\end{figure}

\bigskip In order to illustrate the destruction of a robust barrier, there
exists one example in which a KAM\ surface can be represented just at the
critical point where it is broken into a Cantorus. This is the case of the
standard map \cite{Chiri79} where the critical value of the stochasticity
parameter is known with a very high precision. It is interesting to remind
that the last KAM surface in that case has been identified to be the Golden
KAM with $q$-value equal to $G$, the golden number.\ The breaking of this
surface and its transformation into a Cantorus is known to occur at a
critical value of the stochasticity parameter which is known \cite{Meiss 92}%
\ to be given by $K=K_{C}=0.971635406$. Because such a drawing is not easily
found in the literature for this value of $K$, it appears worthwhile to
present the trajectory following this critical curve. In Fig.(\ref{GoldenKAM}%
) we present part of the phase portrait of the standard map for $K=K_{C}$.
Two series of islands remnants $q=8/5$ and $q=5/3$ can \ be seen,
corresponding to two convergents towards the Golden number $G=\lim $ $%
of\left( \frac{3}{2},\frac{5}{3},\frac{8}{5},\frac{13}{8},...\right) $. \
The last KAM with $q=G$ is exhibited inbetween, and it appears as the last
existing non-chaotized KAM surface, surrounded by chaotic layers around
rational islands.

\bigskip 

This example shows the detailed structure of an irrational KAM barrier
surrounded by chaotic, permeable zones.

\bigskip

\subsection{\label{Perturbedq}Perturbed $q$ -profile in the tokamap}

\subsubsection{Exact perturbed $q$-profile}

In order to understand the radial positions of these barriers on the exact $%
q $ -profile of the tokamap, we have succeeded to draw a one-dimensional $q$%
-profile of the magnetic surfaces, not as function of the radius (since
perturbed surfaces are not circular anymore), but as function of the
distance $X$ \ between the polar axis and their intersection point in the
equatorial plane. We have calculated the intersections of several
trajectories with the equatorial plane (at $\theta =0$ and $\theta =0.5$)
and measured their $q$-value, by computing the average increase of the
poloidal angle according to Eq.(\ref{Iota mesure}).

\bigskip

\begin{figure}
\centering
\includegraphics[width=9.0523cm]{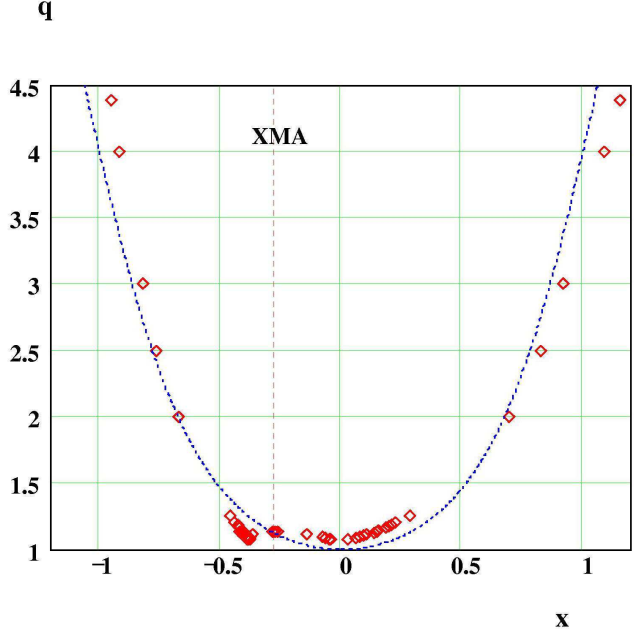}
\caption{Calculated points (diamonds)
of the perturbed $q$-profile (around the magnetic axis) in the equatorial
plane for $L=4.875/2\protect\pi \sim 0.776$, with the plasma extending from $%
X=-1$\ to $+1$, showing overall agreement with the unperturbed profile
(dotted line), except for the the appearance of a local maximum on the
magnetic axis (XMA).}
\label{ExFig17}
\end{figure}

In this way we obtain a rather precise profile of the perturbed $q$-values
of various magnetic surfaces or island chains, represented at the two
(non-symmetrical) points where they cross the equatorial plane, see Fig.(\ref%
{ExFig17}). Let us recall that the radius $r$ represented on the various
figures is defined by $r=\sqrt{2\psi }$ (see Eq.(\ref{newEq35bis}). This
variable $r$ varies from $0$ to $\sqrt{2}$ on the edge of the plasma, while
the variable $x=\sqrt{\psi }=r/\sqrt{2}$ varies from $0$ to $1$ and
represents the \emph{reduced radial coordinate} with respect to the small
radius of the torus. 
\begin{equation}
x=\sqrt{\psi }=r/\sqrt{2}  \label{radii}
\end{equation}%
The abscissa in Fig.(\ref{ExFig17})\textbf{\ }is represented between $%
X\equiv x$ $cos\theta =1$ on the weak field side and $X=-1$ on the strong
field side, as compared with the unperturbed imposed $q$ -profile (Eq. (\ref%
{profilq(psi)})).\bigskip

\begin{figure}
\centering
\includegraphics[width=8.9732cm]{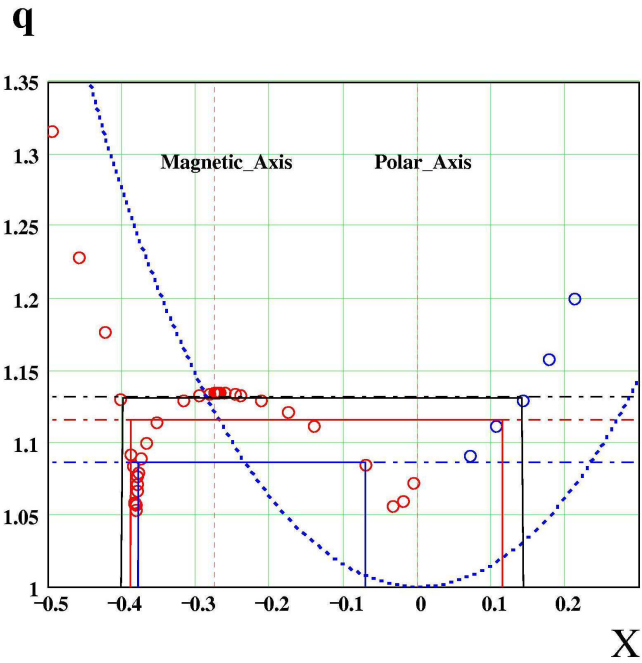}
\caption{Detail of the
perturbed $q$-profile (measured around the magnetic axis) in the equatorial
plane (circles) for $L=4.875/2\protect\pi \sim 0.776$, as compared with the
unperturbed profile (dotted line), showing the appearance of a local maximum
on the magnetic axis, and of two minima. Horizontal levels represent (from
top to botom) the $q$-values $N(1,7)=1.131$ of the upper Cantorus (which
corresponds approximately to the higher $q$-values of the bump), $%
N(1,8)=1.116$ of the lower Cantorus, and $N(1,11)=1.086$ of the KAM barrier
protecting the plasma core, along with the extension of these curves with
the equatorial axis $X$. The value on the magnetic axis is $%
q_{MA}=1.1346>N(1,7)$ and lies at the center of a profile which is locally
hollow at very small distances from the magnetic axis (not visible at this
scale)}
\label{ExFig18}
\end{figure}

\bigskip The measured points roughly agree with the unperturbed curve (Fig.(%
\ref{ExFig17})). But a detailed drawing in Fig.(\ref{ExFig18}) reveals a
rather unexpected point: the perturbed safety factor profile in the
equatorial plane exhibits a \emph{local maximum on the magnetic axis}, and a
minimum on both sides.

\medskip

\medskip The measured $q$-value near the magnetic axis appears to be 
\begin{equation}
q_{MA}\simeq 1.1346\succsim \frac{17}{15}  \label{qMA}
\end{equation}
This \emph{non-monotonous perturbed }$q$\emph{\ -profile }, which is \emph{%
spontaneously created by the magnetic perturbation}, could be the reason for
the appearance of transport barriers in the tokamap.

\bigskip

\subsubsection{\label{SectITBREV}An ITB in the rev-tokamap}

Transport barriers are very frequent in reversed shear situations, and they
also appear in the tokamap model for such cases. We recall indeed that a
transport barrier has already been shown to occur around the minimum of a
reversed shear $q$ -profile (''revtokamap'' \cite{RexTokamap}). In the
discussion of Fig. 4 in Ref. \cite{RexTokamap}, for $L=2.8/2\pi $, it is
observed that ''the chaotic region is sharply bounded from below by a KAM\
barrier'' located at $1/q=0.5772$. We note that this value is only $0.5\%$
away from a noble value $1/q=0.5802$ which corresponds to $q=[1,1,2,\left(
1,\right) ^{\infty }]$ $=1+\left( 1/\left( 1+1/\left( 2+1/G\right) \right)
\right) $\ $=1.7236068...$ which is larger than the Golden number and
appears from (\ref{Queb 4}) to be the most irrational number between $5/3$
and $7/4$. Within a precision of $0.5\%$, the transport barrier found in %
\cite{RexTokamap} could thus well be located on a noble value again, even in
the previous situation of a reversed magnetic shear profile.\bigskip

\subsubsection{Inverse shear in the tokamap}

The inner shell, which has been observed between $q_{C1}=1.080$ and $%
q_{C2}=1.111,$ actually involves $q$-values lower than that measured on the
magnetic axis (Eq. (\ref{qMA})), down to $q\simeq 1.07$ (see Fig.(\ref%
{ExFig18}), precisely because of the non-monotonous profile.

\medskip \bigskip

One can see that the $q$-values decrease from the magnetic axis towards a 
\emph{separatrix} passing through the polar axis (on which two hyperbolic
points are superposed, as long as $L<1$). This whole region inside the
separatrix would appear however to have\textbf{\ }$q=1$ if and only if 
\textbf{\ }$q$ was computed around the polar (geometrical) axis, but this
way of counting has not much physical meaning, besides showing that this
plasma core could appear as an island $m=n=1$ in the reference frame of the
polar axis. As a result, trajectories inside the chaotic shell have actually 
$q-$values (around the magnetic axis) which decrease from the boundary
circle for those trajectories which do not enclose the polar axis $(\psi
=0,\theta =0)$, but which increase for those which encircle the polar axis,
up to the value $N(1,8)$ of the lower Cantorus. In the chaotic sea, the $q$
-profile is monotonously growing : trajectories have $q-$values higher than
the upper Cantorus and lower than the boundary circle on the edge, which is
the expected situation.

\bigskip

\begin{figure}
\centering
\includegraphics[width=10.9765cm]{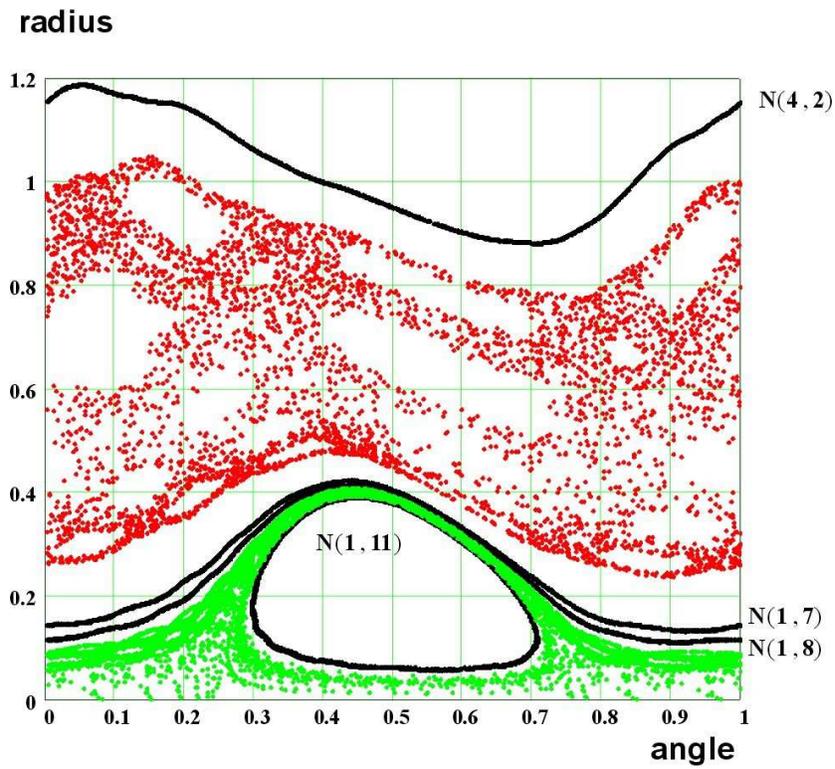}
\caption{The main noble
transport barriers are represented in bold lines in normalized coordinates:
''radius''= $x=\protect\rho /a=\protect\sqrt{\protect\psi }$ and ''angle''= $%
\protect\theta $.\ The robust KAM $N(1,11)$ separating the central protected
plasma core (in white) from the chaotic shell.\ The two semi-permeable
Cantori $N(1,8)$ and $N(1,7)$ form an internal barrier resulting in a very
slow and intermittent motion towards the chaotic sea.\ The robust KAM torus
on the plasma edge $N(4,2)$ has been identified to have a vanishing flux.}
\label{Figb3Quebec}
\end{figure}

In summary, starting from the magnetic axis, we first find a \emph{regular
zone}, with a decreasing perturbed $q$ -profile away from the magnetic axis
(where $q_{MA}\simeq $ $1.13461$), then a robust boundary circle located on $%
q=N(1,11)$ protecting the regular plasma core, then decreasing $q-$values up
to the separatrix, then a regular increase of the $q$ -profile with a $%
q=10/9 $ rational chain, a lower semi-permeable lower Cantorus on $q=N(1,8)$%
, the island remnants $q=9/8$, then the upper Cantorus on $q=N(1,7)$ below
the rational chain $q=8/7$, etc... up to the robust KAM\ torus on the edge
at $N(4,2).$ The two robust barriers and the two Cantori are represented on
Fig.(\ref{Figb3Quebec}) along with a part of each chaotic region.

\bigskip

\subsection{Sticking measurements}

\begin{figure}
\centering
\includegraphics[width=14.1243cm]{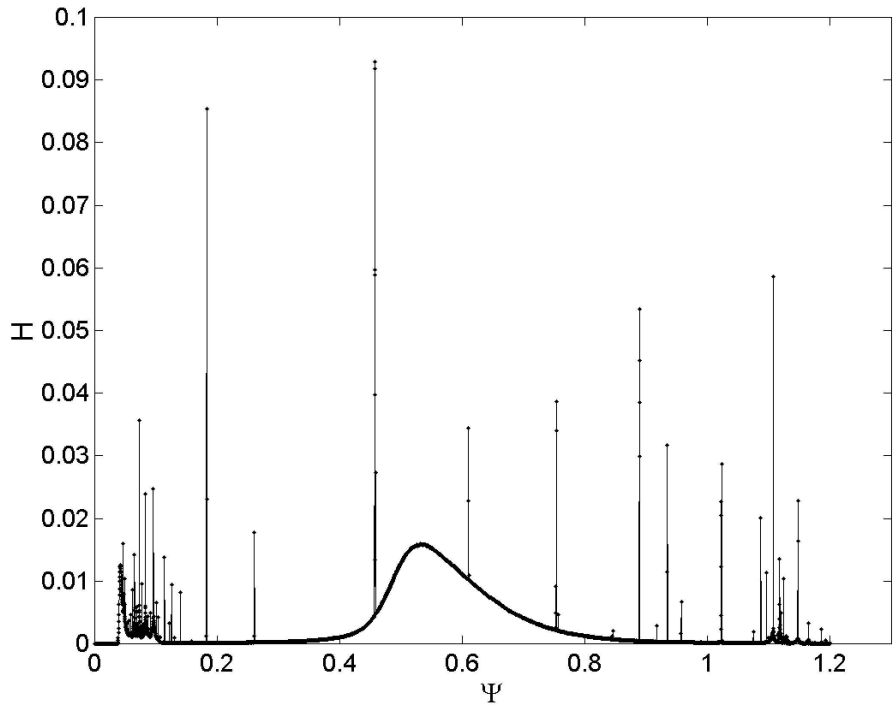}
\caption{Normalized histogram showing the occurrence $H$ of sticking times longer than
$10^5$ (in a run of $10^{11}$\ iterations with $L=4.875/2\protect\pi $), obtained by a sliding average, as function of the radial position $\protect\psi $ of the visited island chain. The values of $\protect\psi $ are taken within small intervals of $2.10^{-4}$. \ This graph exhibits the frequent occurrence of long sticking times in and around the main rational chains \thinspace : the two bumps correspond to widespread $\protect\psi $ values of a line wandering in the main two chaotic zones: the chaotic shell (around $\protect\psi \sim 0.1$) and in the chaotic sea (around $\protect \psi \sim 0.4$ to $0.8$). }
\label{FigMadi}
\end{figure}

\bigskip On the other hand, we have seen\ in Fig.(\ref{ExFig11}) that a long
trapping stage is observed in the chaotic zone surrounding the island
remnants $q=5/2$. For a much longer trajectory ($10^{11}$ iterations) one
observes that such temporary but long trapping can occur around almost all
rational and specially around the main islands known to play a role in the
tokamaks: $q=2/1$, $5/2$ and $3/1$. In Fig.(\ref{FigMadi}) we have computed
the normalized histogram showing the occurrence $H$ of sticking times longer than $10^{5}$ (in a run of $10^{11}$\ iterations with $L=4.875/2\pi $),
obtained by a sliding average, as a function of the radial position $\psi $
of the visited island chain. The values of $\psi $ are taken within small
intervals of $2.10^{-4}$. \ This graph exhibits the frequent occurrence of
long sticking times in and around the main rational chains\thinspace : the
two bumps correspond to widespread $\psi $ values of a line wandering in the
chaotic shell (around $\psi \sim 0.1$) and in the chaotic sea (around $\psi
\sim 0.4$ to $0.8$).\ These are the main two chaotic zones. On the other
hand one remarks sharp peaks representing long sticking events at the
corresponding $\psi $ values.\ In the chaotic sea bump, we identified very
precisely (by measuring the exact average-$\psi $ value of the corresponding 
$q$) long and frequent sticking events around island remnants $q=2$, $5/2$
and $3$ (with satellite peaks at neighboring rationals), etc... This kind of
graph yields a precise measurement of the richness of trapping phenomena.

\bigskip

It would be interesting to study the characteristics of the other \emph{%
secondary barriers}, surrounding the ''holes'' in the phase portrait Fig.(%
\ref{ExFig12}) and to determine if such \emph{local} barriers also
correspond to noble Cantori in the profile of the local $q$-value (around
the local magnetic axis in the center of the island remnant). For instance a
sticking region is easy to observe along the edge on the $q=9/8$ islands.

The interest of this new description of ITB's\ in tokamaks is that
mathematical tools exist for the description of the motion across such
Cantori (''turnstiles'' etc... \cite{turnstile}), and these tools can be
used to study transport across internal transport barriers in confined
plasmas. Calculation of the flux through a Cantorus, defined by the limit of
its convergents, can be performed by using the \emph{turnstile} mechanism %
\cite{turnstile}.\ This subject is presented in the next Section \ref{Dana}.

\bigskip

\section{\label{Dana}Calculation of the flux across noble barriers}

The study of a single long orbit points out some transport phenomena which
occur between some phase-space's zones separated by internal transport
barriers (ITB), but it does not give information about the flux through
these barriers ( i.e. the area of the set formed by points which pass
through a ITB at every iteration). This information will be obtained in this
section from Mather's theorem. Using Greene's conjecture we confirm the
observed existence, for $L=\frac{4.875}{2\pi }\sim 0.776$, of the invariant
circle on the plasma edge having the rotation number $\omega _{1}=\frac{1}{%
N(4,2)}$ and of a Cantorus having the rotation number $\omega _{2}=\frac{1}{%
N(1,7)}.$ \ Both Greene's and Mather's approaches use the intimate
connection between quasiperiodic and periodic orbits.

\subsection{\label{twist maps}Periodic and quasiperiodic orbits. The action
principle}

Let us remind some definitions and useful basic results in the theory of
dynamical systems.

The''\emph{lift}'' of the map $T:[0,1)\times \mathbf{R\rightarrow }%
[0,1)\times \mathbf{R}$, $T\left( \theta ,\psi \right) =\left( \theta
^{\prime }\left( \func{mod}1\right) ,\psi ^{\prime }\right) $\ is the
application $\overline{T}:\mathbf{R\times R\rightarrow R}$ defined by $%
\overline{T}\left( \theta ,\psi \right) =\left( \theta ^{\prime },\psi
^{\prime }\right) $ (because the $\left( \func{mod}1\right) $ is not applied
for $\theta ^{\prime }$, the lift $\overline{T}$ allows to take into account
the number of turns along $\theta $).

\ The map $T$ $:[0,1)\times \mathbf{R\rightarrow }[0,1)\times \mathbf{R}$ is
called a \emph{twist map} if $\frac{\partial \theta ^{\prime }\left( \theta
,\psi \right) }{\partial \psi }\neq 0$ for all $\left( \theta ,\psi \right)
\in \lbrack 0,1)\times \mathbf{R}$. It is a left (respectively right) twist
map if $\frac{\partial \theta ^{\prime }\left( \theta ,\psi \right) }{%
\partial \psi }<0$ (respectively $\frac{\partial \theta ^{\prime }\left(
\theta ,\psi \right) }{\partial \psi }>0$) for all $\left( \theta ,\psi
\right) \in \lbrack 0,1)\times \mathbf{R}$, which means that the perturbed
winding number is a monotonously decreasing (respectively increasing)
function of $\psi $.

\bigskip

A point $\left( \theta _{0},\psi _{0}\right) $ is a \emph{periodic} point of
type $\left( n,m\right) $, with periodicity $m$, if $\overline{T}^{m}\left(
\theta _{0},\psi _{0}\right) $ $=\left( \theta _{0}+n,\psi _{0}\right) $.
The rotation number of a periodic orbit exists and is equal to $\iota /2\pi
=n/m$ (i.e. the limit in Eq.(\ref{Iota mesure}) exists); it is clearly
independent of the initial point on a periodic chain. This global property
persists for invariant curve, under some restriction on the map.\ The
rotation number may also be irrational in this case.\ Beside periodic orbits
and invariant curves (called invariant circles, for topological reasons), an
intermediate case of invariant set exists, on which rotation number exists
and is irrational: the Cantori.\ The dynamics on these three types of
invariant set is generally quasiperiodic: the ''time'' dependence can be
expressed by a generalized Fourier series, with rational frequencies for
rational $\omega $, and irrational incommensurate frequencies in the
remaining cases (see Ref. \cite{m} or Ref.\ \cite{w} for details). Both
periodic and quasiperiodic points can be obtained as stationary points in
the action principle.

\bigskip

The ''action principle'' for maps is the analogous of the Lagrangian
variational principle in continuous dynamics. In the case of Hamiltonian
twist maps, the action generating function $F_{a}$ (defined in Section (\ref%
{SecHmaps})) plays the role of a Lagrangian for discrete systems.

\begin{theorem}
(action principle for periodic orbits) Let $T:S^{1}\times \mathbf{%
R\rightarrow S}^{1}\times \mathbf{R}$ be an area preserving twist map , $%
F_{a}:\mathbf{R}\times \mathbf{R}\rightarrow \mathbf{R}$\textbf{,} its
action generating function and $\left\{ \left( \theta _{0}^{\ast },\psi
_{0}^{\ast }\right) ,...,\left( \theta _{m-1}^{\ast },\psi _{m-1}^{\ast
}\right) \right\} $ a periodic orbit of $T$ of period $m$. Then $\left\{
\theta _{0}^{\ast },...,\theta _{m-1}^{\ast }\right\} $ is a stationary
point of the action 
\begin{equation}
\mathcal{A}\left( \theta _{0},...,\theta _{m-1}\right) =F_{a}\left( \theta
_{0},\theta _{1}\right) +F_{a}\left( \theta _{1},\theta _{2}\right)
+...+F_{a}\left( \theta _{m-1},\theta _{0}\right)  \label{action}
\end{equation}
\end{theorem}

A very simple proof is presented in Ref. \cite{bbk}

\bigskip

In 1927 Birkhoff \cite{birkB} showed that every area-preserving twist map
has at least two periodical orbits of type $\left( n,m\right) $ for each
rational winding number $\frac{1}{q}=\frac{n}{m}$ in an appropriate interval
(called the twist interval). For area-preserving twist map it can be proved
(see Ref. \cite{m} \ p.38 for commentaries) that, for each rational $\frac{n%
}{m}$ in the twist interval described by the map, there exists at least one
periodic orbit of type $\left( n,m\right) $ which extremizes $\mathcal{A}$
(it is called \emph{extremizing orbit}) and at least one periodic orbit of
type $\left( n,m\right) $ which is a saddle point of $\mathcal{A}$ (it is
called a \emph{maxmin orbit}).

\bigskip

In order to study the linear stability properties of a $\left( n,m\right) $-
periodic orbit one computes the multipliers of the orbit (\emph{i.e.} the
eigenvalues $\lambda $ and $\frac{1}{\lambda }$ of the Jacobi matrix
associated to $\underset{m\,\,-\,\,times}{\underbrace{T\circ T\circ
....\circ T}}$ in an arbitrary point of the orbit) or the residue 
\begin{equation}
R=\frac{2-\lambda -1/\lambda }{4}  \label{residue}
\end{equation}
If $R<0$ (i.e. $\lambda >0$ , $\frac{1}{\lambda }>0$) the orbit is formed by 
\emph{direct hyperbolic points} (which are unstable). If $R\in \left(
0,1\right) $ (i.e. $\lambda $ and $\frac{1}{\lambda }$ are complex conjugate
numbers and $\left| \lambda \right| =\left| \frac{1}{\lambda }\right| =1$)
the orbit is formed by \emph{elliptic points} (which are stable). If $R>1$
(i.e. $\lambda <0$, $\frac{1}{\lambda }<0$) the orbit is formed by \emph{%
inverse hyperbolic points }(which are unstable).

\bigskip

\bigskip In Ref.{\large \ }\cite{mmB}\textsc{\ }it\textsc{\ }was proved that
every extremizing orbit has negative residue and that\textsc{\ }every maxmin
orbit has a positive residue.\textsc{\ }It results that \emph{every
extremizing orbit is formed by direct hyperbolic points} and the \emph{%
maxmin orbit is formed by elliptic points} (if $R\in \left( 0,1\right) $) 
\emph{or} \emph{by inverse hyperbolic points} (if $\ R>1$).

\bigskip

\bigskip The \emph{action principle for quasiperiodic orbits} was proposed
by Percival (in Ref.{\large \ }\cite{pB}). In 1920 Birkhoff proved that any
quasiperiodic orbit is the graph of a function (see Ref. \cite{birk1B}). It
can be written in the form 
\begin{equation}
\left\{ \left( \theta \left( t\right) ,\psi (t)\right) ,\,\,t\in \mathbf{R}%
\right\}  \label{Dana1}
\end{equation}%
where $\theta :\mathbf{R\rightarrow R}$ is an increasing function having the
periodicity property $\theta \left( t+1\right) =\theta \left( t\right) +1$
for all $t\in \mathbf{R}$, and where 
\begin{equation}
\psi (t)=-\frac{\partial F_{a}}{\partial \theta }\left[ \theta \left(
t\right) ,\theta \left( t+\omega \right) \right]  \label{Dana2}
\end{equation}%
In these terms the action principle can be written as follows.

\begin{theorem}
( the action principle for quasiperiodic orbits) Let $T:S^{1}\times \mathbf{%
R\rightarrow S}^{1}\times \mathbf{R}$ be an area preserving twist map, $%
F_{a}:\mathbf{R}^{1}\times \mathbf{R}^{1}\rightarrow \mathbf{R}$ its action
generating function, $\omega \in \left( 0,1\right) $ an irrational number
and $\left\{ \left( \theta \left( t\right) ,\psi \left( t\right) \right)
\;\;/\;\;t\in \mathbf{R}\right\} $ a quasiperiodic orbit having the rotation
number $\omega .$ Then $\theta $ is a stationary point of the functional 
\begin{equation}
\mathcal{A}\left( \theta \right) =\overset{1}{\underset{0}{\int }}F_{a}\left[
\theta \left( t\right) ,\theta (t)+\omega \right] \text{ }dt
\label{DanaAction}
\end{equation}
\end{theorem}

For twist maps, Mather (see Refs. \cite{mt} and \cite{mt2B})\textsc{\ }%
proved the existence of a stationary and moreover ''extremizing'' function $%
\theta _{ext}$ which extremizes the functional $\mathcal{A}$ (\ref%
{DanaAction}). It gives rise to an invariant set $\left( C\right) $.\ Its
equation can be obtained from (\ref{Dana1}, \ref{Dana2}). If $\theta _{\text{%
ext}}$ is continuous then $\left( C\right) $ is an \emph{invariant circle. }%
If $\theta _{\text{ext}}$ is not continuous (but has a countable set of
discontinuities because it is monotonous), then the closure \textsc{(}the
set of the limit points\textsc{)} of $\left( C\right) $ would be a Cantor
set, called \emph{''Cantorus''}. A very good survey on this problem is given
in Ref. \cite{m}.

\subsection{Study of quasiperiodic orbits via periodic orbits}

The invariant set (invariant circle or Cantorus) having the irrational
rotation number $\omega $ was obtained using a stationary point $\theta
_{ext}$ of the action functional $\mathcal{A}\left( \theta \right) $ defined
in (\ref{DanaAction}), but it is also the limit circle of a sequence of
periodic orbits having rotation numbers $\frac{n_{\nu }}{m_{\nu }}$ , when $%
\underset{\nu \rightarrow \infty }{\lim }\frac{n_{\nu }}{m_{\nu }}=\omega .$
High order periodic orbits may thus be considered as good enough
approximations for the invariant set and the study of their properties gives
information about the invariant set properties. So, an irrational magnetic
surface can be described as\ the limit of its rational convergents by
observing higher and higher order periodic motions.

\bigskip

There are at least two approaches to study the connection between these
invariant sets and the periodic orbits: the Greene's conjecture (which
relates the existence of the invariant set to the stability of a particular
sequence of periodic orbits) and the Mather's theorem (which gives necessary
and sufficient conditions for the existence of the invariant set and
computes the flux through it).

\bigskip

For a noble number $\omega $ we will denote by $\left( \frac{n_{\nu }}{%
m_{\nu }}\right) _{\nu \in \mathbf{N}}$ the sequence of its convergents (see
Sections \ref{Section 4.2} and \ref{convergents}\textbf{\ }for definitions
and computations). We denote by $R_{\nu }^{+}$ the residue of a maxmin orbit
(passing through elliptic or inverse hyperbolic points) and $R_{\nu }^{-}$
the residue of an extremizing orbit (passing through direct hyperbolic
points) of type $\left( n_{\nu },m_{\nu }\right) $.

\bigskip

\emph{Greene's conjecture} (in Refs.{\large \ }\cite{gr1} and \cite{gr2})
predicts that there would be three situations:

- subcritical ( $R_{\nu }^{+}\rightarrow 0$ and $R_{\nu }^{-}\rightarrow 0$
); in this case there is sequence of island chains converging to a \textbf{%
smooth} invariant circle having the rotation number $\omega $.

- critical ( $R_{\nu }^{+}\nrightarrow 0$ and $R_{\nu }^{-}\nrightarrow 0$
but they are bounded in $\left( 0,\infty \right) $ and in $\left( -\infty
,0\right) $), respectively; in this case there is a sequence of island
chains converging to a \textbf{non smooth} invariant circle having the
rotation number $\omega $.

- supercritical ( $R_{\nu }^{+}\rightarrow \infty $ and $R_{\nu
}^{-}\rightarrow -\infty $); in this case there is no invariant circle
having the rotation number $\omega $, but there is a \textbf{Cantorus} \
with this rotation number $\omega $.

In Ref. \cite{macB}\textsc{\ }a partial proof is presented, but the
conjecture is not yet completely proved.

\bigskip \bigskip

On the other hand, in Ref. \cite{mt1} \emph{Mather} gives an equivalent
condition to the existence of an invariant circle having the rotation number 
$\omega $. For a left twist map and a rational $\frac{n}{m}$ in the twist
interval one can define 
\begin{equation}
\Delta \mathcal{A}_{n/m}=\mathcal{A}_{\text{extremizing}}-\mathcal{A}_{\text{%
maxmin}}  \label{h}
\end{equation}%
the difference of action $\mathcal{A}$ \ between the extremizing orbit and
the maxmin orbit of type $\left( n,m\right) .$ In (\ref{h}) $\left| \Delta 
\mathcal{A}_{n/m}\right| \ $ is the \textbf{escape area} from a domain
bounded by an arbitrary curve $\left( C\right) $ passing by the $m$ points
of the maxmin orbit and by the $m$ points of the extremizing orbit of type $%
\left( n,m\right) .$Let us remind that the escape area \ (the flux) from the
domain $R_{(C)}$ bounded by the closed curve $\left( C\right) $, under the
map $T$, is defined by 
\begin{equation}
L\left( R_{C},T\right) \overset{def}{=}Area\left( T\left( R_{C}\right)
-R_{C}\right)  \label{j6}
\end{equation}

\bigskip

\begin{figure}
\centering
\includegraphics[width=12.2198cm]{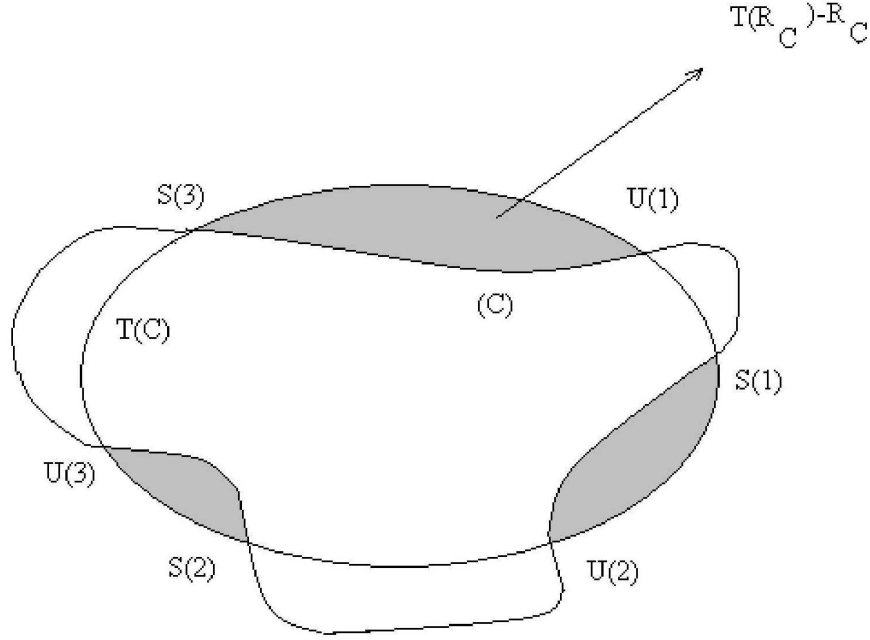}
\caption{The figure
visualizes the escape area from a bounded domain.\ The curve $T(C)$\ is the
image of the closed curve $(C)$\ after one iteration by the map $T$\ . The
domain $R_{C}$\ is the domain bounded by\ $(C)$. The set $T(C)\cap (C)$\
consists in three unstable periodic points (namely $U(1)$, $U(2)$\ and $U(3)$%
) and in three stable points (namely $S(1)$, $S(2)$\ and $S(3)$).}
\label{FigDana}
\end{figure}

\bigskip In Ref.\cite{turnstile} a very interesting result is proved for
right twist maps. It can be easily adapted for the twist maps as follows.\ If%
\textbf{\ }$\left( C\right) \cap T\left( C\right) $ consists in $2m$ points
of an extremizing orbit and of a maxmin orbit of type $\left( n,m\right) $
then \textbf{\ } 
\begin{eqnarray}
L\left( R_{C},T\right) &=&\left| \overset{m-1}{\underset{j=0}{\sum }}\left(
F_{a}\left( \theta _{j}^{\left( u\right) },\theta _{j+1}^{\left( u\right)
}\right) -F_{a}\left( \theta _{j}^{\left( s\right) },\theta _{j+1}^{\left(
s\right) }\right) \right) \right| =\text{ \ \ \ \ \ \ \ \ \ \ \ \ \ \ }
\label{f} \\
\left| \mathcal{A}_{\text{extremizing}}-\mathcal{A}_{\text{maxmin}}\right|
&=&\left| \Delta \mathcal{A}_{n/m}\right|
\end{eqnarray}%
In this formula, $\left( \theta _{j}^{\left( u\right) },\psi _{j}^{\left(
u\right) }\right) $ and $\left( \theta _{j}^{\left( s\right) },\psi
_{j}^{\left( s\right) }\right) $, for $j\in \left\{ 1,2,...,m\right\} $, are
the points of the extremizing orbit and (respectively) of the maxmin orbit
of type $\left( n,m\right) $. An example with periodicity $m=3$ is
illustrated in Fig (\ref{FigDana}).

\bigskip

\bigskip

Mather's theorem states that there is an \emph{invariant circle} having the
rotation number $\omega $ if and only if $\underset{\nu \rightarrow \infty }{%
\lim }\Delta \mathcal{A}_{n_{\nu }/m_{\nu }}=0$. Otherwise, the sequence $%
\left( \Delta \mathcal{A}_{n_{\nu }/m_{\nu }}\right) _{\nu \in \mathbf{N}}$
converges to a \ nonzero limit $\Delta \mathcal{A}_{\omega }$. In this
latter case there is no invariant circle of rotation number $\omega $, but
there is a \emph{Cantorus} with this rotation number, and $\left| \Delta 
\mathcal{A}_{\omega }\right| $ is the \emph{flux through the Cantorus}. It
results that the flux through a Cantorus can be approximated by $\left|
\Delta \mathcal{A}_{n_{\nu }/m_{\nu }}\right| $ \ for large enough values of 
$\nu .$

\bigskip \bigskip

\bigskip The system (\ref{R14}) gives rise to an unique area preserving map%
\textbf{\ }$T:\mathbf{R\times R}_{+}\rightarrow $\textbf{\ }$\mathbf{R\times
R}_{+}$\textbf{\ }such that\textbf{\ }$T(\theta _{\nu },\psi _{\nu })=\left(
\theta _{\nu +1},\psi _{\nu +1}\right) .$ If $T$\textbf{\ }is a twist map
and if the analytical form of $F_{a}$\ can be obtained, the formula (\ref{f}%
) can be used in order to compute $\Delta A_{n_{\nu }/m_{\nu }}$\ (as in the
case of the standard map). But there are a lot of systems (the tokamap-model
for example) whose analytical form of $F_{a}$\ can not be derived.

\bigskip \bigskip

In order to obtain $L(R_{C},T),$\ a direct computation can be done. If we
denote by $f\left( \psi _{\nu +1},\theta _{\nu }\right) $\ $=F_{0}\left(
\psi _{\nu +1}\right) +L\delta F\left( \psi _{\nu +1},\theta _{\nu }\right)
, $\ the generating function (\ref{general gen fct})\ becomes 
\begin{equation}
F\left( \psi _{\nu +1},\theta _{\nu }\right) =\psi _{\nu +1}\cdot \theta
_{\nu }+f\left( \psi _{\nu +1},\theta _{\nu }\right)  \label{d1}
\end{equation}%
and the system (\ref{R14}) can be written as 
\begin{equation}
\left\{ 
\begin{array}{c}
\psi _{\nu +1}=\psi _{\nu }-\frac{\partial f(\psi _{\nu +1},\theta _{\nu })}{%
\partial \theta _{\nu }} \\ 
\theta _{\nu +1}=\theta _{\nu }+\frac{\partial f(\psi _{\nu +1},\theta _{\nu
})}{\partial \psi _{\nu +1}}%
\end{array}%
\right.  \label{d2}
\end{equation}%
One observes that\textbf{\ }$f\left( \psi _{\nu +1},\theta _{\nu }\right) $
in (\ref{d2}) plays the role of a discrete Hamiltonian for the map, which is
periodic in $\theta _{\nu }$.\ This is not the case for $F\left( \psi _{\nu
+1},\theta _{\nu }\right) $ in (\ref{d1}) - where $F$ is the explicit
generating function of the tokamap, given in (\ref{generating fct Tokamap}).

\bigskip \bigskip

By repeating the arguments applied to the calculation of the escape area in
terms of the action function, we deduce the following expression in terms of
the periodic function $f\left( \psi _{\nu +1},\theta _{\nu }\right) $: 
\begin{equation}
\Delta \mathcal{A}_{n/m}=\overset{m-1}{\underset{j=0}{\sum }}\left( \psi
_{j+1}^{(u)}\cdot \frac{\partial f(\psi _{\nu +1}^{(u)},\theta _{\nu }^{(u)})%
}{\partial \psi _{\nu +1}^{(u)}}-f\left( \psi _{\nu +1}^{(u)},\theta _{\nu
}^{(u)}\right) \right)  \label{d3}
\end{equation}%
\begin{equation*}
-\sum_{j=0}^{m-1}\left( \psi _{j+1}^{(s)}\cdot \frac{\partial f(\psi _{\nu
+1}^{(s)},\theta _{\nu }^{(s)})}{\partial \psi _{\nu +1}^{(s)}}-f\left( \psi
_{\nu +1}^{(s)},\theta _{\nu }^{(s)}\right) \right)
\end{equation*}%
Here $\left( \theta _{j}^{\left( u\right) },\psi _{j}^{\left( u\right)
}\right) $\ and $\left( \theta _{j}^{\left( s\right) },\psi _{j}^{\left(
s\right) }\right) $, for $j\in \left\{ 1,2,...,m\right\} $, are the points
of the extremizing orbit and (respectively) of the maxmin orbit of type $%
\left( n,m\right) $\ contained in $C$\ $.$

\bigskip

Because of the relation:%
\begin{equation}
\psi _{j+1}\cdot \frac{\partial f(\psi _{\nu +1},\theta _{\nu })}{\partial
\psi _{\nu +1}}-f\left( \psi _{\nu +1},\theta _{\nu }\right) =  \label{d5}
\end{equation}%
\begin{equation*}
=\psi _{j+1}\left( \theta _{j+1}-\theta _{j}\right) -F\left( \psi
_{j+1},\theta _{j}\right) +\psi _{j+1}\cdot \theta _{j}=
\end{equation*}%
\begin{equation*}
=\psi _{j+1}\cdot \theta _{j+1}-F\left( \psi _{j+1},\theta _{j}\right)
=F_{a}\left( \theta _{j},\theta _{j+1}\right)
\end{equation*}%
it can easily be shown that the\textbf{\ }formula (\ref{d3}) involving only
periodic functions is actually equivalent to the following form:\textbf{\ \ }%
\begin{equation}
\Delta \mathcal{A}_{n/m}=\overset{m-1}{\underset{j=0}{\sum }}\left( \psi
_{j+1}^{\left( u\right) }\cdot \theta _{j+1}^{\left( u\right) }-F\left( \psi
_{j+1}^{\left( u\right) },\theta _{j}^{\left( u\right) }\right) -\psi
_{j+1}^{\left( s\right) }\cdot \theta _{j+1}^{\left( s\right) }+F\left( \psi
_{j+1}^{\left( s\right) },\theta _{j}^{\left( s\right) }\right) \right)
\label{escapeArea}
\end{equation}%
where $F$ is the (non periodic) explicit generating function of the tokamap,
given in (\ref{generating fct Tokamap}). In this formula (\ref{escapeArea}), 
$\left( \theta _{j}^{\left( u\right) },\psi _{j}^{\left( u\right) }\right) $
and $\left( \theta _{j}^{\left( s\right) },\psi _{j}^{\left( s\right)
}\right) $, for $j\in \left\{ 1,2,...,m\right\} $, are the points of the
extremizing orbit and (respectively) of the maxmin orbit of type $\left(
n,m\right) $.

\bigskip

The new formula (\ref{d3}) for the escape area can be used when only the
mixed generating function is given. It works also in the case of the
non-twist maps, when the action generated function involved in (\ref{f})
cannot be defined and the formula (\ref{f}) can not be used. The
representation of mixed generating function $F$ given by (\ref{general gen
fct}) is recommended because $f$\ is periodic in $\theta $, unlike $F$\ and $%
F_{a}$. This explicit periodicity prevents programing mistakes and increases
the numerical stability.

\bigskip

\subsection{Results about the tokamap, a twist map}

The discrete system (\ref{Tok psi})-(\ref{Tok psi explicit 2}) is generated
by the original ''tokamap''\cite{Tmap1}, namely $T:\left[ 0,1\right) \times 
\mathbf{R}_{+}\rightarrow \left[ 0,1\right) \times \mathbf{R}_{+},\,$defined
by$\,T\left( \theta ,\psi \right) =\left( \theta ^{\prime },\psi ^{\prime
}\right) $ {\large \ } 
\begin{equation}
\left\{ 
\begin{array}{c}
\psi ^{\prime }=\frac{1}{2}\left[ \psi -1-L\sin \left( 2\pi \theta \right) +%
\sqrt{\left( \psi -1-L\sin \left( 2\pi \theta \right) \right) ^{2}+4\psi }%
\right] \\ 
\theta ^{\prime }=\theta +\frac{1}{4}\left( 2-\psi ^{\prime }\right) \left[
2-2\psi ^{\prime }+\left( \psi ^{\prime }\right) ^{2}\right] -\frac{L}{2\pi }%
\cdot \frac{\cos \left( 2\pi \theta \right) }{\left( 1+\psi ^{\prime
}\right) ^{2}}\,\,\,\,\,\,\,\,\,\,\,\,(\func{mod}1)%
\end{array}
\right.  \label{j20}
\end{equation}
It can be checked that the tokamap is a \emph{left twist map}. Indeed, a
simple derivation yields 
\begin{equation*}
\frac{\partial \theta ^{\prime }\left( \theta ,\psi \right) }{\partial \psi }%
=\left[ \frac{1}{4}\left[ -6+8\psi ^{\prime }(\theta ,\psi )-3(\psi ^{\prime
}(\theta ,\psi ))^{2}\right] +\frac{L}{\pi }\cdot \frac{\cos \left( 2\pi
\theta \right) }{(1+\psi ^{\prime }(\theta ,\psi ))^{3}}\right] \cdot
\end{equation*}
\begin{equation}
\cdot \frac{\partial \psi ^{\prime }}{\partial \psi }\left( \theta ,\psi
\right) <0  \label{j21b}
\end{equation}

Numerical computations show that the first bracket in (\ref{j21b}) is
negative in $\left[ 0,1\right) \times \left[ 0,1\right] $, which is the
interesting zone in the phase space. On the other hand 
\begin{equation}
\frac{\partial \psi ^{\prime }}{\partial \psi }=\frac{1}{2}\left[ 1+\frac{%
\psi +1-L\sin \left( 2\pi \theta \right) }{\sqrt{\left( \psi -1-L\sin \left(
2\pi \theta \right) \right) ^{2}+4\psi }}\right] >\frac{1}{2}  \label{ltp5}
\end{equation}
for all $\left( \theta ,\psi \right) \in $ $\left[ 0,1\right) \times \left[
0,1\right] $ and $L<1$. As a result, the inequality (\ref{j21b}) is obtained.

\bigskip

The analytical form of the action generating function $F_{a}\left( \theta
_{\nu },\theta _{\nu +1}\right) $ defined in (\ref{eqC})\textsc{\ }for the
Tokamap\textsc{\ }cannot be derived analytically from the map equations (\ref%
{j20}). To obtain it one has to solve the equation (\ref{Tok angle}) for the
unknown $\psi _{\nu +1}$, using the $q$-profile (\ref{W de psi}, \ref%
{profilq(psi)}), and to substitute it in (\ref{eqC}). It can be proved that
the equation (\ref{Tok angle}) has an unique solution $\psi _{\nu +1}=\psi
_{\nu +1}\left( \theta _{\nu },\theta _{\nu +1}\right) $ but this solution
cannot be obtained through analytical methods. However, numerical values of $%
F_{a}\left( \theta _{\nu },\theta _{\nu +1}\right) $ can be computed. In
order to compute $\Delta A_{n/m}$\ , the formula (\ref{d3}) will be used.

\bigskip

\subsection{Numerical results for computing the fluxes in the case $L=4.875/2%
\protect\pi $}

The numerical results presented in this paragraph actually confirm the
existence of a KAM barrier with rotation number $\omega _{1}=1/N\left(
4,2\right) $, as noticed above.{\large \ }We also evaluate the flux through
the upper Cantorus of the internal transport barrier, which has the rotation
number $\omega _{2}=1/N\left( 1,7\right) $.

\bigskip

In computing these fluxes the identification of long periodic orbits is
important. Since the phase portrait of the tokamap has no symmetries it is
quite difficult to localize long periodical orbits. This can however be done
by using an algorithm, described in Appendix B, which is based on the
Fletcher-Reeves method. In spite of the large number of digits used in the
calculations ($18$ digits), the roundoff errors (see a discussion p.276 in %
\cite{LichtenLieberman1983}) still introduce numerical deviations and errors
in computing long periodic orbits, for the well-known reason that hyperbolic
periodic points are unstable and the system exhibits a strongly sensitivity
to the initial conditions.

\bigskip

For the (observed) KAM barrier having the rotation number $\omega
_{1}=1/N\left( 4,2\right) $ the sequence of the convergents was obtained in (%
\ref{Queb 8}). The results of the numerical computations for $R_{\nu }^{+}$, 
$|R_{\nu }^{-}|$ and $\Delta \mathcal{A}_{n_{\nu }/m_{\nu }}$ are presented
in Table (\ref{Table2bNEW}). 
\begin{equation}
\begin{tabular}{lllll}
$\nu $ & $n_{\nu }/m_{\nu }$ & $R_{\nu }^{+}$ & $R_{\nu }^{-}$ & $\Delta 
\mathcal{A}_{n_{\nu }/m_{\nu }}$ \\ 
$1$ & $5/22$ & $0.2321833$ & $-0.2369255$ & $1.7174326e-05$ \\ 
$2$ & $8/35$ & $0.2571793$ & $-0.2626238$ & $4.4808639e-06$ \\ 
$3$ & $13/57$ & $0.2376864$ & $-0.2425847$ & $9.5294768e-07$ \\ 
$4$ & $21/92$ & $0.2446923$ & $-0.2497698$ & $2.2566047e-07$ \\ 
$5$ & $34/149$ & $0.2323534$ & $-0.2370065$ & $4.9617970e-08$ \\ 
$6$ & $55/241$ & $0.2271472$ & $-0.2315720$ & $1.1173445e-08$ \\ 
$7$ & $89/390$ & $0.2146109$ & $-0.2146109$ & $2.4016014e-09$ \\ 
$8$ & $144/631$ & $0.1910067$ & $-0.0612393$ & $-$ \\ 
$9$ & $233/1021$ & $0.1604416$ & $-0.1627059$ & $-$ \\ 
$10$ & $377/1652$ & $0.1219058$ & $-0.1232303$ & $-$ \\ 
$11$ & $610/2673$ & $0.0776218$ & $-0.0781681$ & $-$ \\ 
&  &  &  & $-$%
\end{tabular}
\label{Table2bNEW}
\end{equation}%
The periodic points were computed with an error of the order $10^{-13}$. We
can notice that\ $R_{\nu }^{+}$\ is decreasing from $R_{1}^{+}=0.2321833$\
to $R_{11}^{+}=0.077621821$, and that $R_{\nu }^{-}$\ is increasing from $%
R_{1}^{-}=$\ $-0.326925525$\ to $R_{11}^{-}=-0.078168125$\ in $11$\ steps,
approaching to $0$. The Greene's conjecture (subcritical case) is fulfilled.
Probably because $L=4.875/2\pi $ is near the threshold value for breaking up
the KAM barrier (see Section \ref{acrossEDGE KAM} and Conjecture in Section %
\ref{Interm confinmt}) the convergence is very slow.

\bigskip

\ We observe that\ the sequence of values for the fluxes $(\Delta A_{n_{\nu
}/m_{\nu }})_{\nu \in \mathbf{N}}$\ is rapidly decreasing, hopefully towards 
$0$, \ and this allows us to argue that the flux through this noble KAM
surface\ $\ q=N(4,2)$\ is indeed zero, indicating the existence of a
strongly resistant barrier on the edge of the plasma for $L=4.875/2\pi $, as
observed in the simulations.

\bigskip

The computer errors enable us to evaluate correctly the fluxes through a
curve passing through the points of a Poincar\'{e}-Birkhoff chain containing
more than $2$x$390=780$ points, even the formula for the escape area can be
used. But the error in computing the fluxes increases from $1.862\cdot
10^{-15}$ (for $\Delta A_{n_{1\nu }/m_{1}}$) to $2$.$081\cdot 10^{-13}$ (for 
$\Delta A_{n_{7}/m_{7}}$) and becomes of the same order of magnitude as the
flux for $\nu >7$. From this high periodicity, the computations are not
precise.

\bigskip

For the upper Cantorus having the rotation number $\omega _{2}=\frac{1}{%
N\left( 1,7\right) }$, the sequence of convergents was obtained in (\ref%
{Queb 7}). The results of the numerical computations for $R_{\nu }^{+}$, $%
|R_{\nu }^{-}|$ and $\Delta \mathcal{A}_{n_{\nu }/m_{\nu }}$ are presented
in Table (\ref{Table3NEW}).

\begin{equation}
\begin{tabular}{lllll}
$\nu $ & $n_{\nu }/m_{\nu }$ & $R_{\nu }^{+}$ & $R_{\nu }^{-}$ & $\Delta 
\mathcal{A}_{n_{\nu }/m_{\nu }}\ast 10^{5}$ \\ 
$1$ & $8/9$ & $\ \ \ 0.0272069456$ & $-\ \ \ \ 0.275217416$ & $4.7360e-05$
\\ 
$2$ & $15/17$ & $\ \ \ 0.3082029734$ & $-\ \ \ \ 0.316420328$ & $7.9925e-06$
\\ 
$3$ & $23/26$ & $\ \ \ 0.3646525501$ & $-\ \ \ \ 0.374840518$ & $2.4865e-06$
\\ 
$4$ & $38/43$ & $\ \ \ 0.4459480918$ & $-\ \ \ \ 0.462159188$ & $6.5746e-07$
\\ 
$5$ & $61/69$ & $\ \ \ 0.6621904100$ & $-\ \ \ \ 0.694999295$ & $2.2083e-07$
\\ 
$6$ & $99/112$ & $\ \ \ 1.2004032200$ & $-\ \ \ \ 1.297669846$ & $8.6218e-08$
\\ 
$7$ & $160/181$ & $\ \ \ 3.3011913350$ & $-\ \ \ \ 3.814698002$ & $%
4.6269e-08 $ \\ 
$8$ & $259/293$ & $\ \ 17.1200315801$ & $-\ \ \ 22.284309839$ & $3.4357e-08$
\\ 
$9$ & $419/474$ & $\ 221.306197153$ & $-\ \ 166.19991214$ & $-$ \\ 
$10$ & $678/767$ & $1066.799716960$ & $-\ 546.73330000$ & $-$ \\ 
$11$ & $1097/1241$ & $1971.040117100$ & $-1368.80947500$ & $-$ \\ 
&  &  &  & 
\end{tabular}
\label{Table3NEW}
\end{equation}

\bigskip We notice that Greene's conjecture (the supercritical case) again
is fulfilled: $R_{\nu }^{+}$\ increases quickly (hopefully to $+\infty $)
and $R_{\nu }^{-}$\ decreases rapidly (hopefully to $-\infty $). Due to the
computer's errors, we can not evaluate the fluxes through a curve passing
through the points of a Poincar\'{e}-Birkhoff chain containing more than $2$x%
$293=586$\ points.

\bigskip

\ \ Because the escape areas' sequence is decreasing with a decreasing
slope, we may consider that it goes to a constant, and that its limit\ is
less than $3.4357\cdot 10^{-8}$ and is of the order of $10^{-8}$ $.$ This is
the value of the magnetic field lines' flux through the internal barrier
having the rotation number $N(1,7)$. It means that, at every iteration, the
magnetic field lines passing by points which do occupy a surface (in the
Poincar\'{e} section) having an area of approximately $3.4357\cdot 10^{-8}$\
pass through the Cantorus, in the chaotic sea. In the same time some other
magnetic lines (located\ in a set with the same area in the chaotic sea)
come inside the region bounded by the Cantorus.

\bigskip

We can also observe that we can find elliptic points in every neighborhood
of the KAM barrier, but near the partial transport barrier the high order
convergents are direct or inverse hyperbolic, with exponentially increasing
(with the periodicity order) residues. It results from this that the partial
transport barrier is located on a more chaotic zone, as compared to the KAM
barrier, where there are many zones with regular dynamics (near the elliptic
points) which are embedded in the chaotic zone. The chaoticity of the
chaotic zone is reduced in the vicinity of the KAM, because the residues of
the direct hyperbolic points approach to zero.

\begin{figure}
\centering
\includegraphics[width=11.943cm]{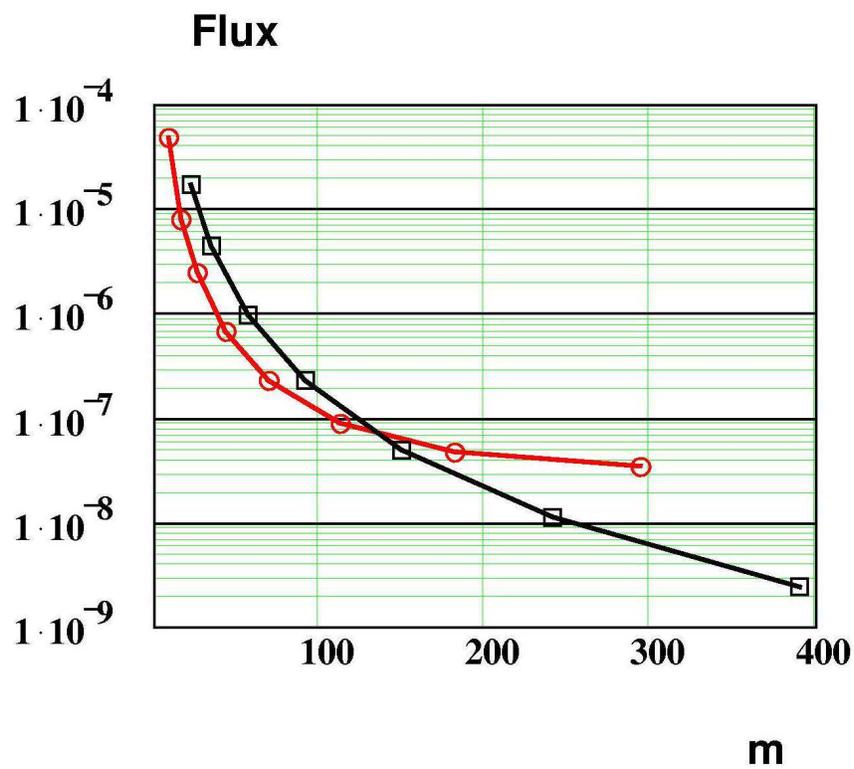}
\caption{Comparison of the decrease of
the flux across convergent island chains towards the upper Cantorus $%
q=N(1,7) $ (circles) and towards the KAM surface on the plasma edge $q=N(4,2)
$ (squares). The former is seen to converge to a finite value of the order
of $10^{-8}$, while the latter is observed with lower values to continue to
decrease hopefully to zero (robust barrier).}
\label{FluxesNEXnew}
\end{figure}

The results can be summarized as follows. For the KAM barrier, the positive
and the negative residues converges slowly to $0$, so the subcritical case
in the Greene's conjecture is verified . The possible explanation for this
slow convergence is that the value of the stochasticity parameter $K=4.875$
is very near the threshold for large scale dispersion due to the breaking up
the edge barrier. By computing the periodic orbit with an error of $10^{-13}$%
, we obtain still decreasing values of the flux of the order of $10^{-9}$,
actually smaller than the flux across the Cantorus. So, we may expect the
convergence towards $0$.

\bigskip

For the Cantorus, the values of the flux are actually comparable with the
values for the KAM for small periods, but the last we were able to compute
is of the order of $10^{-9}$. The Fig.(\ref{FluxesNEXnew}) shows that it
seems to already converge towards a value larger than the one obtained for
the KAM.

\bigskip

In conclusion, the Greene's conjecture and Mather's theorem are verified in
both situations and the numerical results are in agreement with the theory.
From Fig.(\ref{FluxesNEXnew})\textbf{\ }we can thus conclude that the
existence of \emph{(i) }a KAM surface, the plasma edge barrier, \ having the
rotation number $\frac{1}{N\left( 4,2\right) }$\ and of \emph{(ii) }a
partial barrier, a Cantorus, with the rotation number $\frac{1}{N\left(
1,7\right) }$\ was indeed confirmed by using numerical methods.

\bigskip

\section{\label{AsymptRadial}Asymptotic radial subdiffusion}

Dispersive motion of magnetic lines in a stochastic zone is often described
very roughly as a random walk. We stress the fact that the radial motion
observed here is \emph{non-diffusive}. In order to determine more precisely
the average dispersion properties of magnetic lines in a situation of
incomplete chaos (diffusion, subdiffusion or superdiffusion) and their
scaling properties with the perturbation, we have calculated the asymptotic
properties of a set of $5000$ magnetic lines. Initial conditions are chosen
in the chaotic shell on an initial circle $\psi =0.001$, $r\sim 0.0316$,
outside of the protected plasma core, either as a \emph{bunch of lines}
within a small angular interval $\Delta \theta =0.001$, or on a complete 
\emph{circle} with regularly spaced initial values of the initial angle $%
\theta $ between $0$ and $1$. The values of the stochasticity parameter $L$
are varied in a wide range between $0.776$ and $6.68$, beyond the threshold
for large scale chaos above which all magnetic lines from the chaotic shell
finally cross the edge barrier, transformed into a permeable Cantorus. Here
we also study the radial position $x=\sqrt{\psi }$ of a magnetic line, which
is unity on the unperturbed plasma edge.

\bigskip

We have to stress the fact that with such large values of the stochasticity
parameter $L$, almost all magnetic lines in the central chaotic shell are
crossing the plasma edge. This situation is of course of less interest for
confinement studies.\ It would rather be characteristic, instead, of a
situation of plasma disruption due to the rapid escape of the magnetic
lines. The time behavior of this escape is however interesting in order to
study the scaling properties of this dispersive motion, as compared with
other maps like the standard map, where a similar flux diffusion has been
intensely studied since many years.

\subsection{Long time behavior}

Although the individual motion of one magnetic line appears to be discrete
and random, average properties like\ the \emph{mean poloidal flux} $\psi
_{m} $\ and the \emph{mean radius}, averaged over initial conditions, can be
described by ''continuous'' functions of ''time'' (the number of large turns
around the torus) : 
\begin{equation}
\psi _{m}(t)\equiv \left\langle \psi (t)\right\rangle \text{ \ \ \ \ , \ \ \
\ \ \ }x_{m}(t)\equiv \left\langle \sqrt{\psi (t)}\right\rangle \text{, \ \
\ \ \ \ \ }  \label{averages}
\end{equation}
In order to deduce dispersion properties, we also calculate a third
independent quantity : the mean square displacements (MSD) of the\ flux,
defined in the reference frame of the average motion : 
\begin{equation}
MSD_{\psi }(t)\equiv \left\langle \left[ \psi (t)-\psi _{m}(t)\right]
^{2}\right\rangle  \label{msdpsi}
\end{equation}
The MSD of the radius is also computed, and we check that the following
exact relation is verified at each time in terms of the average motion : 
\begin{equation}
MSD_{x}(t)\equiv \left\langle \left[ \sqrt{\psi (t)}-x_{m}(t)\right]
^{2}\right\rangle =\psi _{m}(t)-x_{m}(t)^{2}  \label{msdx}
\end{equation}

The intuitive picture emerging from these results is the dispersion of
magnetic lines from an initial circle of intersection points, passing local
weak barriers, coming back an forth and expanding radially in average. The
asymptotic state of course strongly depends on the value of the
stochasticity parameter $L$ , as well as all characteristic times which
appear to scale like $L^{-2}$ (see Eqs.(\ref{asymptoticformulas}) below).

\bigskip

\subsubsection{Asymptotic motion in the confined domain}

\bigskip

In the confinement domain, the global diffusion of course vanishes, but the
details of the time behavior of an ensemble of magnetic lines is
characteristic of the phase portrait presented in Fig.(\ref{ExFig12})\textbf{%
. }For $L=4.875/2\pi \sim 0.776$, the orders of magnitude of the
characteristic times can be inferred from Fig.(\ref{ExFig11}) and tested in
the present average evolution : residence times inside the ITB would be of
the order of $\tau _{ITB}\approx 10^{6}$ (downwards peaks), residence times
in the chaotic shell would be of the order of $\tau _{shell}\approx 10^{7}$
and in the chaotic sea of the order of $\tau _{sea}\approx 10^{8}$
iterations.

\begin{equation}
\bigskip 
\begin{array}{cc}
\tau _{ITB}\approx & 10^{6} \\ 
\tau _{shell}\approx & 10^{7} \\ 
\tau _{sea}\approx & 10^{8}%
\end{array}
\label{ResidTIMES}
\end{equation}

\begin{figure}
\centering
\scalebox{0.80}[0.61]{\includegraphics{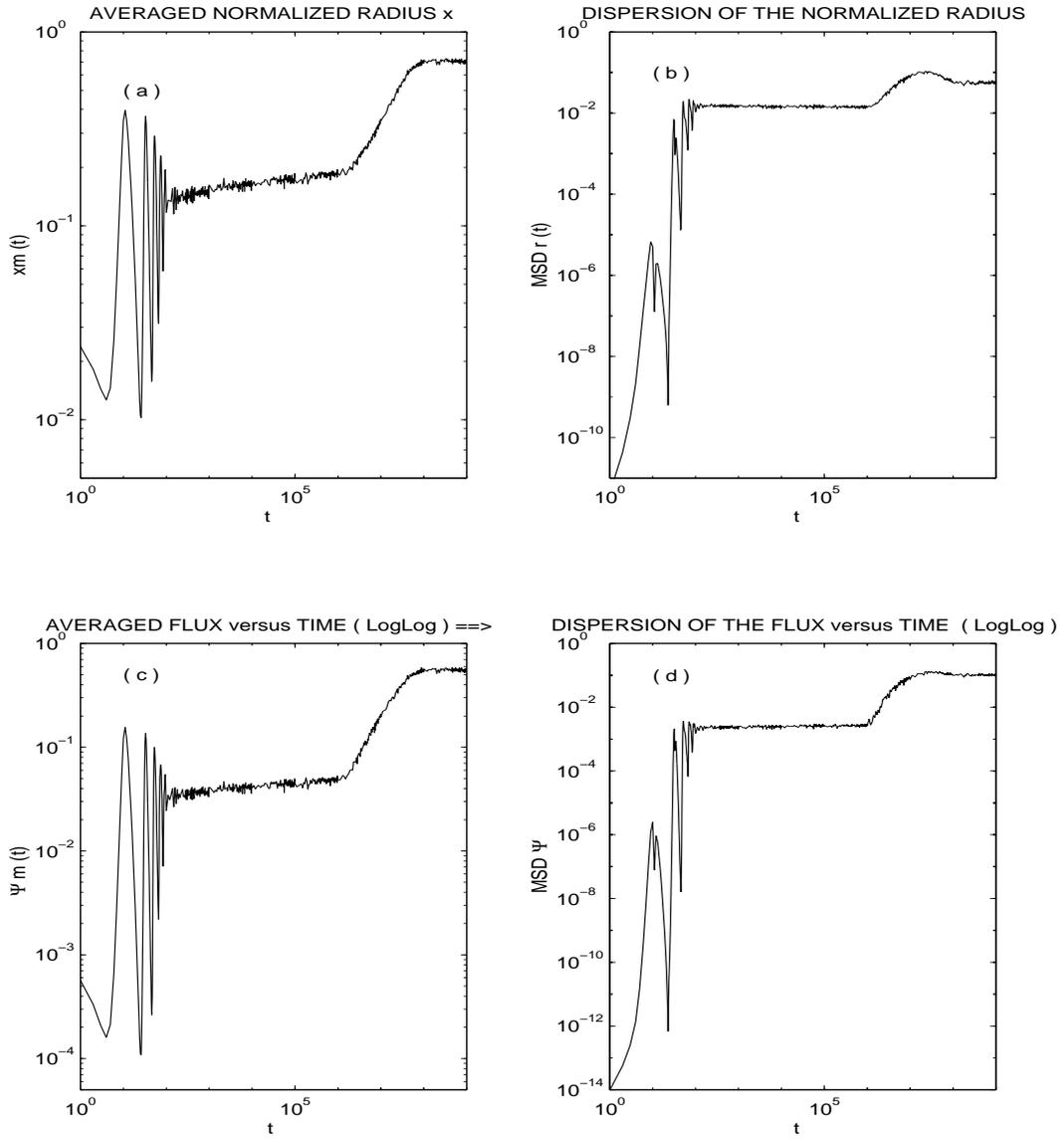}}
\caption{Average radius (a), mean
square radial dispersion (b), average flux (c) and mean square flux
dispersion (d) of an initial bunch of $500$ lines at $L=4.875/2\protect\pi $
followed during $10^{9}$ iterations.}
\label{FigDif1}
\end{figure}

\medskip

We first consider a bunch of $500$ initial conditions within $\Delta \theta
=0.001$ followed over $10^{9}$ iterations for $L=4.875/2\pi \sim 0.776$. We
observe that none of the $500$ considered magnetic lines succeeds to escape
from the plasma edge even after $10^{9}$ iterations, and the average radius
saturates around a value $x_{m}(t)$ $\sim 0.7$ which proves that the robust
KAM barrier on the plasma edge is still resistant (see Figs.(\ref{ExFig11})
and (\ref{ExFig12})). The detailed form of the time dependence is however
interesting, especially concerning the average radius $x_{m}(t)$\ and the
radial dispersion $MSD_{x}(t)$ (see Fig.(\ref{FigDif1}). In Fig.(\ref%
{FigDif1}.a)\textbf{\ }which represents\textbf{\ \ }$x_{m}(t)$,\textbf{\ }we
see at short times large oscillations starting from the initial condition $%
x\sim 0.0316$ and saturating after about $10^{2}$ iterations; this can be
explained by the motion of the bunch of lines as a whole, along the non
circular magnetic surfaces making an excursion towards larger $x$ values at
each poloidal turn in $\theta $.\ 

\bigskip

For larger times, up to $10^{6}$, a slow increase of the average value is
observed up to a value $x_{m}(t)$ $\sim 0.2$, while the dispersion remains
constant (Fig.(\ref{FigDif1}.b)). This value $x_{m}(t)$ $\sim 0.2$ is
important since it roughly corresponds to the position of the transport
barrier (averaged along $\theta $). The resulting interpretation is the
following : the ensemble of magnetic lines describe a CTRW with long sojourn
times inside (or below) the ITB, so that the average position does not
exceed that of the ITB. The dispersion remains constant (zero diffusion)
since the chaotic motion occurs inside a restricted region below the upper
Cantorus of the barrier.

\bigskip

Understanding the global motion for longer times, between $10^{6}$ and $%
10^{8}$, appears as a challenging question. It is seen indeed in \ Fig.(\ref%
{FigDif1}.a)\textbf{\ }that\ the average radius suddenly begins to increase
in time (roughly like $t^{1/2}$), while the dispersion $MSD_{x}(t)$ (Fig.(%
\ref{FigDif1}.b)) first increases (roughly like $t^{1}$) from $10^{6}$ to $%
10^{7}$, has a maximum and then decreases toward a constant value. An
interpretation of this overshooting can be tentatively proposed as follows.

- For times shorter than $\tau _{ITB}\approx 10^{6}$ the dispersion is
dominated by the trapping processes inside the ITB, and the very few lines
which succeed to escape across the barrier actually create a slow increase
of the curve.

- Then from $\tau _{ITB}\approx 10^{6}$ to $\tau _{shell}\approx 10^{7}$
usual diffusion occurs in the chaotic shell and $MSD_{x}(t)$ grows linearly
in time.

- Around $\tau _{shell}\approx 10^{7}$ saturation of the dispersion is
observed because of the finite size of this chaotic shell limited by the ITB.

- After $\tau _{shell}\approx 10^{7}$ a non negligible part of the magnetic
lines are in the chaotic sea, and long sticking events occur around the main
rational chains (see Fig.(\ref{FigMadi})) which result in a decrease of the
dispersion.\ 

- Final saturation\ of the dispersion is obtained around $\tau _{sea}\approx
10^{8}$ with an average value $x_{m}(t)$ $\sim 0.7$.

\bigskip

This average position of the magnetic lines can also be roughly explained in
terms of the estimated average residence times : each region (chaotic shell,
ITB and chaotic sea) has indeed an average radial position (averaged over \ $%
\theta $), and when these average positions are weighted by the percentage
of time spent in each region, one finds indeed a result of the order of the
asymptotic radius observed in Fig.(\ref{FigDif1}.a): $x_{m}(t)$ $\rightarrow
0.7$. Such an agreement should not be taken too seriously, and very long
iterations should be necessary in order to compute accurate residence times
for this presumed CTRW, but the orders of magnitude found yield a first
explanation.

\bigskip

\begin{figure}
\centering
\scalebox{0.80}[0.61]{\includegraphics{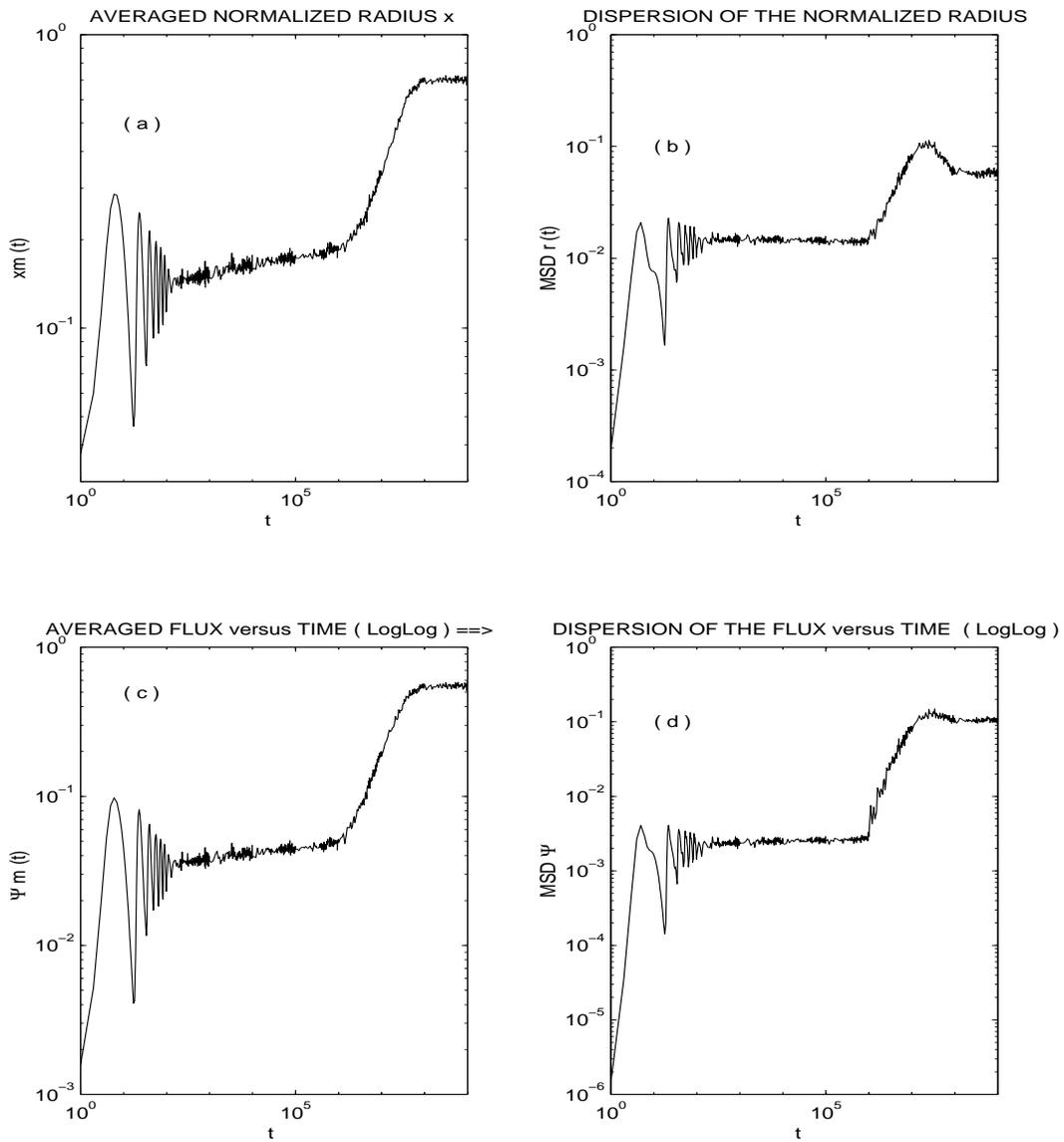}}
\caption{Average radius (a),
mean square radial dispersion (b), average flux (c) and mean square flux
dispersion (d) of an initial circle of $5000$ lines at $L=4.875/2\protect\pi 
$ .}
\label{FigDif7}
\end{figure}

When the same calculation is performed for an initial \emph{circle} of
magnetic lines starting from the same value $\psi =0.001$, $r\sim 0.0316,$
analogous results are obtained in Fig.(\ref{FigDif7}) except for the short
time oscillations which are of smaller amplitude but which however remain.

\bigskip

\subsubsection{Asymptotic motion in the transition domain}

\bigskip

For a slightly higher value of the stochasticity parameter $L=5/2\pi \sim
0.796$, the same series of transitions occurs at small value of the ''time''
: we may thus consider shorter asymptotic times and a larger number of
initial conditions. For such $L$ value, all magnetic lines actually go
beyond the plasma edge, indicating that the KAM on the edge is already
broken, transformed in some Cantorus. The latter still represents some
barrier, and the average motion is different inside and outside this
permeable barrier.

\bigskip 
\begin{figure}
\centering
\scalebox{0.80}[0.61]{\includegraphics{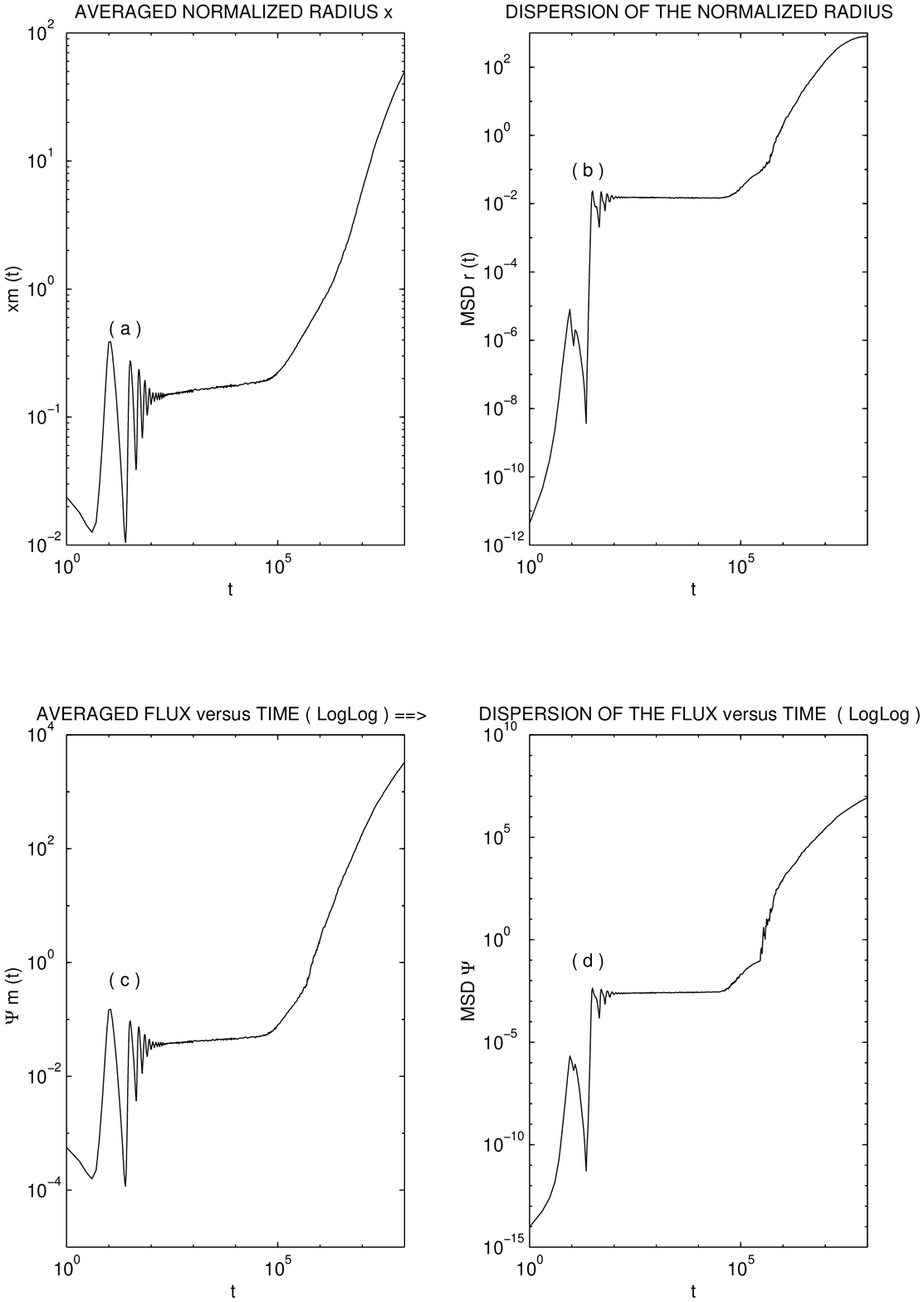}}
\caption{Average radius (a),
mean square radial dispersion (b), average flux (c) and mean square flux
dispersion (d) of an initial circle of $5000$ lines at $L=5/2\protect\pi $.}
\label{FigDif2}
\end{figure}

\bigskip In Figs.(\ref{FigDif2}.a to d) we have represented the four
functions $x_{m}(t)$, $MSD_{x}(t)$, $\psi _{m}(t)$ and $MSD_{\psi }(t)$ for
a \emph{bunch of }$5000$\emph{\ initial lines}. Clearly a similar transition
region is obtained, as in Figs. (\ref{FigDif1}) and (\ref{FigDif7}), but
here for smaller values of time (roughly from $10^{5}$ to $10^{6}$). In this
time domain a straight line is observed in each graph, corresponding to the
following transient time behaviors: $x_{m}(t)\sim t^{1/2}$, $MSD_{x}(t)\sim
t^{1}$, $\psi _{m}(t)\sim t^{1}$ and $MSD_{\psi }(t)\sim t^{2}$ which
represent a transient diffusion of the radius and a superdiffusion of the
flux $\psi $. For such an initial bunch of lines, the initial oscillations
are well pronounced. Here the asymptotic behavior is not yet reached after $%
10^{8}$ iterations and it seems that the dispersion of the radius $%
MSD_{x}(t) $ could tend to a constant value, probably because of the
presence of a resistant barrier very far out of the plasma.\ The latter is
of no physical relevance for the tokamak, but is however characteristic of
the present process of destruction of successive barriers in an incomplete
chaos situation.

\bigskip

\subsubsection{Asymptotic motion in the large scale dispersion domain}

\begin{figure}
\centering
\scalebox{0.80}[0.61]{\includegraphics{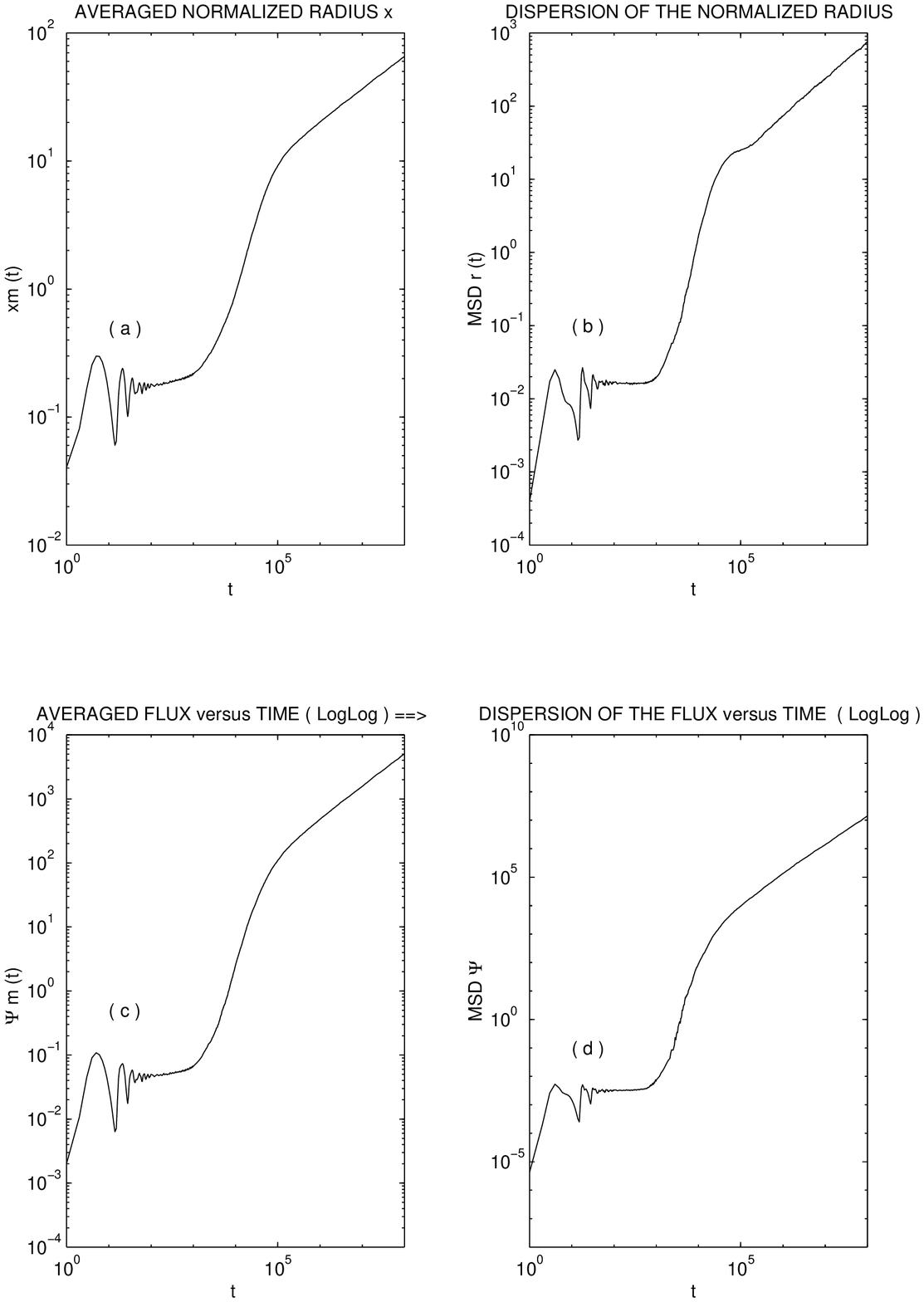}}
\caption{Average radius (a),
mean square radial dispersion (b), average flux (c) and mean square flux
dispersion (d) of an initial circle of $5000$ lines at $L=5.5/2\protect\pi $.}
\label{FigDif3}
\end{figure}

Beyond such transition region, for $L=5.5/2\pi \sim 0.875$ the time behavior
is more regular. In Fig.(\ref{FigDif3}.a to d), we have represented the four
function $x_{m}(t)$, $MSD_{x}(t)$, $\psi _{m}(t)$ and $MSD_{\psi }(t)$. As
expected the initial oscillations are less pronounced, but still exist. Some
transient behavior can hardly be seen (around $10^{3}$ in $MSD_{\psi }(t)$)
and a rapid growth leads to a final asymptotic state reached after $%
10^{5}-10^{6}$ iterations, in which a simple time behavior is measured.

\bigskip 

\bigskip

One notes a common (but not fully understood) feature of all the graphs in
that domain of $L$-values: all these functions seem to present an inflection
point when reaching the absolute value unity. In the graphs for $x_{m}(t)$
this occurs at a time of physical relevance, called the escape time $%
t_{\varphi }$ (the time to reach the edge of the plasma at $\psi =1$, $x=1$,
such that $x_{m}(t_{\varphi })=1$) which is here $t_{\varphi }\approx 10^{4}$%
.

\bigskip

In Fig.(\ref{FigDif3}.b) for $MSD_{x}(t)$ we remark that some overshooting
of the curve appears around $10^{5}$, before the asymptotic state is
reached. Such bump does not appear on the other functions. Is it simply the
effect of a distant barrier, out of the plasma edge, which temporarily traps
magnetic lines ?

\bigskip

\begin{figure}
\centering
\scalebox{0.80}[0.61]{\includegraphics{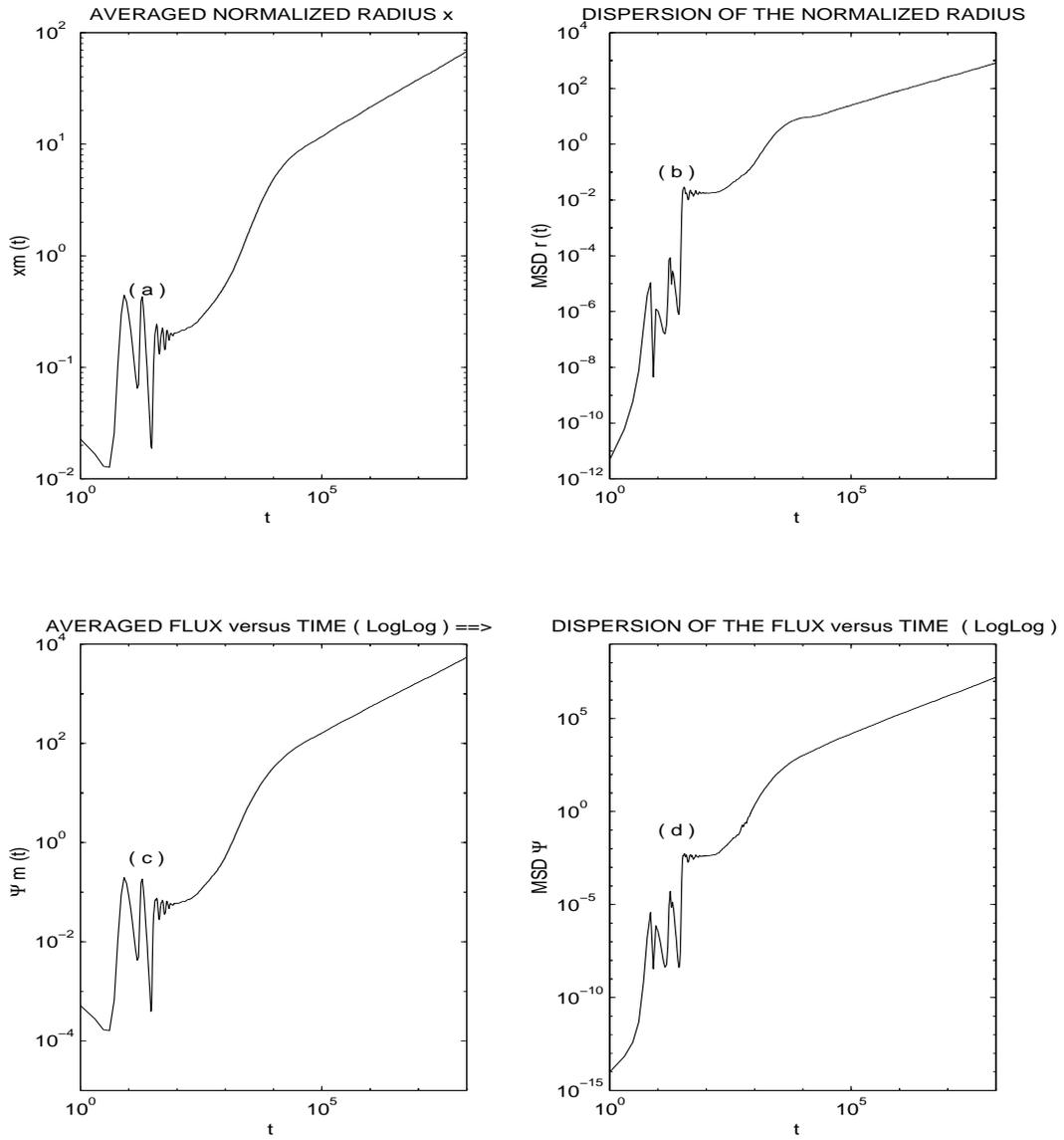}}
\caption{Average radius (a),
mean square radial dispersion (b), average flux (c) and mean square flux
dispersion (d) of an initial bunch of $5000$ lines at $L=6/2\protect\pi $.}
\label{FigDif4}
\end{figure}

Such a bump also appears in Fig.(\ref{FigDif4}.b) for $MSD_{x}(t)$ at a
higher value $L=6/2\pi \sim 0.955$ for an initial bunch of $5000$ lines. In
each of the Figs.(\ref{FigDif4}.a to d), all the features remain unchanged,
except that the escape time is decreased \ $t_{\varphi }\approx 10^{3}$\ \
and that asymptotic state is reached more rapidly, around $10^{4}$.

\bigskip 

\bigskip

\begin{figure}
\centering
\scalebox{0.80}[0.61]{\includegraphics{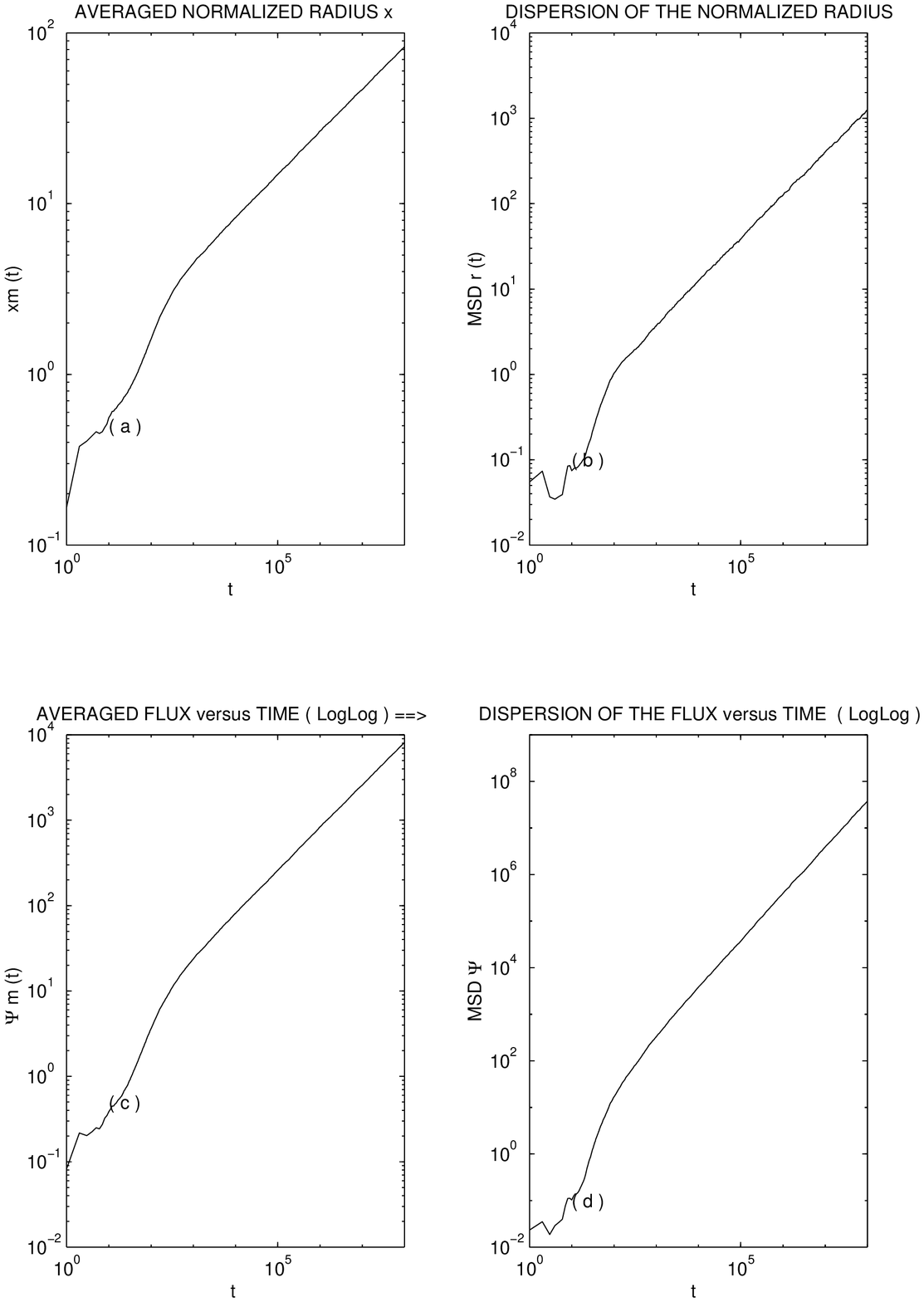}}
\caption{Average radius (a),
mean square radial dispersion (b), average flux (c) and mean square flux
dispersion (d) of an initial circle of $5000$ lines at $L=9/2\protect\pi $.}
\label{FigDif5}
\end{figure}

For higher $L$ values $L=9/2\pi $, dispersion is still more rapid and the
asymptotic state is reached more rapidly.\ Here we follow a circle of $5000$
initial lines. One note however in Figs.(\ref{FigDif5}.a to d) that the
curve in log-log scale has a long time slope which is slowly decreasing
towards what we consider as the asymptotic state, so that long iterations
are necessary to evaluate accurately the time behavior and the dispersion
coefficients introduced below (see Eqs. (\ref{asymptoticformulas}) ).

\bigskip

\bigskip

For a set of values of the stochasticity parameter $L$ in the large scale
dispersion domain, we have measured the asymptotic slope and dispersion
coefficient of each quantity (\ref{averages} to \ref{msdx}). Our
measurements indicate that the average radius is found to increase
asymptotically in time as $t^{1/4}$ : 
\begin{equation}
x_{m}(t)\rightarrow a\text{ }t^{1/4}  \label{xt1o4}
\end{equation}
In a consistent way we find 
\begin{equation}
\begin{array}{ccc}
\psi _{m}(t)\rightarrow b\text{ }t^{1/2} & \text{,} & MSD_{\psi
}(t)\rightarrow 2\text{ }D_{\psi }\text{ }t%
\end{array}
\label{psi1o0}
\end{equation}
i.e. \emph{diffusion of the flux}, like in the standard map, and of course 
\begin{equation}
MSD_{x}(t)\rightarrow 2\text{ }D_{x}\text{ }t^{1/2}  \label{msdx1o2}
\end{equation}
which means \emph{radial sub-diffusion}. Although surprising, this is
however not a fully new result: similar phenomena of \emph{\ ''strange''
transport} $^{14}$\ \cite{MSchles} (where $MSD_{r}(t)\sim t^{\mu }$ with $%
\mu \neq 1$) have been measured, namely for charged particles sticking
around an island remnant in a chaotic three-wave chaotic model \cite%
{Nakach91}\ resulting in a $t^{1/3}$ behavior \cite{PGeisel 91}.

\bigskip \bigskip

It has to be noted that, with the same unperturbed magnetic field (\ref%
{BKnorr}), the direct integration of the magnetic line equations of motion,
in presence of Fourier series magnetic perturbations or a distribution of
current filaments, has already yielded a fully equivalent $t^{1/4}$
asymptotic behavior the quantity $\sqrt{MSD_{x}(t)}$\ rapidly called by
these authors as being some ''average displacement'' of magnetic lines in
chaotic layers \cite{deRover99}.\ This $t^{1/4}$ law is\ in total agreement
with the results (\ref{msdx1o2}) obtained here in a simple map.

\bigskip

\bigskip The above subdiffusive result (\ref{msdx1o2}) is however reached
asymptotically for long times.\ In the present incomplete chaotic regime of
the tokamap, it shows a \emph{slower dispersion} ($\mu <1$) which is
different from the usual diffusion in a completely chaotic situation \cite%
{Haida}. A comparison is performed below, after Eq.(\ref{QLdiffusion}) and
in Fig.(\ref{ExFig24}), in Section \ref{ClassicQL}.

\bigskip

For physical times, of the order of the average time $t_{\varphi }$\
necessary to escape from the boundary circle, different \emph{transient
regimes }are observed instead, with a rapid \emph{radial} \emph{%
super-diffusion} ($\mu >1$) observed when crossing the plasma edge. In other
words, although the asymptotic, long time properties are diffusion of the
flux and sub-diffusion of the radius, several other behaviors are observed
along the time evolution, with a slower or faster time variation.

The numerical results\ measured for the coefficients $a$, $b$, $D_{x}$ and $%
D_{\psi }$ are the following, as function of the stochasticity parameter $L:$%
\begin{equation}
\begin{array}{ccccc}
L & a & b & D_{x} & D_{\psi } \\ 
0.836 & 0.625 & 0.460 & 0.035 & 0.060 \\ 
0.875 & 0.651 & 0.500 & 0.038 & 0.070 \\ 
0.955 & 0.673 & 0.534 & 0.041 & 0.081 \\ 
1.321 & 0.795 & 0.745 & 0.057 & 0.158 \\ 
1.671 & 0.899 & 0.952 & 0.072 & 0.255 \\ 
2.149 & 0.972 & 1.150 & 0.103 & 0.409 \\ 
2.387 & 1.029 & 1.276 & 0.108 & 0.496 \\ 
2.626 & 1.119 & 1.478 & 0.112 & 0.621 \\ 
3.342 & 1.209 & 1.789 & 0.163 & 1.005 \\ 
4.138 & 1.403 & 2.327 & 0.180 & 1.551 \\ 
4.775 & 1.498 & 2.657 & 0.207 & 2.040 \\ 
5.252 & 1.579 & 2.947 & 0.226 & 2.484 \\ 
6.048 & 1.691 & 3.393 & 0.267 & 3.373 \\ 
6.685 & 1.638 & 3.421 & 0.369 & 4.133%
\end{array}
\label{TableResults}
\end{equation}

\subsection{Scaling properties}

Another important feature of transport processes is the \emph{scaling
property} of the corresponding dispersion coefficients (\ref{TableResults})
in terms of the perturbation parameter $L$. In order to draw the scaling
laws for these coefficients, we define arbitrarily the first values by $%
L_{o},a_{o},b_{o},D_{xo}$ and $D$ and we plot the above numerical results
reduced to these first values as $L/L_{o},a/a_{o},b/b_{o},D_{x}/D_{xo}$ and $%
D_{\psi }/D_{\psi 0}$, respectively.

\begin{figure}
\centering
\scalebox{0.67}[0.625]{\includegraphics{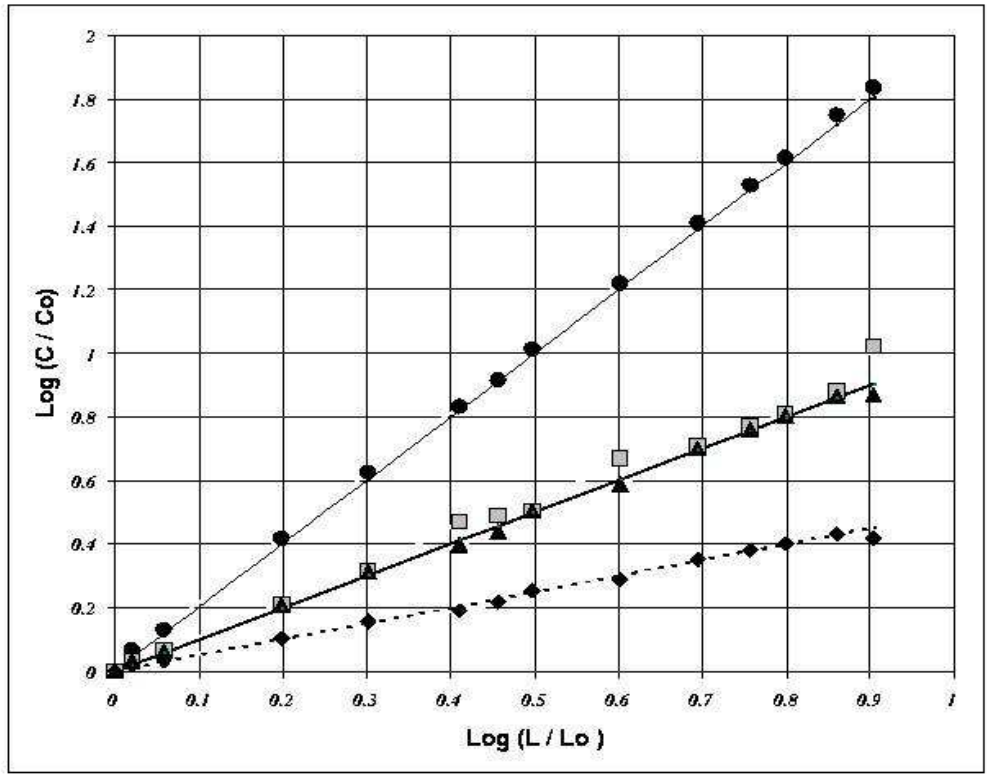}}
\caption{Graphical presentation of the
scaling dependence in the stochasticity parameter $L$ of the asymptotic
coefficients of flux diffusion $D_{\protect\psi }\sim L^{2}$ (black
circles),\ of radius subdiffusion $D_{x}\sim L$ (grey squares), of average
flux growth $b\sim L$ (black triangles) and average radius growth $a\sim
L^{1/2}$ (black diamonds). \ For a lowest value $L_{0}=5.25/2\protect\pi %
\sim 0.836$ this graph represents these various coefficients $C$ in a
logarithmic plot of $\log (C/C_{0})$ as function of $\log (L/L_{0})$. The
three straight lines correspond, from top to bottom, to the expected
behaviors in $L^{2}$, $L$ and $L^{1/2}$, respectively.}
\label{FigScalingLaws}
\end{figure}

\bigskip We clearly see in Fig.(\ref{FigScalingLaws})\textbf{\ }that the
various coefficients scale as follows : $a(L)\sim L^{1/2}$, $b(L)\sim L\sim
D_{x}(L)$ and $D_{\psi }(L)\sim L^{2}$.

\bigskip

Summarizing the results obtained for the both dependences in $L$ and $t$
(see Section (\ref{LocalisationITB})), we can write down the following
asymptotic formulas describing both the long time dependence (Eq.(\ref{xt1o4}%
) to (\ref{msdx1o2})) and the scaling properties of the so-defined
dispersion coefficients $a(L)$, $b(L)$, $D_{x}(L)$ and $D_{\psi }(L)$ : 
\begin{equation}
\begin{array}{c}
\lim_{t\rightarrow \infty }x_{m}(t)=a(L).t^{1/4}\simeq \left[ L^{2}t\right]
^{1/4} \\ 
\lim_{t\rightarrow \infty }\psi _{m}(t)=b(L).t^{1/2}\simeq \left[ L^{2}t%
\right] ^{1/2} \\ 
\lim_{t\rightarrow \infty }MSD_{x}(t)=2\text{ }D_{x}(L).t^{1/2}\simeq \left[
L^{2}t\right] ^{1/2} \\ 
\lim_{t\rightarrow \infty }MSD_{\psi }(t)=2\text{ }D_{\psi }(L).t\simeq %
\left[ L^{2}t\right]%
\end{array}
\label{asymptoticformulas}
\end{equation}%
This clearly shows a global scaling in $\left[ L^{2}t\right] $\ . Such
scaling laws can be simply interpreted as resulting from a simple intuitive
law of the type $\delta \psi (L,t)\approx \left[ \delta x^{2}\right] \simeq
L $.

\bigskip

In the standard map, which is periodic in both variables angle and action,
it is well known \cite{DQLSM}, \cite{Fourier}\ (see also p.299 in \cite%
{LichtenLieberman1983}) that computation of the diffusion coefficient of the
action yields the result $D(L)=D_{ql}(L).S(L)$, where $S(L)$ involves a
series of Bessel function $J(L)$, and where the simple coefficient 
\begin{equation}
D_{ql}(L)=L^{2}/4  \label{old14}
\end{equation}%
can be obtained from a random phase argument. This classical result for the
standard map is actually modified by the presence of accelerator modes,
which do not occur here. Beyond the scaling relation $D_{\psi }(L)\simeq
L^{2}$ (Eq. (\ref{asymptoticformulas})) found to be satisfied in the
tokamap, it is interesting to compare the exact numerical results with Eq. (%
\ref{old14}). We find 
\begin{equation}
D_{\psi }(L)\simeq D_{ql}(L).\left[ 1\pm 0.1\right]  \label{old15}
\end{equation}%
which shows not only that the same quasi-linear diffusion coefficient $%
D_{ql}(L)$ applies both in the standard map and in the tokamap, but moreover
that no Bessel functions seem to appear here, probably due to the lack of
periodicity of the map in the flux variable.

\bigskip

Some methods for the calculation of diffusion coefficients in maps where
variables are clearly separated are given in \cite{Yanna}. An analytical
derivation of even a simple result like Eq.(\ref{old15}) for the tokamap
(Eqs. (\ref{Tok psi}), (\ref{Tok angle}) where variables are not separated)
seems however to be difficult to prove in general using the usual Fourier
path diagram method \cite{Fourier}. It is however very simple to check that
the same random phase argument can be applied to the tokamap and the same
result (\ref{old14}) is obtained, as for the standard map but without any
Bessel functions here. Unlike the standard map, this asymptotic behavior (%
\ref{old15}) is valid for all values of the stochasticity parameter $L$.

\bigskip

\subsection{\label{finite times}Asymptotic times and transient regime inside
the plasma beyond escape threshold.}

\medskip

The asymptotic state for $L=5.5/2\pi \sim 0.796$ (beyond escape threshold)
has been obtained after a rather long relaxation times 
\begin{equation}
t_{R}\approx 10^{6}.  \label{trelaxat}
\end{equation}
We have no theoretical explanation for such slow phenomenon in the tokamap,
but in similar, simpler systems, relaxation times are also very long. For
the standard map \cite{Dif-Sta-Map} \cite{SymbolDynam98}, very long
iteration sequences have been necessary to perform measurements of the
transition probability between different basins and to characterize the
corresponding CTRW. A detailed kinetic theory has been elaborated very
recently \cite{RaduKinThSt Map 2000} for the standard map, showing also very
long relaxation times.

Since asymptotic behavior is only reached after such long relaxation times,
of course magnetic lines are already far beyond the physical domain of the
plasma ($\psi \leqslant 1$). For instance in Fig.(\ref{FigDif3}.a), for a
value $L=5.5/2\pi $ just after breaking of the edge KAM, we can see that the
average radius $r_{m}$ reaches the edge of the (unperturbed) plasma after a
''physical'' or ''escape time'' $t_{\varphi }$ such that $r_{m}(t_{\varphi
})=1$, which yields 
\begin{equation}
t_{\varphi }\approx 10^{4}  \label{tphysical}
\end{equation}
At least for values of $L\gtrsim 5/2\pi \sim 0.796$ , beyond the escape
threshold, this time represents an estimate of the escape time, or more
precisely, an estimation of the number of turns (the long way) around the
torus after which a magnetic line has reached the plasma edge.

\medskip

It is interesting to translate the number of iterations in the tokamap (thus
the number of turns) into an average time for thermal particles to reach the
plasma edge by following the magnetic lines (in absence of collisional or
magnetic drift effects). One iteration time step in the tokamap represents
actually the length of one large turn of length $2\pi R_{0}$ around the
torus. For a thermal particle following the field line the physical time (%
\ref{tphysical}) can thus be evaluated as the time necessary for a particle
to travel $t_{\varphi }$ times the long way along the torus, \emph{i.e.} $%
2\pi R_{0}\ast t_{\varphi }/V_{th}$. In Tore Supra, this time is of the
order of $10^{-2}$ $s.$ for ions and $10^{-4}$ $s.$ for electrons in usual
conditions ($T=5$ $keV$) 
\begin{equation}
\begin{array}{lll}
t_{\varphi \text{ }i}=10^{-2}s. & \text{,} & t_{\varphi \text{ }e}=10^{-4}s.%
\end{array}
\label{tphisec}
\end{equation}%
In the same way one can deduce that the relaxation time, corresponding to $%
10^{6}$ iterations steps of the tokamap, actually represents $20$ $s$ for
ions and $0.5$ $s.$ for electrons in Tore Supra. 
\begin{equation*}
\begin{array}{lll}
t_{R\text{ }i}=20\text{ }s. & \text{,} & t_{R\text{ }e}=0.5\text{ }s.%
\end{array}%
\end{equation*}%
These are the times necessary to reach the slow asymptotic subdiffusion
regime at $L=5.5/2\pi \sim 0.796$ . The transient superdiffusive regime
occurring for $t<t_{R}$ is thus not excluded from the domain of practical
interest, as well as asymptotic subdiffusion regime obtained here.

\medskip

These values of the escape time in the tokamap (\ref{tphisec}) are thus much
shorter than the \emph{particle confinement time} $\tau _{P}\succeq \tau
_{E}\sim 0.20$ $s.$ which is admitted to be roughly equal to the energy
confinement time $\tau _{E}$, in the absence of additional heating for Tore
Supra for instance.{\LARGE \ }Such escape times (\ref{tphysical}) are thus
relevant for particle confinement losses in tokamaks.

\medskip

For times of the order of the escape time $t\lessapprox t_{\varphi }$, the
average radius is growing like $x_{m}(t_{\varphi })\sim t^{1.}$, much faster
than in the asymptotic state (\ref{xt1o4}). The general shape of the curve
Fig.(\ref{FigDif3}.a) indeed indicates a slow initial growth, followed by a
rapid increase of the slope, similar to an exponential growth, and a final
saturation with a lower asymptotic slope. We note that the
pseudo-exponential growth occurs precisely around the physical time (\ref%
{tphysical}).

\medskip

In other words, although the asymptotic behavior of magnetic line motion in
the tokamap is a rather regular one (even if a ''strange'' one) described by
(\ref{asymptoticformulas}), nevertheless for finite times of the order of
the physical time we observe the following\ extremely rapid \emph{transient
time-behavior} : 
\begin{equation}
\begin{array}{l}
MSD_{\psi }(t_{\varphi })\sim t^{5.88} \\ 
MSD_{x}(t_{\varphi })\sim t^{2.75} \\ 
x_{m}(t_{\varphi })\sim t^{1.}%
\end{array}
\label{transitbehaviour}
\end{equation}
In the \emph{transient regime occurring inside the plasma}, for $L=5.5/2\pi $%
, we have thus a \emph{superdiffusion} $MSD_{x}(t_{\varphi })\sim t^{2.75}$
of the magnetic lines, around an average radius growing like $%
x_{m}(t_{\varphi })\sim t^{1.}$ \textit{i.e.} in a ballistic way. This last
observation would seem \textit{a priori} much less favorable for plasma
confinement in perturbed magnetic structures, but it is actually not the
case.

\medskip

\subsection{\label{ClassicQL}Comparison with classical diffusive predictions}

\medskip

We have to compare the present result (\ref{transitbehaviour}) with the
classical analytical prediction, \textit{i.e}. the \emph{diffusion
coefficient of magnetic lines in a completely stochastic magnetic field}.
This process is parametrized by the amplitude of the magnetic perturbation $%
\beta =\mid \delta B\mid /B$ and by parallel and perpendicular coherence
lengths of the perturbation, $\lambda _{\shortparallel }$ and $\lambda
_{\perp }$, respectively. This naturally introduces the control parameter of
the problem : the Kubo number 
\begin{equation}
\alpha \equiv \beta \frac{\lambda _{\shortparallel }}{\lambda _{\perp }}
\label{Kubo}
\end{equation}
which roughly measures the parallel coherence of the fluctuating magnetic
field : magnetic diffusion will be small if $\alpha $ is small. The
quasilinear approximation $D_{QL}$ for the diffusion coefficient in a
completely chaotic field is simply : 
\begin{equation}
D_{QL}=\frac{\lambda _{\shortparallel }}{\lambda _{\perp }^{2}}.\sqrt{\frac{%
\pi }{2}}.\alpha ^{2}=\sqrt{\frac{\pi }{2}}\beta ^{2}\lambda
_{\shortparallel }  \label{DQL}
\end{equation}
It is actually a quadratic approximation in the Kubo number $\alpha $ ,
where the first factor takes the dimensionality into account (the units of
the magnetic diffusion coefficients are in $m^{2}/m$.). Higher order terms
have been derived from a more general description of diffusion in a
completely stochastic fields in \cite{Haida}. This coefficient (\ref{DQL})
has been originally derived in \cite{RR78}, \cite{KP78}. The radial mean
square displacement $<x^{2}(z)>=2D_{QL}.z$ is growing linearly with the
toroidal distance $z$ .

\medskip

In order to make a connection with physical parameters of the tokamak, we
have to know the relation between the stochasticity parameter $L$ and $\beta 
$ the relative strength of the perturbation introduced for instance in the
non-axisymmetric external coils. From the alternative derivation of the
tokamap from canonical variables \cite{Bo-map 1}, this relation can be
derived as : 
\begin{equation}
L=\frac{2\pi }{\varepsilon _{T}}\left( \frac{\delta B}{B}\right)
\label{KdedB}
\end{equation}%
where $R_{0}/a=1/\varepsilon _{T}$ is the aspect ratio of the torus.{\LARGE %
\ }

\medskip

\begin{figure}
\centering
\includegraphics[width=12.1144cm]{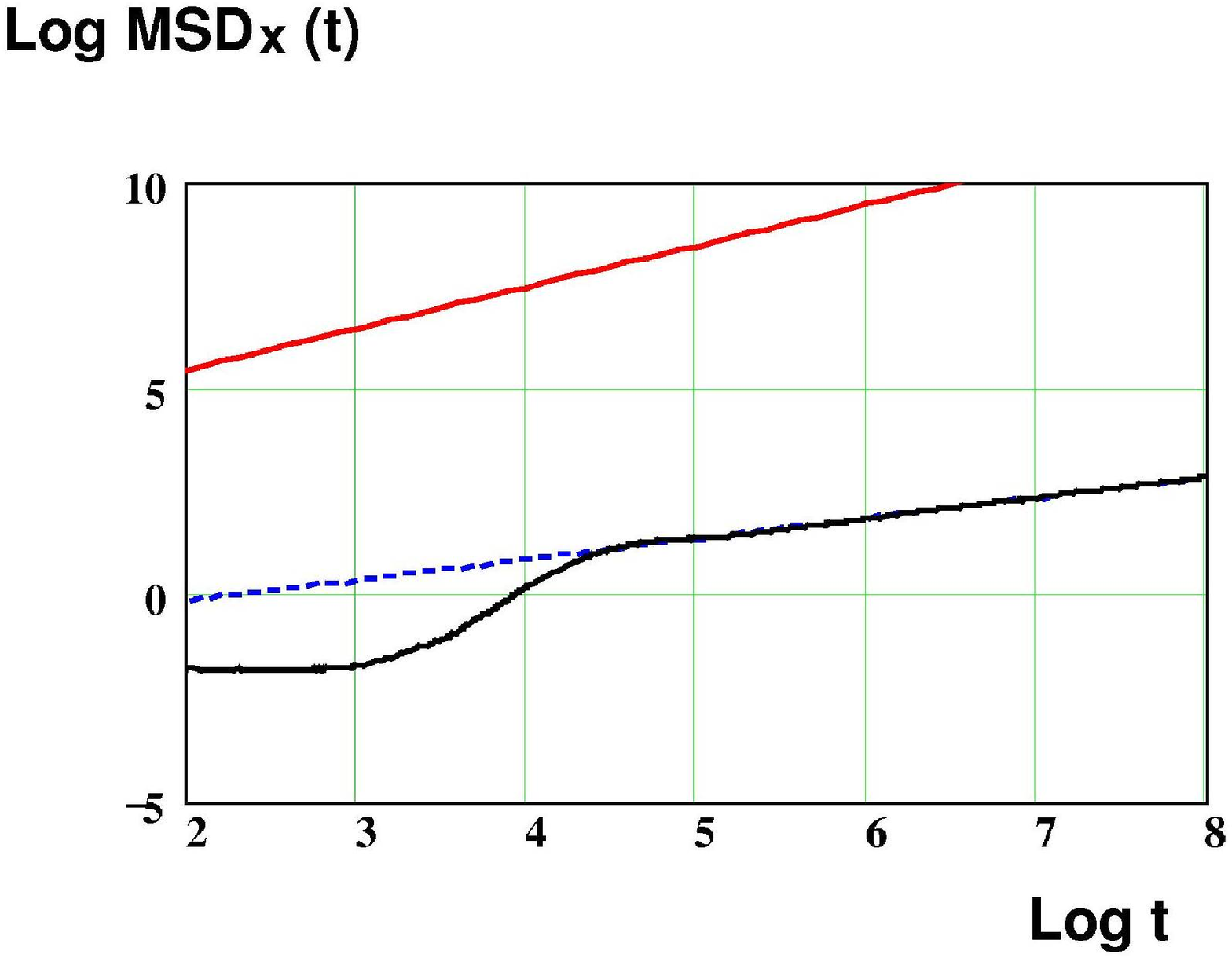}
\caption{Dispersion
measurements of the radial position in the present situation of \emph{%
incomplete chaos} are represented (lower black curve) for $L=5.5/2\protect%
\pi \sim 0.876$ (as in Fig.Dif 3b) with\ its asymptotic slope (dotted line)
which exhibits an \emph{asymptotic radial subdiffusion.} In the domain
between $t=10^{3}$ and $10^{4}$ (around the escape time $t_{\protect\varphi %
} $ , see Eq. 99), one clearly observe a superdiffusive regime. All these
measured behaviors of $MSD_{x}(t)$ nevertheless appear at all relevant times
to remain much smaller than the quasi-linear diffusion in complete chaos as
represented by the upper straight line described by the classical eq. 105 of
Ref.[\textbf{69}] (Rechester \& Rosenbluth 1978).}
\label{ExFig24}
\end{figure}

Within the present dimensionless units (reduced to the minor radius $a$ of
the plasma column), and with $t$ being as previously the number of long
turns around the torus, this classical result writes (with $z=2\pi R_{0}.t$, 
$\lambda _{\shortparallel }\approx 2\pi R_{0}$, $r^{2}=2x^{2}$ and $L=2\pi
\beta /\varepsilon _{t}):$%
\begin{equation}
MSD_{QL}(t)=\sqrt{2\pi }L^{2}t  \label{QLdiffusion}
\end{equation}%
It can easily be shown that this linear function (\ref{QLdiffusion}) remains
for all times much above the numerical result of Fig.(\ref{FigDif3}.b)%
\textbf{, }even for physical times of the order of $t_{\varphi }$ where the
growth rate is like $t^{2.75}$: in Fig.(\ref{ExFig24})\textbf{\ }we draw%
\textbf{\ }the dispersion measurements of the radial position in the present
situation of \emph{incomplete chaos} for $L=5.5/2\pi \sim 0.876$ (as in Fig.(%
\ref{FigDif3}.b)) along with\ its asymptotic slope (dotted line) which
exhibits an \emph{asymptotic radial subdiffusion.} In the domain between $%
t=10^{3}$ and $10^{4}$ (around the escape time $t_{\varphi }$ , see Eq. (\ref%
{tphysical})), one clearly observe a superdiffusive regime. All these
measured behaviors of $MSD_{x}(t)$ nevertheless appear at all relevant times
to remain much smaller than the quasi-linear diffusion in complete chaos as
represented by the upper straight line described by the classical Eq. (\ref%
{QLdiffusion}) \cite{RR78}.

\bigskip 

In conclusion, the transient superdiffusion observed in the tokamap model
for times of the order of the physical or escape time is actually less
effective than the classical quasilinear prediction. This means that
classical transport in a completely stochastic magnetic field as studied in %
\cite{RR78}, \cite{KP78} and \cite{Haida} is actually more dramatic for
plasma confinement than the strange transport observed in the present model
for incomplete chaos. A transient, very rapid growth, although
''superdiffusive'' is not always ''more dangerous'' than a linear one...

\medskip

\section{\label{Conclus}CONCLUSIONS:}

The chaotic motion of magnetic lines in a toroidal perturbed geometry has
been analyzed using a Hamiltonian map called tokamap \cite{Tmap1} defined by
(\ref{Tok angle}, \ref{Tok psi explicit}, \ref{Tok psi explicit 2}) in terms
of the stochasticity parameter $L$. It describes the successive intersection
points of a magnetic line in a poloidal plane $(\psi ,\theta )$ . From the
previous analysis of fixed points bifurcations, new results have been
derived for the Shafranov shift in this map (Fig.(\ref{Shafran})). We
restrict ourselves to the case of an monotonous unperturbed safety factor
profile with a central value $q(0)=1$.

\medskip

The generic phase portrait in the poloidal plane $(\psi ,\theta )$ consists
in a protected \emph{non-chaotic plasma core}, surrounded by a \emph{chaotic
shell} limited by a semi-permeable \emph{Cantorus} which has been
identified. For increasing values of $L$ magnetic lines finally cross this
first barrier and wander in a wide chaotic sea, extending up to the plasma
edge (Fig.(\ref{ExFig12})). This domain of the phase portrait has the
realistic structure with remnants of island chains on the dominant rational $%
q$-values (Fig.(\ref{ExFig4})), and sticking process of magnetic lines
around these island remnants.

\medskip

The important point we want to stress is that the different barriers found
in this configuration always correspond to \emph{irrational values of the
safety factor}, and not to rational values where island chains are formed.
The golden number $G$ is known to play no role in this map \cite{RexTokamap}%
. It is thus quite natural from KAM theory that the ''next most irrational''
numbers, the \emph{''noble'' numbers} $N(1,k)=1+1/(k+G^{-1})$ \cite{Percival}
could actually play an important role. The wandering motion in the inner
shell is indeed limited by a KAM surface (or a robust Cantorus) which is
located in the case $L=4.875/2\pi \sim 0.776$ downward at $q=N(1,11)=1.086..$%
, and upward by a semi-permeable Cantorus at $q=N(1,8)=1.116...$. The
wandering motion in chaotic sea is limited downward by another identified
Cantorus, which is located slightly above the previous one at $%
q=N(1,7)=1.131 $ : we thus find the existence of a \emph{double-sided
internal transport barrier}. Although this barrier includes a rational
surface ($q=9/8=1.125$ in the case $L=4.875/2\pi \sim 0.776$) we have shown
that the two Cantori, which are actually the \emph{two sides of the barrier,
are located on noble values of the safety factor}. The outer edge of the
chaotic sea remains a strong external transport barrier (a KAM surface or a
robust Cantorus) at $L=4.875/2\pi $ ; its radial position in the $q$
-profile appears to be located with a rather good precision at the most
irrational number between $q=4$ and $q=4.5$, \emph{i.e.} at $%
q=N(4,2)=3+N(1,2)=4+1/(2+G^{-1})=4.382...$ The transport barriers observed
in this map have been proved (from flux argument) to be \emph{noble Cantori. 
}The flux of lines across permeable Cantori has been discussed in Section %
\ref{Dana}, by using very precise calculations of high order periodic points
along rational chains convergent toward the Cantorus. The method used for
determining these periodic point is summarized in \textbf{Appendix B.}

\medskip

The motion along a very long trajectory shows moreover that the magnetic
line motion is \emph{intermittent }( Fig.(\ref{ExFig11})), with long stages
in the inner shell and in the \emph{outer chaotic sea}, with occasional 
\emph{sticking periods} around island remnants (Fig.(\ref{ExFig13})). These
features are characteristic of a continuous time random walk (CTRW). The
external barrier actually confines magnetic lines inside the plasma up to
the appearance of a large scale motion above a threshold of the
stochasticity parameter around $L\sim 5/2\pi $.

\medskip

In order to describe the motion of magnetic lines beyond this large scale
motion threshold, we have considered a bunch of magnetic lines in a small
area of phase space and study their statistical average evolution. We have
computed, the average radial position (\ref{averages}) of the bunch of
lines, and the mean square deviation of the radius (\ref{msdx}) compared to
this average radius : this is the quantity defining the \emph{spatial
diffusion properties} in the asymptotic stage (subdiffusion or
superdiffusion). We have also calculated the diffusion (\ref{msdpsi}) of the
poloidal flux $\psi $ : the latter appears to describe asymptotically a
classical diffusion (in $\psi $ !) and behaves as traditionally admitted, 
\emph{i.e. }like the standard map in the quasi-linear stage, with a
diffusion coefficient in action space which scales like $D\sim L^{2}$ in the
domain $L=\left[ 5/2\pi ,21/2\pi \right] $. This diffusive behavior does not
hold for magnetic lines in physical space.

\medskip

Since the action of poloidal flux $\psi $ is proportional to the square of
the radius $x$, this \emph{diffusion process in action} is equivalent to a
fourth moment of $r$. It is thus not surprising at all that the radial $r$%
-motion appears to be asymptotically a \emph{spatial subdiffusion } 
\begin{equation}
\begin{array}{llll}
<\left[ x(t)-x_{m}(t)\right] ^{2}>\Longrightarrow 2D_{x}(L)\text{ }t^{1/2} & 
& \text{with} & D_{x}(L)\sim L%
\end{array}
\label{subdifspatiale}
\end{equation}
In contradistinction with usual results for dispersion in completely chaotic
situations (quasi-linear description) , the radial motion of stochastic
magnetic lines in the present tokamap model, with incomplete chaos, thus
appears as a \emph{radial ''strange diffusion'' }and not classical
diffusion. In (\ref{subdifspatiale}) $x_{m}(t)$ is the average radius 
\begin{equation}
\begin{array}{llll}
x_{m}(t)=<\sqrt{\psi (t)}>\Longrightarrow a(L)\text{ }t^{1/4} &  & \text{with%
} & a(L)\sim L^{1/2}%
\end{array}
\label{raymoyen14}
\end{equation}
which grows asymptotically as $(L^{2}t)^{1/4}$. The scaling laws (\ref%
{asymptoticformulas}) for the coefficients of these three average processes
reveal a nice \emph{scaling} in $(L^{2}t)$ for the asymptotic diffusion
properties of the tokamap, even for strange diffusion.

\medskip

These general and asymptotic properties of magnetic line motion in a
perturbed toroidal geometry have been obtained in the tokamap model.
However, the threshold value of the stochasticity parameter $L\sim 1$ , for
which particles begin to escape from the plasma to describe large scale
motion and a true diffusion in action space, is sufficiently high, and the 
\emph{relaxation time} $t_{R}$ (\ref{trelaxat}), necessary to reach the
asymptotic stage, is sufficiently long for this asymptotic stage to be
reached when the average radial position of the bunch of lines has already
crossed the plasma edge $\psi =1$. As a consequence, the magnetic line
motion inside the plasma is \emph{either} confined by a KAM external
barrier, and the asymptotic motion can only be \emph{subdiffusive }when $L<1$
(like in Ref. \cite{SymbolDynam98}), \emph{or} escaping from the plasma when 
$L>1$, but this occurs in average at a physical time $t\varphi $ (\ref%
{tphysical}) much shorter than the relaxation time $t_{R}$, so that the
motion inside the plasma is not yet the asymptotic one (see Eq. \ref%
{transitbehaviour}), and actually describes a \emph{transient regime of
radial superdiffusion}. This last observation could seem much less favorable
for plasma confinement in perturbed magnetic structures with \emph{%
incomplete chaos} when $L>1$, but actually the resulting motion appears to
be \emph{much less dramatic} \emph{for plasma confinement} than the usual
classical expectation from the quasilinear description of classical
diffusion in \emph{completely stochastic} magnetic fields \cite{RR78}, \cite%
{KP78}, \cite{Haida}.

\bigskip

\textbf{Acknowledgements: }\emph{We want to thank R.\ MacKay, Y.\ Elskens
and E.\ Petrisor for many fruitful discussions about mathematical aspects.
Four of the authors (D.C., G.S., M.V. and F.S.) have benefitted from grants
from the french Minist\`{e}re des Affaires Etrang\`{e}res through C.E.A.
Partial support is acknowledged from NATO, Linkage Grants PST.CLG.971784 and
977397.}

\bigskip

\section{\textbf{APPENDIX A: Standard safety factor profile.}}

\medskip

Standard radial profiles can be derived, within a circular cross section, by
taking into account elementary observations in cylindrical geometry $(\rho
,\theta ,z)$. The basic assumption of the model consists in assuming a
parabolic density profile with vanishing condition on the edge: 
\begin{equation}
n(x)=n(0)\left[ 1-x^{2}\right]  \label{A1}
\end{equation}%
where the radius $x=\rho /a$ is reduced to the small radius $a$ of the
plasma. Such a profile is not inconsistent with early observations of
ohmically heated plasmas \cite{TFR 87}.

\smallskip

Next, a general relation is assumed between temperature and density
profiles. A systematic fitting of many profiles on various early tokamaks %
\cite{Capes 85} indicates that the temperature profile can adequately be
represented by squaring the density profile. Such a result can also be
obtained from simple analytical models of energy balance (see \emph{e.g.} p.
41 in \ \cite{Mis1970}).\ We thus consider : 
\begin{equation}
T(r)=T(0)\left[ 1-x^{2}\right] ^{2}  \label{A2}
\end{equation}%
which also implies $\eta _{e}\equiv d\ln T_{e}/d\ln n_{e}=2.$

\smallskip

In order to obtain the density current profile, we assume a Spitzer
dependence of the resistivity $\eta \sim T^{3/2}$ as function of the
temperature \cite{Spitzer 53} . For a stationary magnetic field, Faraday's
law implies indeed an irrotational electric field $\frac{\partial \mathbf{B}%
}{\partial t}=-\mathbf{\nabla }\times \mathbf{E}=0$ hence in an
axisymmetrical system $\frac{\partial E_{z}}{\partial \rho }=0$, \emph{i.e.}
a constant electric field along the radius. From a simple expression of the
Ohm's law we thus find that $E_{z}(\rho )=\eta (\rho )j_{z}(\rho )=cst.$ is
actually a constant along the radius $\rho $. From the temperature
dependence of the Spitzer resistivity one thus deduces in this simple model
that the electric current density profile is 
\begin{equation}
j_{z}(x)=j_{z}(0)\left[ 1-x^{2}\right] ^{3}  \label{A3}
\end{equation}
where z is the toroidal direction in cylindrical geometry. We note that Eqs.
(\ref{A1}, \ref{A2} and \ref{A3}) imply 
\begin{equation}
\frac{\left\langle j_{z}\right\rangle _{\psi }}{j_{z}(0)}=\frac{\left\langle
p\right\rangle _{\psi }}{p(0)}  \label{Minardi Eq.}
\end{equation}
where $p$ is the plasma pressure and $\left\langle ...\right\rangle _{S}$
denotes the average inside a magnetic surface $\psi $. This relation (\ref%
{Minardi Eq.})\ has been demonstrated to hold in ohmically heated discharges
of the TCV tokamak, as quoted by Minardi \cite{Minardi}\ . He proved that
the relation (\ref{Minardi Eq.}) follows from the stationarity of the
magnetic entropy, in general conditions where the ohmic tokamak is
considered as a ''dissipative open system in equilibrium which interacts
externally with the ohmic transformer''.

\medskip

We may then use Amp\`{e}re's law in cylindrical geometry 
\begin{equation}
j_{z}=\left( \mathbf{\nabla }\times \mathbf{B}\right) _{z}=\frac{1}{\rho }%
\frac{d}{d\rho }\left( \rho B_{\theta }\right)  \label{A4}
\end{equation}
and integrate the current density profile (\ref{A3}) to obtain 
\begin{equation}
B_{\theta }(x)=B_{\theta }(1)x\left( 2-x^{2}\right) \left(
x^{4}-2x^{2}+2\right)  \label{A5}
\end{equation}
where the value $B_{\theta }(1)$ on the edge is given by the total electric
current $I$ in the plasma : $B_{\theta }(1)=aj_{z}(0)/8=I/2\pi a$.

\smallskip

In the standard model for the toroidal magnetic field the safety factor $%
q(\rho )$ is defined in terms of $B_{\theta }$ and on the central value $B_{0%
\text{ }}$of the toroidal magnetic field on the magnetic axis by $%
q(x)=\varepsilon _{T}xB_{0}/B_{\theta }(x)$ so that we finally obtain the
safety factor profile in cylindrical geometry as 
\begin{equation}
q(x)=\frac{4q(0)}{\left( 2-x^{2}\right) \left( x^{4}-2x^{2}+2\right) }
\label{A6}
\end{equation}%
This standard profiles indicates a value on the edge four times larger than
on the axis $q(1)=4q(0)$ which is rather reasonable in most usual ohmic
cases. By denoting $x^{2}\Rightarrow $ $\psi $, we obtain the expression
actually used in \cite{Tmap1} : 
\begin{equation}
q(\psi )=\frac{4q(0)}{\left( 2-\psi \right) \left( \psi ^{2}-2\psi +2\right) 
}  \label{A7 qdePsi}
\end{equation}%
where $\psi $ varies from $0$ in the center to $1$ on the edge.

\smallskip \bigskip

In order to make use of traditional canonically conjugated coordinates $%
\left( r,\theta \right) $, see Eq.(\ref{newEq35bis}), we may also introduce $%
r^{2}/2=x^{2}$ 
\begin{equation}
\psi =\frac{r^{2}}{2}  \label{A8 Psider}
\end{equation}%
where the modified coordinate $r\equiv x\sqrt{2}$ varies from $0$ in the
center to $\sqrt{2}$ on the edge. The difference between $x=\rho /a$ and $%
r=\rho \sqrt{2}/a$ may simply be interpreted in terms of the $r$ as
occurring in a plasma of radius $\sqrt{2\text{.}}$ In either way the edge of
the plasma is located at $\psi =1$, $x=1$, $r=\sqrt{2}$ and we generally use
plots in terms of the radius $x$ varying from $0$ to $1$ inside the
unperturbed plasma edge.

\medskip \bigskip

\section{\textbf{APPENDIX\ B}: \textbf{Numerical method for finding periodic
points in discrete maps}}

In this appendix we explain how to determine hyperbolic and elliptic
periodic points, by a numerical algorithm derived from a generalization of
the Fletcher-Reeves method, involving the Jacobi matrix of the tokamap. More
details are given in \cite{RepGyury}.

\bigskip

The existence of points with everywhere dense trajectories and the \
possibility to approximate every point by a periodic point are two important
and apparently contradicting aspects of chaotic systems. Nevertheless, the
localization of many periodic points appears highly necessary in order to
describe the structure of the phase space. This is because a typical
Hamiltonian system appears to belong to an intermediate category between
completely integrable and globally chaotic systems. Their phase space
decomposes in a well organized architecture of invariant subdomains. In this
architecture the periodic points play the role of a skeleton. The knowledge
of their relative positions and residues, permits to determine, at least
qualitatively, the whole phase space structure.

\bigskip

The residue's absolute value (see Eq.(\ref{residue})) of hyperbolic periodic
points increases exponentially with the order of periodicity $m$. This
exponential increase imposes clear restrictions on the practical possibility
of numerical localization of the higher order hyperbolic periodic points.
The localization of periodic points approaching an invariant circle is
facilitated by a - let us say a \emph{continuity principle}, as follows.
Consider the residue of a sequence of chain of elliptic or hyperbolic
points, convergents of an irrational, generic, KAM barrier : then the
residues of periodic orbits will approach the residues of the invariant
circle, i. e. will \ approach to zero (Greene's conjecture \cite{RMacKayBook}%
). In any neighborhood of an invariant circle there are elliptic \ and
direct hyperbolic points.

A quite inverse phenomenon occurs at half-permeable barriers, the Cantori,
where the typical values of the residues tend to $-\infty $\ for direct
hyperbolic points and to $+\infty $\ for inverse hyperbolic points.

\subsection{Methods for computing periodic points (theoretical aspects)}

We precise that in the following, the \emph{angular coordinate} (denoted by $%
x_{1}$) will not be a priori reduced modulo $1$, despite the fact that in
the numerical calculations, for an increased numerical accuracy, we
separately represent the integer and the factional part of the angular
variables. After each iteration, the integer and fractional parts of the
angle are \ computed and stored separately.

\bigskip

For our purpose, in order to find a periodic point of order of periodicity $%
m $ and $q=m/n$, we must solve a system of a nonlinear equation, like 
\begin{eqnarray}
x_{1} &=&TM_{1}(\ x_{1},x_{2})-n  \label{ae} \\
\ x_{2} &=&TM_{2}(\ x_{1},x_{2})  \notag
\end{eqnarray}%
where $TM=T^{m}$\textbf{\ }is the $m$ times iterate of the original map $T$.
The first equation simply expresses the fact that the $m^{th}$ iterate $%
TM_{1}(x_{1},x_{2})$ of the angular coordinate has the same value as the
initial one ($x_{1})$ but increased by the integer number \ $n$\ of periodic
points.

\emph{The problem of localizing periodic orbit is reduced to solving a
system of nonlinear equations.}

\bigskip

\bigskip More generally, we can consider the following problem: elaborate an
algorithm for solving the system of $N$ nonlinear equations 
\begin{equation}
F_{i}(\mathbf{x})=F_{i}(x_{1},x_{2},..x_{N})=0\ \text{\ \ \ \ \ \ \ \ \ \ \
\ \ \ \ \ \ \ \ \ \ (}1\leq i\leq N\text{)}  \label{eq1}
\end{equation}
by using the values of the functions $F_{1},$\ $F_{2},...,$\ $F_{N}$.

\bigskip

There are a lot of numerical methods for solving systems of nonlinear
equations. Because of the large roundoff errors we are obliged to choose a
very rapidly convergent method. We chose the Fletcher-Reeves minimizing
method because it has a very rapid convergence (double exponential) and a
basin of attraction which is large enough.

\bigskip

Well known iterative methods seem to be not appropriate for our aim.\
Consider for instance the traditional \emph{Newton iterative method} \cite%
{NumRecipe}.\ It is known to converge quadratically (near the exact root,
the number of significant digits approximately doubles at each step) if the
initial point is chosen in the basin of attraction of the root, but the
global convergence properties are very poor and unpredictable \cite%
{NumRecipe}. The basin of attraction is usually a fat fractal set whose
boundary can be very near the root, so that the choosing the initial point
is a real problem.

In our explicit computations, for higher periodic points (of periodicity $%
m>100$), the domain of attraction (convergence) reduces so much that it
appears very difficult to localize it (as difficult as to find the periodic
point itself).

There also exist numerical methods which converge exponentially but which
have a much larger basin of attraction than the Newton method: minimizing
methods for instance.

\subsubsection{Minimizing methods}

The initial problem, that of solving the system (\ref{eq1}), is equivalent
to minimizing the function $f$ defined by 
\begin{equation}
\ f(\mathbf{x})=\sum_{j=1}^{N}(F_{j}(\mathbf{x}))^{2}.w_{j}  \label{eq3}
\end{equation}
The weights $w_{j}$ \ are positive and are chosen according to
dimensionality and relative precision criteria. In our case we simply choose 
$w_{i}=1$ .

\bigskip

Like in the case of solving nonlinear, non-algebraic equations, there exist
only iterative minimization methods, which find the relative minimum closest
to the starting point. The determination of an appropriate starting point
will be a separate problem.

\bigskip

The most efficient minimization methods make use as much as possible of the
differentiability properties of the functions to be minimized (called the 
\emph{''objective function''}).

\ 

We will consider only this class of methods. In all of this class of
methods, the important step is considered to be the choice of a direction,
along which the one-dimensional minimization\ is performed, starting from a
previous approximation. How to make the most efficient one directional
minimization, is usually not treated in the literature. In contrast, we will
treat these two problem together.

\bigskip

Let $\mathbf{d}=(d_{1},...,d_{N})$ be a direction, and let $\mathbf{a}%
=(a_{1},...,a_{N})$ be some approximation of the minimum point, to be
improved. Let us consider the following function of a single variable $t$: 
\begin{equation}
G(t)=f(\mathbf{a+d}.t)  \label{eq4}
\end{equation}
where$\ f$ is given by (\ref{eq3}) and 
\begin{equation}
\mathbf{x=a+d}.t  \label{xdt}
\end{equation}
The derivative at $t=0$ is: 
\begin{equation}
G^{\prime }(t=0)=\sum_{j=1}^{N}d_{j}.\left[ \frac{\partial f(\mathbf{x})}{%
\partial x_{j}}\right] _{\mathbf{x}=\mathbf{a}}=\langle \mathbf{d,(\mathbf{%
\nabla }f)}_{\mathbf{x}=\mathbf{a}}\rangle  \label{eq5}
\end{equation}

\subsubsection{The Laplace method}

The oldest method\emph{\ }is the \emph{Laplace}, or the so-called \emph{%
steepest descent method}.\ \ The Laplace method consists in the following
choice 
\begin{equation}
\mathbf{d=-(\nabla }f)_{\mathbf{x}=\mathbf{a}}  \label{eq6}
\end{equation}
such that $G^{\prime }(t=0)<0$ and\ that $G$ decreases along positive values
of $t$, at least at the beginning. A Laplace iteration consists in finding
the minimum $t_{1}$ along this direction, in moving to this new point $%
\mathbf{a}_{1}\mathbf{=a+d.}t_{1}$, in computing a new direction $\mathbf{d}%
_{1}\mathbf{=-(\nabla }f)_{\mathbf{x}=\mathbf{a}_{1}}$\ by (\ref{eq6}) in
the new point and in repeating this procedure up to desired precision is
reached.

\ \ 

Clearly this method cannot be the best one because it is not invariant with
respect to the coordinate changes, even for linear transforms: from (\ref%
{eq4}) $\mathbf{d}$ must be a contravariant vector, while from (\ref{eq6})
it is a covariant one. This lack of invariance appears more clearly in the
case where the $x_{i}$ variable components are physical quantities of
different kind, so that this Laplace method cannot be optimal. Nevertheless,
the method works exactly, if $f(\mathbf{x})$ \ is of the form 
\begin{equation}
f(\mathbf{x})=g_{1}(x_{1})+..+g_{N}(x_{N})  \label{eq7n}
\end{equation}
where the $g_{i}$ are functions of a single variable.

\bigskip\ 

Another drawback of the Laplace method is related with the fact that even in
the case where $f(\mathbf{x})$ is a polynomial of second order, it remains
approximative, i.e. it requires an infinite number of iterations to reach
the exact result; \emph{excepted in the case where it is of the form}\ (\ref%
{eq7n}). Even worse, the rate of convergence is of the order $O((\Lambda
-\lambda )/(\Lambda +\lambda ))$ \ where $\Lambda $ \ and $\lambda $ \ are
the largest and smallest eigenvalues of the associated quadratic form \
(they are nonnegative, otherwise the minimum would be $-\infty $ ). If \ $%
\Lambda \gg \lambda $ the convergence rate is very slow.

\bigskip

The above mentioned problems can be solved by another class of methods,
known as \emph{''conjugated direction method'' }( the etymology will be
explained below).

\subsubsection{Conjugated direction method}

The methods in this class have the following properties:

\emph{A)} If $f(\mathbf{x})$\ is a second order polynomial, then the method
is \emph{exact in N steps} \ .

\emph{B)} For computing the directions of successive one-dimensional
minimizations, at most the first order derivatives are used.

\bigskip An important consequence of property \emph{A)} is the very quick
rate of convergence, in the general, nondegenerate case: the error decreases
like 
\begin{equation}
O(\exp (-k.2^{p}))\   \label{eq8n}
\end{equation}
where\textbf{\ }$k$ is a positive constant depending on initial guess and $%
p.N$ is the total number of iterations.

\bigskip

This \ conjugated direction method consists in :

\emph{(i)} approximating locally, at each iteration step, the objective
function by a second order polynomial

\emph{(ii)}\ finding a linear change of variables so that this polynomial
becomes of the form of (\ref{eq7n}).

\bigskip

\ The step \emph{(ii)} is equivalent to finding a new coordinate system,
generated by the conjugated directions $\mathbf{d}_{1}\mathbf{,d}_{2}\mathbf{%
,...,d}_{N}$ obtained by the $N$ successive minimizations, such that 
\begin{equation}
\langle \mathbf{d}_{i}\text{ },\mathbf{Q.d}_{j}\rangle =0\ \ \text{if \ \ }%
i\neq j\ \ \   \label{eq9n}
\end{equation}%
where $\mathbf{Q}$ is the Hessian matrix, i.e. is the matrix of the second
order derivatives of the objective function. The relation (\ref{eq9n}) can
be reinterpreted in the following way. The Taylor approximation of the
objective function to be minimized can be written as 
\begin{equation}
f(\mathbf{x}_{0}\mathbf{+h})=f(\mathbf{x}_{0})+\langle \mathbf{b}\text{ }%
\mathbf{,h\rangle +\langle h}\text{ }\mathbf{,Q.h\rangle /}2\ +O(|h|^{3})\ \
\ \   \label{eq10n}
\end{equation}%
where $\mathbf{b=}\nabla f(\mathbf{x}_{0})=\mathbf{(\nabla }f\mathbf{)}_{%
\mathbf{x}_{0}}$ and $\mathbf{Q=}\left\| \frac{\partial ^{2}f}{\partial
x_{i}\partial x_{j}}\right\| _{\mathbf{x}_{0}}$. In Eq.(\ref{eq10n})\textbf{%
\ }the quadratic term $\mathbf{\langle h}$ $\mathbf{,Q.h\rangle }$\ defines
a new scalar product denoted by square brackets: 
\begin{equation}
\ \left[ \mathbf{x,y}\right] =\langle \mathbf{x}\text{ }\mathbf{,Q.y\rangle }
\label{af}
\end{equation}%
The relation (\ref{eq9n}) expresses the orthogonality $\left[ \mathbf{d}_{i}%
\mathbf{,d}_{j}\right] $ $=0$ of the conjugated directions obtained at the $%
i-th$ and $j-th$ minimization, relative to this new scalar product.
Geometrically this means that in the \emph{new coordinate system, defined by
the conjugated directions} (see below how to generate these ones), the level
surfaces (let us say in $3$ dimensions) are homothetic ellipsoids, with the%
\textbf{\ }same center, proportional semi-axis, and axes along coordinate
axes. The conjugated directions are the coordinate axes. Clearly, the
minimum (the center) is reached by a succession of $3$ steps, minimizing
along coordinates.

\ 

From the above explanation it follows that the \ $2^{d}$ order Taylor
expansion term in Eq. (\ref{eq10n})\textbf{\ }of the objective function
generates (locally) a natural, canonical Euclidean geometry, in the space of
variables which can be of very different \ significance.

\bigskip

\emph{Remark: }Unfortunately, all of these general purpose algorithms are 
\emph{local}. They require an initial point to start in order to find some
local minimum in the neighborhood ( in fact a \emph{basin of attraction} or
'' hydrographic basin'' \ to which the initial point belongs; their
boundaries are regular, not fractal like the basin of convergence of the
Newton iterations).

\bigskip

\subsubsection{The Fletcher-Reeves method}

There are different ways to generate the \emph{conjugated directions}. \ For
most of applications the optimal one is the so called \emph{Fletcher-Reeves
method } \cite{FR2}. The optimality refers to the reduced number of
operations and storage requirement of order $O(N)$ to generate the
conjugated directions, by a recursive process, using the gradients of the
objective function.

\bigskip\ 

One\ main\ iterative step (which minimizes exactly a quadratic function)
consists in $N$ sub-steps. Each \emph{sub- step}, indexed by $k$, denoted by 
$S_{k}$ consists in the following operations. Denote by $\mathbf{x}_{0}$ the
point obtained in previous step or just the initial guess of the minimum
point, by $\mathbf{d}_{0}$, the first conjugated direction.

\paragraph{Sub-step $S_{0}$ :}

At the first substep ($k=0$ ) we take 
\begin{equation}
\mathbf{d}_{0}=-\nabla f(\mathbf{x}_{0})  \label{aff}
\end{equation}
Compute the point $\mathbf{x}_{1}$ by minimizing the function $f(\mathbf{x}%
_{0}+t.\ \mathbf{d}_{0})=\psi _{0}(t)$ with respect to variable $t$. This is
performed here by using simply the Newton method (see next Section)\textbf{. 
}Let the optimal value be$\ t^{\ast }$. Then $\mathbf{x}_{1}=\mathbf{x}%
_{0}+t^{\ast }.\ \mathbf{d}_{0}$

Suppose we performed the sub-steps $S_{0}$ , ..., $S_{k}$, let $\mathbf{d}%
_{k}$ the direction computed in sub-step $S_{k}$ and let $\mathbf{x}_{k+1}$
be the point reached at the end of sub-step $S_{k}.$

\paragraph{Sub-step $S_{k+1}$ ($k<N$)}

In the point reached in the previous sub-step, a new conjugated direction, $%
\mathbf{d}_{k+1}$, is computed, according to:

\begin{equation}
\mathbf{d}_{k+1}=\ -\nabla f(\mathbf{x}_{k+1})+\beta _{k}.\ \mathbf{d}_{k}
\label{ag}
\end{equation}
where $\beta _{k}$ is defined by 
\begin{equation}
\ \beta _{k}.=\frac{\left\| \nabla f(\mathbf{x}_{k+1})\right\| ^{2}}{\left\|
\nabla f(\mathbf{x}_{k})\right\| ^{2}}  \label{ah}
\end{equation}
in which the norm is the standard Euclidean one.

\ \ \ \ \ \ \ \ \ \ \ \ \ 

\bigskip Then we find the new point \textbf{\ }$\mathbf{x}_{k+2}$ by
minimizing the function 
\begin{equation}
f(\mathbf{x}_{k+1}+t.\ \mathbf{d}_{k+1})=\psi _{k+1}(t)  \label{eq11n}
\end{equation}
with respect to the $t$ variable, and using the optimal value$\ t^{\ast }$.\
Then we compute \textbf{\ }$\mathbf{x}_{k+2}\mathbf{\ }$by$\ $ 
\begin{equation}
\mathbf{x}_{k+2}=\mathbf{x}_{k+1}+t^{\ast }.\ \mathbf{d}_{k+1}  \label{ahh}
\end{equation}

These sub-steps are continued recursively, while the sub-steps $S_{0}$ , $%
S_{1},...$ , $S_{N-1}$ are executed. After performing the sub-step $S_{N-1}$
the iteration step is completed and the approximate minimum point $\mathbf{x}%
_{N}$ \ is reached. If the objective function is a quadratic one, the point $%
\mathbf{x}_{N}$ is the exact minimum point, reached just in ONE iteration
step, with $N$ sub-steps. If the objective function is not quadratic, then
we go to the next iteration step, by using for starting point at sub-step $%
S_{0}$ the last obtained point $\mathbf{x}_{N}$, until the required
precision is reached.

\bigskip

\ It was proved in \cite{FR2} that the so-computed directions are really
conjugated to each other, i.e. they satisfy (\ref{eq9n}).

\bigskip

\ We have used this method to find periodic points, but by applying moreover
some improvement in the most frequent (consequently the most time-consuming)
steps like the one-dimensional minimization.

\subsubsection{The necessity to improve the one-dimensional minimization}

\ 

\ By Eq.(\ref{eq8n}) - where $p$ is the number of full iteration steps (each
step\ involving $N$ substeps)-, the speed of convergence is very high, let
us say \textbf{\ }like\textbf{\ }a double exponential (exponential of
exponential in the number of iterations\textbf{)}. In contrast, if we use
for the one-dimensional minimization only the values of the function ($f(%
\mathbf{x}_{k}+t.$\ $\mathbf{d}_{k})=\psi (t)$) and of the first order
derivatives, the speed of convergence in the one-directional Newton
minimization is only (simply)\ exponential. Consequently \ the overall
convergence will be slowed down to simple exponential rate.

\bigskip

\ The general minimization problem for the function $f(\mathbf{x}_{k}+t.$\ $%
\mathbf{d}_{k})=\psi _{k}(t)$ (\ref{eq11n}) can be reduced to solving the
equation 
\begin{equation}
\psi _{k}^{\prime }(t)=0  \label{psiprime}
\end{equation}
and by Newton method one obtains $\psi _{k}^{\prime }(t_{i})+\psi
_{k}^{\prime \prime }(t_{i}).(t_{i+1}-t_{i})=0$ which requires to compute
the second order \ partial derivatives of the objective function. If we can
compute $\psi "(t)$, then the one-dimensional optimization can be
accelerated by the one-dimensional Newton method which gives: 
\begin{equation}
t_{i+1}=t_{i}-\psi _{k}^{\prime }(t_{i})/\psi _{k}^{\prime \prime }(t_{i})
\label{ai}
\end{equation}
where the index $i$ labels the Newton iterations at substep $k$\textbf{. }In
the neighborhood of an optimal point its convergence is very fast, i.e
double exponential with respect to the number of Newton iterations.

\bigskip

Nevertheless in our problem, concerning high order periodic points, this
method is very difficult to use: in order to compute $\psi ^{\prime \prime
}(t_{k})$ we must compute the second order derivatives of the $m-$times
iterated map $T$ (the Tokamap for instance \cite{Tmap1}), which leads not
only to a complicated program, but rather to a very CPU time-consuming
program in the case of large values of $m$.

\subsection{The improved one-dimensional minimization}

\ We will see how to avoid the computation of second order derivatives,
without destroying the double exponential convergence. This is the main
improvement we bring to the Fletcher-Reeves method. Such a ''trick'' works
only in the case of the minimization problems arising from the reduction
method, discussed\ above: we will suppose that the functional to be
minimized is of the form of Eq. (\ref{eq3}).$\ $Let $\psi (t)$ be one of the
functions $\psi _{k}(t)$, and $\mathbf{d}$ one of the directions $\mathbf{d}%
_{k}.$%
\begin{equation}
\ \ \psi (t)=f(\mathbf{x}+t.\ \mathbf{d})  \label{ak}
\end{equation}
\ where the vector $\mathbf{d}$ is one of the conjugated directions\textbf{. 
}Then 
\begin{equation}
\psi ^{\prime }(t)=\sum_{i,j=1}^{N}2\text{ }w_{j}\left[ \frac{\partial F_{j}%
}{\partial x_{i}}\right] _{\mathbf{x}+t.\mathbf{d}}.F_{j}(\mathbf{x}+t.\ 
\mathbf{d}).d_{i}  \label{eq14a}
\end{equation}
and

\begin{equation}
d^{2}\psi /dt^{2}=2\left[ g(t)+\sum_{i,j,k=1}^{N}w_{j}\left[ \frac{\partial
F_{j}}{dx_{i}}.\frac{\partial F_{j}}{\partial x_{k}}\right] _{\mathbf{x}+t.%
\mathbf{d}}.d_{i}\ d_{k}\right]  \label{eq12n}
\end{equation}%
where 
\begin{equation}
g(t)\equiv \sum_{i,j,k=1}^{N}w_{j}\left[ \frac{\partial ^{2}F_{j}}{\partial
x_{i}.\partial x_{k}}\right] _{\mathbf{x}+t.\mathbf{d}}.F_{a}(\mathbf{x}+t.%
\mathbf{d}).d_{i}\ d_{k}  \label{eq17a}
\end{equation}%
For the case of searching periodic points, the computation of first order
derivatives reduces to the computation of the products of Jacobi matrices.
On the other hand, the second term (\ref{eq17a}) $\frac{\partial ^{2}F_{j}}{%
\partial x_{i}\partial x_{k}}.F_{j}$\ is small near the solution, because of
its last factor, and will be neglected. It can be proved that the
so-obtained iterative process still has a double exponential convergence.

\subsubsection{Summary of the improved Fletcher Reeves method.}

Shortly\textbf{, }the essence of our method is the following: The optimal
value of $t^{\ast }$ where the minimum of

\begin{equation}
f(\mathbf{x}_{k+1}+t.\ \mathbf{d}_{k+1})=\psi _{k+1}(t)\equiv \psi (t)
\end{equation}
is reached, is approximated by

\begin{equation}
t^{\ast }=-\frac{\psi ^{\prime }(0)}{\Delta (0)\ }\ \   \label{eq13}
\end{equation}
where $\psi ^{\prime }(t)$ is given by (\ref{eq14a}) and where we have
introduced 
\begin{equation}
\Delta (t)=\sum_{i,j,l=1}^{N}w_{j}.\frac{\partial F_{j}}{\partial x_{i}}.%
\frac{\partial F_{j}}{\partial x_{k}}.d_{k+1,i}\ d_{k+1,l}  \label{eq15n}
\end{equation}
where the $d_{k+1,l}$ are defined from the $l^{th}$ component ($l=1,2,...,N$%
) of the vector at step $k+1$, i.e. from: 
\begin{equation}
\mathbf{d}_{k+1}=\left( d_{k+1,1},...,d_{k+1,l},...,d_{k+1,N}\right)
\label{dindexdouble}
\end{equation}

In the rest, the steps follows the standard Fletcher-Reeves strategy. Our
modification:\emph{\ }

\emph{(a)} preserves the same double exponential convergence speed and

\emph{(b)} saves computer CPU time, for the following reasons:

\qquad \emph{(b1)} we do not compute the second derivatives of $F_{j}$.
These computations in the case of periodic point search require CPU times
which increase quadratically with $m$, the periodicity order

\ \ \ \ \ \ \emph{(b2)} we perform SINGLE iterations, instead of many (at
least four) Newton iterations at each substep. The CPU time is reduced at
least by a factor of four.

\emph{(c) }is not unstable, do not produce overflow and has larger domain of
convergence, compared to Newton method.

We observed a single drawback: the tendency to remain ''suspended'' in some
local minima, for larger order periodic points (hundreds).

This problem was corrected by refining the output (i.e. periodic point
coordinates) of the conjugated gradient program by another new program. This
program now uses the classical Newton method. The instability of the Newton
method, mentioned above, does not show up, because the starting point
(initial guess for the Newton method) is already very close to the solution
and it is in the domain of convergence (or attraction basin).

\subsubsection{General strategy for periodic point search (programming
details)}

In order to localize a periodic points of periodicity $m$ and having a given
type $\left( n,m\right) $ we must realize the following steps:

\emph{1) to find an initial point for the minimizing method.}

This is achieved in our program by a preliminary search for the absolute
minimum of the objective function on a finite grid inside the input
tetragon. The coordinates of the grid's point giving the lowest value serve
as input for conjugated gradient search.

The vertices of this tetragon, for the lowest order periodic points, are
obtained from preliminary graphics of tokamap phase portrait. For highest
order, the input tetragon was obtained recurrently form previous lowest
order periodic points.

\emph{2) to minimize the objective function}, using the improved
Fletcher-Reeves method, with the starting point determined by \emph{1)}.

\bigskip \bigskip

For this we used a \textbf{C++ }program which\textbf{\ } contains the
improved Fletcher-Reeves method\textbf{.}

If the output error of the conjugate gradient program is too large (as
compared with a maximal imposed error), the results can be refined by the
Newton iteration program, the input of which is the output of previous
program.

The output of the program contain: the coordinates of the searched periodic
point.

The program computes the safety factor related to the magnetic axis (i.e. $m$
and $n$) and the error (by comparing the initial and final positions of the
chain of periodic points). The cross-check of numerical floating-point
accuracy is the deviation of the value of the determinant of the Jacobi
matrix from\ its known unity value (in the output\ it is named as
''determinant'' error). The complete program files can be obtained by
sending a mail to one of the authors \footnote{%
stein@ns.central.ucv.ro}.

The first program computes the safety factor related to the magnetic axis,
i.e. $m$ and $n$. This program, with simple modifications, can be used in
different cases, in order:

\emph{a)} to find periodic points of 2 dimensional maps,

$\emph{b)}$ to solve general nonlinear system of equations,

\emph{c) }in case of field-line tracking, to find the periodic field lines.
In this case the differential equations of the field lines must be completed
with the so called Jacobi variational equations (i.e. the linearized
equations), or more generally

\emph{d)} to find the periodic trajectories of the dynamical systems with
finite degree of freedom, described by discrete iterative maps or by
differential equations.

\end{document}